\documentclass[10pt]{iopart}
\usepackage{iopams}
\usepackage{psfrag}
\usepackage{pstricks}
\usepackage{array}
\usepackage{lscape}
\usepackage{graphics}
\usepackage{epsfig}

\newcommand{\sgn}{\mathrm{sgn}}
\renewcommand{\Re}{\mathrm{Re}}
\renewcommand{\Im}{\mathrm{Im}}
\renewcommand{\vec}{\boldsymbol}
\newcommand{\veccal}[1]{\pmb{\mathcal{#1}}}

\newcommand{\op}[1]{\overline{\vec{#1}}}

\newcommand{\opcal}[1]{\overline{\pmb{\mathcal{#1}}}}

\newcommand{\ep}{\epsilon}

\newcommand{\p}[1]{\vec{p}_{#1}}
\newcommand{\Ps}[1]{\vec{P}_{#1}}
\newcommand{\Psd}[1]{\frac{\vec{P}_{#1}}{2}}

\newcommand{\hvec}[1]{\hat{\vec{#1}}}
\newcommand{\ehp}[1]{\hvec{e}_H(\p{#1})}
\newcommand{\evp}[2]{\hvec{e}_V^{#1}(\p{#2})}

\newcommand{\ex}{\hvec{e}_x}
\newcommand{\ey}{\hvec{e}_y}
\newcommand{\ez}{\hvec{e}_z}

\newcommand{\Der}[2]{\frac{\partial #1 }{\partial #2}}

\newcommand{\iint}{\int\!\!\!\int}

\newcommand{\intr}[1]{\int  \rmd^3 \rr_{#1}\,}
\newcommand{\iintr}[2]{\int  \rmd^3 \rr_{#1}\, \rmd^3 \rr_{#2}\,}

\newcommand{\intp}[1]{\int \frac{\rmd^2 \vec{p}_{#1}}{(2\pi)^2}\,}

\newcommand{\intpV}[2]{\int_{\scriptstyle{#2}} \frac{\rmd^2 \vec{p}_{#1}}{(2\pi)^2}\,}
\newcommand{\iintpV}[3]{\int \!\!\! \int_{\scriptstyle{#3}}\frac{\rmd^2 \vec{p}_{#1}}{(2\pi)^2}\frac{\rmd^2 \vec{p}_{#2}}{(2\pi)^2}\,}

\newcommand{\intk}[1]{\int \frac{\rmd^3 \vec{k}_{#1}}{(2\pi)^3}\,}

\newcommand{\intK}[1]{\int \frac{\rmd^3 \vec{K}_{#1}}{(2\pi)^3}\,}

\newcommand{\no}{\nonumber}

\newcommand{\x}{\vec{x}}

\newcommand{\rr}{\vec{r}}
\newcommand{\RR}{\vec{R}}
\newcommand{\E}{\vec{E}}

\newcommand{\kk}{\vec{k}}
\newcommand{\KK}{\vec{K}}

\newcommand{\alp}[2]{\alpha_{#1}(\vec{p}_{#2})}
\newcommand{\alpime}[2]{\alpha'_{#1}(\vec{p}_{#2})}

\newcommand{\Gue}[1]{\op{G}^{#1}_{S}}
\newcommand{\GVu}[1]{\op{G}^{#1}_{SV}}

\newcommand{\RpS}[3]{\op{S}^{\,#1}\left(\p{#2}\right|\left.\p{#3}\right)}

\newcommand{\RS}[1]{\op{S}^{\,#1}}

\newcommand{\Ical}{\veccal{I}}

\newcommand{\Gcal}{\opcal{G}}

\newcommand{\Gamcal}{\op{\Gamma}}
\newcommand{\Pcal}{\opcal{P}}

\newenvironment{vecteur}{\left( \begin{array}{c}}{\end{array} \right)}
\newenvironment{matrice}{\left(\begin{array}{cc}}{\end{array} \right)}

\newlength{\Hlarger}
\begin{document}

\title[Electromagnetic wave scattering from a random layer with rough interfaces II]{Electromagnetic wave scattering from a random layer with rough interfaces II:  Diffusive intensity}

\author{Antoine Soubret\dag 
\footnote[3]{asoubret@hms.harvard.edu} and G\'erard Berginc\ddag}

\address{\dag\ NOAA, Environmental Technology Laboratory, 325 Broadway, Boulder, CO 80303}

\address{\ddag\ Thal\`es Optronique, Bo\^{\i}te Postale 55, 78233 Guyancourt Cedex, France}

\begin{abstract}
A general approach for the calculation of the incoherent  intensity scattered by
a random medium with rough boundaries has been developed using a Green function formalism.
The random medium consists of spherical particles whose  physical repartition 
is described by a pair-distribution function. The boundary contribution is included in the Green 
functions with the help of scattering operators which can represent any existing theory of scattering by rough surfaces.
By using a standard procedure, we derive the integral Bethe-Salpeter equation under the ladder approximation and, by differentiation,
the vectorial radiative transfer equation. Furthermore, with the help of our formalism, the boundary conditions necessary to solve 
the radiative transfer equation are expressed in terms of the scattering operators of the 
rough surfaces. Finally, using the reciprocity  properties of the Green functions, we are able to include the enhanced 
backscattering contributions to take into account every state of polarization of the incident and the scattered waves. 

\end{abstract}



\maketitle
\section{Introduction}
In the preceding paper~\cite{Soubret3} (referred as I), we have developed a
general formalism based on Green functions to calculate the electromagnetic field
scattered by a random medium with rough boundaries. 
 In using these Green functions, we have 
determined  the average electric
field, also named  the coherent field. In this paper, we use the formalism developed in I to calculate
the diffusive intensity also named incoherent intensity. 

The study of electromagnetic wave propagation through random media has been an intensive field of research
for many decades~\cite{Chandra,Kourganoff,Sobolev,Case,Ishi1,Ulaby3,Lenoble,Kong,Duder,Hulst2,Fung,Apresyan,Thomas,Kozlov,Kong2001-1}.
Many aspects of the transport of waves in such media are well described by the phenomenological radiative
transfer theory~\cite{Chandra,Case,Ishi1,Ulaby3,Kong,Duder,Thomas,Kong2001-1}.
In this approach, the main quantity, used to describe the propagation, is the specific intensity $\mathcal{I}(\vec{R},\hvec{k})$ (if we take into account the polarization of the wave, the specific intensity can be defined as a Stokes vector or a tensor) which gives the power flux per unit area and solid angle at the point $\vec{R}$ which goes in the direction $\hvec{k}$. 
In writing   a balance equation on the energy, it is shown that the specific intensity satisfies a Boltzmann type equation called radiative transfer equation. If the particles
are inside a slab with a permittivity different from the surrounding medium, boundary conditions must be added to the radiative transfer equation in order to calculate the specific intensity. For rough surfaces, these boundary conditions are expressed with scattering  operators, where several approximate analytical expressions exist
depending on the roughness of the surface. Numerical calculations of the radiative transfer
equation taking into account rough boundaries can be found in references~\cite{Ulaby3,Fung,Lam1,Lam3}

However, all previous studies on the radiative transfer theory are based on heuristic
principles since in these works, the specific intensity is  a fundamental
quantity which is not defined from the
electromagnetic field $\vec{E}$ and $\vec{B}$.
The link between the electromagnetism and the classical radiometry theory is due to  Walther~\cite{Walther} who has first recognized that the specific
intensity can be deduced from the Wigner function 
$\veccal{I}(\vec{R},\vec{k})=Const\times\intr{} \exp(-\rmi\vec{k}\cdot\rr)\,\vec{E}(\vec{R}+\rr/2)\otimes \vec{E}^*(\vec{R}-\rr/2)$,
 where the tensorial product $\otimes$ permit to take into account the different polarizations of the waves. If we suppose that the electric field depends slightly on the coordinates $\vec{R}$ compared to $\rr$ (quasi-uniform fields hypothesis~\cite{Apresyan}), it is shown that the Wigner function is written as $\veccal{I}(\vec{R},\vec{k})=Const\times\delta(||\vec{k}||-K_0)\,\veccal{I}(\vec{R},\hvec{k})$, where $\hvec{k}=\vec{k}/||\vec{k}||$ and $K_0$ the wavenumber. In this case, the function $\veccal{I}(\vec{R},\hvec{k})$ can be identified with the specific intensity. Nevertheless, there are some differences between
 this definition of the specific intensity with the radiometric one. In fact, this electromagnetic definition of the specific intensity is not always a positive function, which
 is in contradiction with the radiometric interpretation in terms of power.
This problem has been the subject of much study~\cite{Apresyan,Friberg,Dragoman} where it has been demonstrated that in the geometrical limit, the electromagnetic definition of the specific intensity is always positive. In other cases, the negative value of the specific intensity are due to interference effects.

As soon as we have determined the relationship between the electromagnetic field and the specific intensity, we can derive the radiative transfer equation from the Maxwell equations. The procedure  
is to write the Maxwell equations in an integral form with the help of Green functions and
to apply the Wigner transform to the equation derived. Then, in differentiating this equation
we  obtain the radiative transfer equation~\cite{Ishi2,Rytov4,Sheng1,Sheng2,POAN,Rossum,Lag1,Kuz,Bara1,Mark,Tig1,Kong2001-2,Kong2001-3,Mish}. Furthermore, in starting from the wave equations,  we are able to take into account
new contributions to the scattered intensity such as the enhanced backscattering and the correlations between the scatterers that can not be introduced in  the phenomenological
radiometric approach.
The objective of this   paper is to derive the radiative transfer equation from the wave equation by taking into account rough boundaries. Several works have investigated
this topics, but they describe the scattering by the rough surfaces either by using the small-perturbation method~\cite{Mudaliar1,Mudaliar2,Mudaliar3} or in an  unconventional fashion~\cite{Furutsu1,Furutsu2}. In our approach, we use  scattering operators which are a versatile and unified way to describe how the wave interacts with the boundaries ~\cite{Voro}.
To use these operators, we have introduced two kinds of Green functions (I). The first one $\op{G}_{SV}(\rr,\rr_0)$ 
describes the field scattered by the volume (V), which contains the scatterers, and by the rough surfaces (S). The second type of  Green function is $\op{G}_{S}(\rr,\rr_0)$, which describes the field scattered
by a slab with rough boundaries where the scatterers have been replaced by an \emph{homogeneous} medium of permittivity $\ep_e$ named effective permittivity. As demonstrated in I, the Green functions $\op{G}_{S}(\rr,\rr_0)$ are easily expressed as a function of the rough surface scattering operators. In using these kind of Green functions, we were able 
to separate the contribution coming from the surface and the volume.
The main advantage of our approach is that the equations obtained are similar to the equations generally used to describe the wave scattered  by an infinite random medium~\cite{Kong,Sheng1,Sheng2,Rossum,Kong2001-3}. 
To take into account the boundaries, we replace the Green function of an infinite
homogeneous random medium with $\op{G}_{S}(\rr,\rr_0)$.
Yet, there is a slightly difference from the classical procedure. Usually, we  use 
the normalized vector $\hvec{k}$ ($||\hvec{k}||=1$) to describe the propagation direction of the wave
since the  generalized Dirac function in the specific intensity $\opcal{I}(\rr,\vec{k})$ imposes that $\vec{k}=K_0\,\hvec{k}$. In this study, we will use the two-dimensional vector $\p{}=k_x\ex+k_y\ez$ to describe the propagation direction, and we will recover the vector $\hvec{k}$ by following decomposition:
$K_0\hvec{k}=\p{}+a\alp{0}{}\ez$, where $\alpha_0(\p{})=(K_0^2-\p{}^2)^{1/2}$, and $a$ is the sign of the vertical component of $\hvec{k}$ ($a=\sgn(\hvec{k}\cdot\ez)$). This choice, which is unusual for the radiative transfer theory, is in fact the standard  in  scattering by rough boundaries
where the vertical axis $z$ has to be differentiated from the axis $\vec{x}=x\ex+y\ey$ for a  surface profile defined by $z=h(\x)$~\cite{Voro}. Furthermore, the distinction between upward waves for $a=+$ and downward waves for $a=-$ is useful to write the boundary conditions. At the end of this paper, we will explain how to rewrite the equations obtained in the usual form with $\hvec{k}$ vectors.
\section{Cross-section}
\label{MullerDef}
The geometry of the problem and the notation are described in  paper I.
In order to characterize the scattered intensity by an object, we usually introduce
the bistatic cross-section, which is  the power scattered per solid angle normalized
by the incident power flux. In this paper, we will use a generalization of this concept called
Muller bistatic cross-section, which permits an accounting for every  state of polarization of the incident and scattered waves. First, as an intermediate of calculation, we have to introduce the scattering operators describing the field scattered by the rough surfaces and the random media~\footnote{Notice that in the paper I, we have introduce the scattering operators describing the field scattered by an \emph{homogeneous}  slab with rough boundaries. However, we can also define, in the same way, scattering operators
when the medium is inhomogeneous.}.
For an incident plane wave,
\begin{equation}
\vec{E}^{0i}(\rr)=\vec{E}^{0i}(\p{0})\,\e^{\rmi\,\p{0}\cdot\x-\rmi\,\alp{0}{0}z}\,,
\end{equation}
 the field scattered by the random medium and the rough boundaries is:
\begin{equation}
\vec{E}^{0s}_{SV}(\rr)=\intp{}\e^{\rmi\,\p{}\cdot\x+\rmi\,\alp{0}{}z}\,\op{R}_{SV}(\p{}|\p{0})\cdot\E^{0i}(\p{0})\,,\label{defRSV-1}
\end{equation}
where $\alp{0}{}=\sqrt{K_0^2-\p{}^2}$, and  using the notation defined in I, we decompose the vector $\vec{E}^{0i}(\p{0})$ and the dyad $\op{R}_{SV}(\p{}|\p{0})$ on the following basis:
\begin{eqnarray}
\fl\vec{E}^{0i}(\p{0})&=\sum_{\beta=H,V} E^{0i}(\p{0})_\beta\,\hvec{e}^{0-}_\beta(\p{0})\\
\fl\op{R}_{SV}(\p{}|\p{0})&=\sum_{\beta,\beta_0=H,V}R_{SV}(\p{}|\p{0})_{\beta,\beta_0}\,\hvec{e}^{0+}_{\beta}(\p{})\hvec{e}^{0-}_{\beta_0}(\p{0})\,,\label{dyadRSV}
\end{eqnarray}
where the polarization vector $\hvec{e}^{0\pm}_{\beta_0}(\p{})$ for the polarization TM ($\beta=V$) and the polarization TE ($\beta=H$) are defined by 
\begin{equation}
\fl \evp{0\pm}{}=\pm\frac{\alpha_0(\p{})}{K_0}\hvec{p}-\frac{||\p{}||}{K_0}\ez\,,\qquad
\ehp{}=\ez\times\hvec{p}\,.
\end{equation}
Far from the scattering medium ($K_0||\rr||\gg 1$), we can obtain an asymptotic expression of the integral in  equation \eref{defRSV-1} by using the stationary phase approximation~\cite{Born}:
\begin{equation}
\vec{E}^{0s}_{SV}(\rr)=\frac{\e^{\rmi K_0||\rr||}}{||\rr||}\,\op{f}(\p{}|\p{0})\cdot\E^{0i}(\p{0})\,,\label{eq1}
\end{equation}
with
\begin{eqnarray}
 \p{}= K_0\,\frac{\x}{||\rr||}\,,\label{pfar}\\
 \op{f}(\p{}|\p{0})= \frac{\alp{0}{}}{2\pi\rmi}\op{R}_{SV}(\p{}|\p{0})\,.
\end{eqnarray}
From expression \eref{eq1}, we can decompose the vector $\vec{E}^{0s}_{SV}(\rr)$ on the following
basis:
\begin{equation}
\vec{E}^{0s}_{SV}(\rr)=E^{0s}_{SV}(\rr)_V\,\evp{0+}{}+E^{0s}_{SV}(\rr)_H\,\ehp{}\,,
\end{equation}
where the vector $\p{}$ is defined by \eref{pfar}.
The incident and scattered intensity can be described with the help of Jones tensors~\cite{Apresyan}:
\begin{eqnarray}\fl
\veccal{J}^{0s}(\rr)&=e_0\,\E^{0s}_{SV}(\rr)\otimes\E^{0s\,*}_{SV}(\rr)\,,\\\fl
&=e_0\,\sum_{\beta,\beta'=H,V} E^{0s}_{SV}(\rr)_{\beta}E^{0s\,*}_{SV}(\rr)_{\beta'}\,\hvec{e}^{0+}_{\beta}(\p{})\otimes\hvec{e}^{0+}_{\beta'}(\p{})\,,\\\fl
\veccal{J}^{0i}(\p{0})&=e_0\,\E^{i}(\p{0})\otimes\E^{0i\,*}(\p{0})\,\\\fl
&=e_0\,\sum_{\beta_0,\beta'_0=H,V} E^{0i}(\p{0})_{\beta}E^{0i\,*}(\p{0})_{\beta'_0}\,\hvec{e}^{0-}_{\beta_0}(\p{0})\otimes\hvec{e}^{0-}_{\beta'_0}(\p{0})\,,
\end{eqnarray}
often written in matrix form~\cite{Hecht}:
\begin{eqnarray}
\fl [\veccal{J}^{0s}(\rr)]=e_0\,\begin{matrice}
|E^{0s}_{SV}(\rr)_V|^2 & E^{0s}_{SV}(\rr)_V [E^{0s}_{SV}(\rr)_H]^*\\
E^{0s}_{SV}(\rr)_V [E^{0s}_{SV}(\rr)_H]^* & |E^{0s}_{SV}(\rr)_V|^2
\end{matrice}\,,\\
\fl [\veccal{J}^{0i}(\rr)]=e_0\,\begin{matrice}
|E^{0i}(\rr)_V|^2 & E^{0i}(\rr)_V [E^{0i}(\rr)_H]^*\\
E^{0i}(\rr)_V [E^{0i}(\rr)_H]^* & |E^{0i}(\rr)_V|^2
\end{matrice}\,.
\end{eqnarray}
Here 
\begin{equation}
e_0=\frac{\ep_{vac}\,c_{vac}\,n_0}{2}
\end{equation}
and $\ep_{vac}$, $c_{vac}$  are, respectively, the permittivity and the speed of light in the 
vacuum, and $n_0=\sqrt{\ep_0}$ the optical index of the medium 0.
The incident and scattered Jones tensors are related by the tensorial cross-section defined by 
\begin{equation}
\op{\sigma}(\p{}|\p{0}):\veccal{J}^{0i}(\p{0})=\frac{4\pi}{A}||\rr||^2\,\veccal{J}^{0s}(\rr)\,,
\label{deftensorcross}
\end{equation}
where $A$ is the area illuminated by the incident wave and $:$ is the product between two tensors as defined in \ref{AppTensor}. In the vectorial basis $[\evp{0\pm},\ehp{}]$, we have
\begin{equation}
\fl \sum_{\beta_0,\beta_0'=H,V}\sigma(\p{}|\p{0})_{\beta\beta';\beta_0\beta_0'}\,E^{0i}_{\beta_0}(\p{0})E^{0i\,*}_{\beta_0'}(\p{0})=\frac{4\pi}{A}||\rr||^2\,E^{0s}_{SV}(\rr)_{\beta} E^{0s\,*}_{SV}(\rr)_{\beta'}
\end{equation}
where 
\begin{equation}
\fl \op{\sigma}(\p{}|\p{0})=\sum_{\beta,\beta';\beta_0,\beta_0'=H,V}\,\sigma(\p{}|\p{0})_{\beta\beta';\beta_0\beta_0'}(\hvec{e}^{0+}_\beta(\p{})\otimes\hvec{e}^{0+}_{\beta'}(\p{}))(\hvec{e}^{0-}_{\beta_0}(\p{0})\otimes\hvec{e}^{0-}_{\beta_0'}(\p{0}))\,.
\end{equation}
Definition \eref{deftensorcross} is obviously a generalization of the usual scattering cross-sections since
the elements $\op{\sigma}_{VV;VV}$, $\op{\sigma}_{VV;HH}$, $\op{\sigma}_{HH;VV}$, and $\op{\sigma}_{HH;HH}$ of the tensor $\op{\sigma}_{\beta\beta';\beta_0\beta_0'}$  are given by
\begin{eqnarray}
\op{\sigma}_{VV;VV}=\frac{4\pi}{A}\,\frac{||\rr||^2|E^{0s}_{SV}(\rr)_V|^2}{|E^{0i}(\p{0})_V|^2}\,,\\
\op{\sigma}_{VV;HH}=\frac{4\pi}{A}\,\frac{||\rr||^2|E^{0s}_{SV}(\rr)_V|^2}{|E^{0i}(\p{0})_H|^2}\,,\\
\op{\sigma}_{HH;VV}=\frac{4\pi}{A}\,\frac{||\rr||^2|E^{0s}_{SV}(\rr)_H|^2}{|E^{0i}(\p{0})_V|^2}\,,\\
\op{\sigma}_{HH;HH}=\frac{4\pi}{A}\,\frac{||\rr||^2|E^{0s}_{SV}(\rr)_H|^2}{|E^{0i}(\p{0})_H|^2}\,,
\end{eqnarray}
which are the usual scattering cross-sections per unit area $\sigma_{VV}$, $\sigma_{VH}$, $\sigma_{HV}$ and $\sigma_{HH}$.
From equation \eref{eq1}, we deduce that
\begin{equation}
\veccal{J}^{0s}(\rr)=\frac{1}{||\rr||^2}(\op{f}(\p{}|\p{0})\otimes\op{f}^*(\p{}|\p{0})):\veccal{J}^{0i}(\p{0})
\end{equation}
if the medium 0 is non-absorbing ($\ep_0\in \mathbb{R}^{+}$).
The cross-section $\op{\sigma}$ is then
\begin{eqnarray}
\op{\sigma}(\p{}|\p{0})&=\frac{4\pi}{A}\op{f}(\p{}|\p{0})\otimes\op{f}^*(\p{}|\p{0})\,,\\
&=\frac{\alpha^2_0(\p{})}{A\,\pi}\,\op{R}_{SV}(\p{}|\p{0})\otimes\op{R}^*_{SV}(\p{}|\p{0})\,.
\end{eqnarray}
For  a random medium and rough surfaces described statistically,  we usually separate the scattering contribution into a coherent $\op{\sigma}^{coh}$ and an incoherent part $\op{\sigma}^{incoh}$:
\begin{eqnarray}
\fl \op{\sigma}^{coh}(\p{}|\p{0})=\frac{\alpha^2_0(\p{})}{A\,\pi}\,\ll\op{R}_{SV}(\p{}|\p{0})\gg_{SV}\otimes\ll[\op{R}_{SV}(\p{}|\p{0})]\gg_{SV}\,,\\
\fl\op{\sigma}^{incoh}(\p{}|\p{0})=\frac{\alpha^2_0(\p{})}{A\,\pi}\,\Big[\ll\op{R}_{SV}(\p{}|\p{0})\otimes\op{R}_{SV}(\p{}|\p{0})
\gg_{SV} \nonumber\\
\lo  -\ll\op{R}_{SV}(\p{}|\p{0})\gg_{SV}\otimes\ll\op{R}_{SV}(\p{}|\p{0})\gg_{SV}\Big]\,
\end{eqnarray} 
where $\ll \gg_{SV}$ is the average over the surface and the volume disorder.
 For statistically homogeneous random medium and surfaces,
the average of the scattering operators $\op{R}_{SV}(\p{}|\p{0})$ contains a Dirac distribution (\ref{AppStat}), and we define
a tensor $\op{R}_{\ll SV\gg}$ by
\begin{equation}
(2\pi)^2\delta(\p{}-\p{0})\,\op{R}_{\ll SV\gg}(\p{0})=\ll \op{R}_{SV}(\p{}|\p{0})\gg_{SV}\,.\label{defRcoh}
\end{equation}
Similarly, the average of the tensorial product  also contains  a Dirac distribution, and we
introduce the tensor $\opcal{R}^{incoh}_{\otimes\ll SV\gg}$ such that
\begin{eqnarray}
\fl (2\pi)^2\delta(\vec{0})\,\opcal{R}^{incoh}_{\otimes\ll SV\gg}(\p{}|\p{0})=& \ll\op{R}_{SV}(\p{}|\p{0})\otimes \op{R}_{SV}(\p{}|\p{0})\gg_{SV}\nonumber\\
&  -\ll\op{R}_{SV}(\p{}|\p{0})\gg_{SV}\otimes \ll\op{R}_{SV}(\p{}|\p{0})\gg_{SV}\,.
\label{defodotRSV}
\end{eqnarray}
Accordingly, the coherent and incoherent bistatic cross-section are
\begin{eqnarray}
\fl \op{\sigma}^{coh}(\p{}|\p{0}) =\frac{\alpha_0^2(\p{})}{\pi}\,(2\pi)^2\delta(\p{}-\p{0})\op{R}_{\ll SV\gg}(\p{0})\otimes\op{R}_{\ll SV\gg}(\p{0})\,,\\
\fl \op{\sigma}^{incoh}(\p{}|\p{0}) =\frac{\alpha_0^2(\p{})}{\pi}\,\opcal{R}^{incoh}_{\otimes\ll SV\gg}(\p{}|\p{0})\,,\label{expCross}
\end{eqnarray} 
where we have used the fact that $\delta(\vec{0})=A/(2\pi)^2$ for a finite patch of area A \eref{AppStat}. 
The coherent component $\op{\sigma}^{coh}$ is directed only in the specular direction due to the Dirac distribution.
In paper I, we have described how to calculate the coherent electric field,
and we have obtain that
\begin{equation}
\ll \vec{E}^{0s}_{SV}(\rr)\gg_{SV}=\op{S}^{coh}(\p{0})\cdot\vec{E}^{0i}(\p{0})\,\e^{\rmi\,\p{0}\cdot\x+\rmi\alp{0}{0}\,z}\,.\label{eqE0s}
\end{equation}
Comparing equation \eref{eqE0s} with the average of  equation \eref{defRSV-1} and by using  definition \eref{defRcoh}, we obtain
\begin{equation}
\op{R}_{\ll SV\gg}(\p{0})=\op{S}^{coh}(\p{0})\,.
\end{equation}
and the coherent cross-section is
\begin{equation}
\op{\sigma}^{coh}(\p{}|\p{0})=\frac{\alpha_0^2(\p{})}{\pi}\,(2\pi)^2\delta(\p{}-\p{0})\op{S}^{coh}(\p{0})\otimes\op{S}^{coh}(\p{0})\,.
\end{equation}
As was demonstrated in I, the dyad  $\op{S}^{coh}(\p{0})$ is related to the average
scattering matrix $\op{S}^{0+0-}(\p{}|\p{0})$ by the following relationship:
\begin{equation}
<\op{S}^{0+0-}(\p{}|\p{0})>_S=(2\pi)^2\,\delta(\p{}-\p{0})\,\op{S}^{coh}(\p{0})\,.
\label{eqS0+0-}
\end{equation}
The dyad $\op{S}^{0+0-}(\p{}|\p{0})$ describes the fields scattered by an homogeneous
slab of permittivity $\ep_e$ with rough boundaries. The Dirac distribution in  equation \eref{eqS0+0-} is due to the  statistical homogeneity of the rough surfaces.
The coherent component $\op{\sigma}^{coh}$ is thus totally determined by the dyad $\op{S}^{coh}(\p{0})$ and the effective medium $\ep_e$ described in I.
Accordingly, in the rest of this paper, we will focus only on the calculation of the incoherent part of the scattering.
As mentioned in the introduction, the fundamental quantity in the radiative transfer equation  is the specific intensity. We define $\Ical^{incoh}(\RR,\kk)$ as the Wigner transform of the incoherent intensity,   in the medium 0, scattered by the random medium and the surfaces:
\begin{eqnarray}
\fl \Ical^{incoh}(\RR,\kk)=
\frac{\ep_{vac}\,c_{vac}\,n_0}{2}\int_{V_0}\rmd^3\rr\,\e^{-\rmi\kk\cdot\rr}\Big[
\ll\E^{0\,s}_{SV}(\RR+\frac{\rr}{2})\otimes\E^{0\,s\,*}_{SV}(\RR-\frac{\rr}{2})\gg_{SV}\nonumber\\
\lo -\ll\E^{0\,s}_{SV}(\RR+\frac{\rr}{2})\gg_{SV}\otimes\ll\E^{0\,s\,*}_{SV}(\RR-\frac{\rr}{2})\gg_{SV}\Big]\,.\label{defIncoh}
\end{eqnarray}
In introducing definition \eref{defRSV-1} in equation \eref{defIncoh}, and by using  equation \eref{expCross} with the quasi-uniform field approximation~\cite{Apresyan},  we obtain:
\begin{equation}
\fl \frac{\alpha_0(\p{})^2}{\pi}\veccal{I}^{incoh}(\RR,\kk)=(2\pi)\,\delta(k_z-\alp{0}{})\,\op{\sigma}^{incoh}(\p{}|\p{0}):\veccal{J}^{0i}(\p{0})\,,\label{I00incohR00incoh0}
\end{equation}
with $\kk =\p{}+k_z\,\ez$.
We have supposed that the medium 0 is non-absorbing, and thus, the intensity $\veccal{I}^{incoh}(\RR,\kk)$ does not depend on $\RR$ as it appears in equation \eref{I00incohR00incoh0}.
From the result in \eref{I00incohR00incoh0}, we can determine the bistatic cross-section $\op{\sigma}^{incoh}$ by calculating the specific incoherent intensity defined by
the Wigner transform \eref{defIncoh}.
The Dirac distribution in  equation \eref{I00incohR00incoh0} insures that the vector 
$\kk$ has a fixed normed $||\kk||=K_0$.
The tensor $\Ical^{incoh}(\RR,\kk)$ is not exactly the specific intensity used in the radiometry
theory since it is not homogeneous to a power per unit of area and solid angle. 
The usual specific intensity $\veccal{I}^{incoh}(\RR,\hvec{k})$ can be deduce from 
 $\veccal{I}^{incoh}(\RR,\kk)$  from the following relationship:
 \begin{equation}
\veccal{I}^{incoh}(\RR,\kk)=\frac{(2\pi)^3}{K_0^2}\delta(||\kk||-K_0)\,\veccal{I}^{incoh}(\RR,\hvec{k})\,,
 \end{equation}
where $\hvec{k}=\kk/||\kk||$. The Dirac  distribution $\delta(||\kk||-K_0)$ insure that the wave propagation direction $\kk$ has a fixed norm given by $K_0$. 
As demonstrated by  equation \eref{I00incohR00incoh0}, this property can also be written under the following form: 
\begin{equation}
\veccal{I}^{incoh}(\RR,\kk)=(2\pi)\delta(k_z-\alp{0}{})\veccal{I}^{incoh}(\RR,\p{})\,,
\end{equation}
where 
\begin{equation}
\veccal{I}^{incoh}(\RR,\p{})=\frac{\pi}{\alpha_0(\p{})^2}\op{\sigma}^{incoh}(\p{}|\p{0}):\veccal{J}^{0i}(\p{0})\,.\label{I00incohR00incoh0-2}
\end{equation}
It can be easily checked that we
have the following relationship between $\veccal{I}^{incoh}(\RR,\kk)$, $\veccal{I}^{incoh}(\RR,\hvec{k})$ and $\veccal{I}^{incoh}(\RR,\p{})$:
\begin{equation}
\fl \int\frac{\rmd^3\kk}{(2\pi)^3}\,\veccal{I}^{incoh}(\RR,\kk)=\int \rmd^2\hvec{k} \,\,\veccal{I}^{incoh}(\RR,\hvec{k})=\intp{}\veccal{I}^{incoh}(\RR,\p{})\,,\label{eqwigner}
\end{equation}
where  $\rmd^2\hvec{k}$ is the elementary solid angle.
If we decompose  vector $\vec{k}$, on a spherical basis
\begin{equation}
\vec{k}=K_0\begin{vecteur}
\sin \theta \cos \phi \\
\sin \theta \sin \phi\\
\cos \theta
\end{vecteur}\,
\end{equation}
then
\begin{eqnarray}
\frac{\rmd^2\p{}}{(2\pi)^2}&=||\p{}||\rmd||\p{}||\rmd\phi\\
&=\frac{K_0^2\cos \theta}{(2\pi)^2}\,\rmd^2\hvec{k}\,,
\end{eqnarray}
since $\kk=\p{}+\alpha_0(\p{})\ez$ implies that $||\p{}||=K_0\,\sin \theta$, $\alpha_0(\p{})=K_0\cos\theta$, and $\rmd^2\hvec{k}=\sin \theta \rmd \theta\rmd \phi$.
From equality \eref{eqwigner} we find
\begin{equation}
\veccal{I}^{incoh}(\RR,\hvec{k})=\frac{K_0^2\cos \theta}{(2\pi)^2}\veccal{I}^{incoh}(\RR,\p{})\,,
\end{equation}
and from \eref{I00incohR00incoh0-2} we write the cross-section as a function of $\veccal{I}^{incoh}(\RR,\hvec{k})$:
\begin{equation}
4\pi\cos \theta \,\veccal{I}^{incoh}(\RR,\hvec{k})=\op{\sigma}^{incoh}(\hvec{k}|\hvec{k}_0):\veccal{J}^{0i}(\hvec{k}_0)
\end{equation}
with $K_0\hvec{k}=\p{}+\alpha_0(\p{})\ez$, $K_0\hvec{k}_0=\p{0}+\alpha_0(\p{0})\ez$, and 
\begin{eqnarray}
\veccal{J}^{0i}(\hvec{k}_0) & =\veccal{J}^{0i}(\hvec{p}_0)\,,\\
&=\frac{\ep_{vac}c_{vac} n_0}{2}\,\vec{E}^{0i}(\p{0})\otimes\vec{E}^{0i\,*}(\p{0})\,,
\end{eqnarray}
We recover here the generalization of the definition used in the scalar radiative transfer,
where the cross-section is given by
\begin{equation}
\sigma^{incoh}(\hvec{k}|\hvec{k}_0)=\frac{4\pi \cos \theta \,\mathcal{I}(\RR,\hvec{k})}{\mathcal{J}(\hvec{k}_0)}\,.
\end{equation}
where $\mathcal{I}(\RR,\hvec{k})$ is the usual scalar specific intensity, and $\mathcal{J}(\hvec{k}_0)$ the incident power flux.
\section{Bethe-Salpeter equation and specific intensity}
\label{Bethe1}
To obtain the bistatic cross-section $\op{\sigma}^{incoh}$, we are now going to calculate the specific intensity $\opcal{I}^{incoh}$ by using the Green functions defined in I.
We have shown, in particular, the following relationships:
\begin{eqnarray}
\op{G}_{SV}^{00}&=\op{G}_{S}^{00}+\op{G}_{S}^{01}\cdot\op{T}_{SV}^{11}\cdot\op{G}_{S}^{10}\,,\label{Lip00}\\
\op{G}_{SV}^{11}&=\op{G}_{S}^{11}+\op{G}_{S}^{11}\cdot\op{T}_{SV}^{11}\cdot\op{G}_{S}^{11}\,,\label{Lip11}
\end{eqnarray}
where the operator $\op{T}_{SV}^{11}$ satisfies the coherent potential approximation $<\op{T}_{SV}^{11}>_V=0$.
In using this approximation, the average tensorial products of the Green functions are given by
\begin{eqnarray}
\fl \ll\GVu{00}\otimes\GVu{00\,*}\gg_{SV}=<\Gue{00}\otimes\Gue{00\,*}>_S \nonumber\\
\qquad \qquad +<\Gue{01}\otimes\Gue{01\,*}>_S:
<\op{\Gamma}^{11}_S>_S:<\Gue{10}\otimes\Gue{10\,*}>_S\,, \label{G00eq}\\
 \fl \ll\GVu{11}\otimes\GVu{11\,*}\gg_{SV} =<\Gue{11}\otimes\Gue{11\,*}>_S\nonumber\\
 \qquad \qquad +<\Gue{11}\otimes\Gue{11\,*}>_S:
<\op{\Gamma}^{11}_S>_S:<\Gue{11}\otimes\Gue{11\,*}>_S\,,\label{G11eq}
\end{eqnarray}
where we have defined $\op{\Gamma}^{11}_S=<\op{T}_{SV}^{11}\otimes\op{T}_{SV}^{11}>_V$.
In these equations, we use the following convention for the tensorial product of two dyads
$\op{f}(\rr,\rr_0)$ and $\op{g}(\rr,\rr_0)$:
\begin{equation}
\op{f}\otimes\op{g}(\rr,\rr'|\rr_0,\rr_0')=\op{f}(\rr,\rr_0)\otimes\op{g}(\rr',\rr_0')\,,
\end{equation}
and for the product between two tensors,
\begin{equation}
\fl [\opcal{M}^1:\opcal{M}^2](\rr,\rr'|\rr_0,\rr_0')=\int_{V_1}\rmd^3\rr_1\rmd^3\rr_1'\, 
\opcal{M}^1(\rr,\rr'|\rr_1,\rr_1'):\opcal{M}^2(\rr_1,\rr'_1|\rr_0,\rr_0')\,.
\end{equation}
We now  introduce the intensity operator $\Pcal^{11}$ and the Bethe-Salpeter equation satisfied by $<\GVu{11}\otimes\GVu{11\,*}>_V$: 
\begin{eqnarray}
\fl \ll\GVu{11}\otimes\GVu{11\,*}\gg_{SV}=<\Gue{11}\otimes\Gue{11\,*}>_S\nonumber\\
\lo +<\Gue{11}\otimes\Gue{11\,*}>_S:
\Pcal^{11}:\ll\GVu{11}\otimes\GVu{11\,*}\gg_{SV}\,.
\label{Betheeq}
\end{eqnarray}
In the next section, we will give an approximate expression for the intensity operator
$\opcal{P}^{11}$
which will not depend on the rough surface profiles $z=h_1(\x)$ and $z=h_2(\x)$ but
only on the properties of the scatterers. It will appear that the tensor $\opcal{P}^{11}$
describes the intensity scattered by a particle taking into account the correlations
with other particles.
To simplify the notations, we introduce two new tensors $\opcal{G}_{\ll SV\gg}^{11}$ and 
$\opcal{G}_{\infty <S>}^{11}$ defined by
\begin{eqnarray}
\opcal{G}_{\ll SV\gg}^{11}=\ll\GVu{11}\otimes\GVu{11\,*}\gg_{SV}\,,\\
\opcal{G}_{\infty\,<S>}^{11}=<\Gue{11}\otimes\Gue{11\,*}>_S\,,\label{defGinftyS}
\end{eqnarray}
making the Bethe-Salpeter equation \eref{Betheeq} now
\begin{equation}
\opcal{G}_{\ll SV\gg}^{11}=\opcal{G}_{\infty\,<S>}^{11}+\opcal{G}_{\infty <S>}^{11}:\opcal{P}^{11}:\opcal{G}_{\ll SV\gg}^{11}\,.\label{Betheeq2}
\end{equation}
In definition \eref{defGinftyS}, we have introduced the symbol $\infty$ to emphasize that the
propagator $\opcal{G}_{\infty\,<S>}^{11}$ between two scattering events by the particles describes either a wave propagating directly between the two scatterers (which is taken into account by the term $\op{G}^{\infty}_{1}$ in $\op{G}^{11}_{S}$) or a wave reflected by the boundaries (which is taken into account by the terms $\op{G}_{S}^{1\pm1\pm}$ in $\op{G}^{11}_{S}$).
In iterating the Bethe-Salpeter equation \eref{Betheeq} and comparing it with equation \eref{G11eq}, we  express the operator $<\op{\Gamma}^{11}_S>_S$ as a function of the intensity operator
$\Pcal^{11}$:
\begin{eqnarray}
\fl <\Gamcal^{11}_{S}>_S &=\Pcal^{11}+\Pcal^{11}:\opcal{G}_{\infty\,<S>}^{11}:\Pcal^{11}
+\Pcal^{11}:\opcal{G}_{\infty\, <S>}^{11}:\Pcal^{11}:\opcal{G}_{\infty\,<S>}^{11}:\Pcal^{11}+\dots\,,\nonumber\\
\fl  &= \Pcal^{11}+\Pcal^{11}:\left[ 
\opcal{G}_{\infty\,<S>}^{11} +
 \opcal{G}_{\infty\,<S>}^{11}:\Pcal^{11}:\opcal{G}_{\infty\,<S>}^{11}+\dots
\right]:\Pcal^{11}\,.\label{devGamma}
\end{eqnarray}
The term in the bracket is identical to the right-hand side of  equation \eref{Betheeq2}, and
equation \eref{devGamma} is written
\begin{equation}
<\Gamcal^{11}_{S}>_S=\Pcal^{11}+\Pcal^{11}:\opcal{G}_{\ll SV\gg}^{11}:\Pcal^{11}\,.\label{gamcalGueGue}
\end{equation}
By introducing equation \eref{gamcalGueGue} in \eref{G00eq}, we have
\begin{eqnarray}
\fl \opcal{G}_{\ll SV\gg}^{00} &=\opcal{G}_{\infty\,<S>}^{00}+\opcal{G}_{<S>}^{01}:
\Pcal^{11}:\opcal{G}_{<S>}^{10} +\opcal{G}_{\infty\,<S>}^{01}:\Pcal^{11}:\opcal{G}_{\infty\,\ll SV\gg}^{11}:\Pcal^{11}:\opcal{G}_{<S>}^{10}\,,\nonumber \\ \fl \label{eqG00dev}
\end{eqnarray}
where we have defined
\begin{eqnarray}
\opcal{G}_{\ll SV \gg}^{00}=\ll \GVu{00}\otimes\GVu{00\,*}\gg_{SV}\,\\
\opcal{G}_{\infty \ll SV \gg}^{00}=<\Gue{00}\otimes\Gue{00\,*}>_{S}\,\\
\opcal{G}_{<S>}^{10}=<\GVu{10}\otimes\GVu{10\,*}>_{S}\,,\\
\opcal{G}_{<S>}^{01}=<\Gue{01}\otimes\Gue{01\,*}>_S\,.
\end{eqnarray}
In equation \eref{eqG00dev}, we have expressed the intensity in the medium 0 produced by a source in the medium 0, which is described
by the tensor $\opcal{G}_{\ll SV \gg}^{00}$ as a function of the tensor $\opcal{G}_{\ll SV\gg}^{11}$. As the intensity operator $\Pcal^{11}$ represents
the intensity scattered by one particle,  equation \eref{eqG00dev} decomposes
the scattering process in three terms:  the first one  where  no scattering in the volume happens, the second one is the single scattering term, and the third 
term contains higher  scattering contribution by particles.

To calculate $\veccal{I}^{incoh}$ in equation \eref{defIncoh}, we need to connect the scattered intensity
$\vec{E}^{0s}_{SV}$ with the Green function $\GVu{00}$. For a source of radiation produced
by a current $\vec{j}_{source}$, the incident field is given by
\begin{eqnarray}
\E^{0i}(\rr)&=\rmi\,\omega\,\mu_{vac}\int_{V_0} \rmd^3\rr_0
\,\op{G}_{0}^{\infty}(\rr,\rr_0)\cdot\vec{j}_{source}(\rr_0)\,.\label{Chap3ES0r}
\end{eqnarray}
The field $\vec{E}^{0}_{SV}$ in the medium 0 produced by the source and the waves scattered by the random medium and the rough surfaces leads to
\begin{eqnarray}
\vec{E}^{0}_{SV}(\rr)&=\E^{0i}(\rr)+\E^{0s}_{SV}(\rr)\,,\\
&=\rmi\,\omega\,\mu_{vac}\int_{V_0} \rmd^3\rr_0
\,\op{G}^{00}_{SV}(\rr,\rr_0)\cdot\vec{j}_{source}(\rr_0)\,.
\end{eqnarray}
If we define a new function $\op{G}_{SV}^{0+0-}$ by
\begin{equation}
\op{G}_{SV}^{00}=\GVu{0+0-}+\op{G}_{0}^{\infty}\,,\label{defG0+0-}
\end{equation}
then the scattered intensity is given by
\begin{eqnarray}
\E^{0s}_{SV}(\rr)&=\rmi\,\omega\,\mu_{vac}\int_{V_0} \rmd^3\rr_0
\,\op{G}^{0+0-}_{SV}(\rr,\rr_0)\cdot\vec{j}_{source}(\rr_0)\,.\label{Chap3ES0rbis}
\end{eqnarray}
The tensorial product  appearing in definition \eref{defIncoh} of the Wigner transform of the specific 
intensity $\veccal{I}^{incoh}$ is thus
\begin{eqnarray}
\fl\E^{0\,s}_{SV}(\RR+\frac{\rr}{2})\otimes\E^{0\,s\,*}_{SV}(\RR-\frac{\rr}{2})=(\omega\mu_{vac})^2\nonumber \\
\fl \times\iint_{V_0}\frac{\rmd^3 \rr_0}{(2\pi)^3}\frac{\rmd^3 \rr'_0}{(2\pi)^3}
\op{G}_{SV}^{0+0-}(\vec{R}+\frac{\rr}{2},\rr_0)\cdot\vec{j}_{source}(\rr_0)\otimes
\op{G}_{SV}^{0+0-\,*}(\vec{R}-\frac{\rr}{2},\rr'_0)\cdot\vec{j}^*_{source}(\rr'_0)\,.\nonumber\\
\label{exptensor}
\end{eqnarray}
After some calculation and by using  development \eref{eqG00dev},  definition
\eref{defG0+0-},
and the decomposition
$\op{G}_{S}^{00}=\op{G}_{0}^{\infty}+\op{G}_{S}^{0+0-}$
introduced in I, 
in equation 
\eref{exptensor}, we obtain the three following  contributions  for the specific intensity
defined by \eref{defIncoh}:
\begin{equation}
\opcal{I}^{incoh}=\opcal{I}^{incoh}_{L=0}+\opcal{I}^{incoh}_{L=1}+\opcal{I}^{incoh}_{Ladder}\,,\label{decIncoh}
\end{equation}
with
\begin{eqnarray}
\fl\veccal{I}^{incoh}_{L=0}(\RR,\kk)= & e_0\int_{V_0}\rmd^3\rr\e^{-\rmi\,\kk\cdot\rr}\,
\left[<\E^{0\,s}_{S}(\RR+\frac{\rr}{2})\otimes\E^{0\,s\,*}_{S}(\RR-\frac{\rr}{2})>_{S}\right.\nonumber\\
\fl &  \left.\qquad \qquad \qquad -<\E^{0\,s}_{S}(\RR+\frac{\rr}{2})>\otimes<\E^{0\,s\,*}_{S}(\RR-\frac{\rr}{2})>_{S}
\right]\,,\label{I00incoh0}\\
\fl\veccal{I}^{incoh}_{L=1}(\RR,\kk)= & e_0\int_{V_0}\rmd^3\rr\e^{-\rmi\,\kk\cdot\rr}\,
\opcal{G}^{01}_{<S>} :\Pcal^{11}:<\vec{E}^{1\,t}\otimes\vec{E}^{1\,t}>_{S}({\RR+\frac{\rr}{2}},{\RR-\frac{\rr}{2}})\,,\label{I00incoh1}\\
\fl\veccal{I}^{incoh}_{Ladder}(\RR,\kk)= & e_0\int_{V_0}\rmd^3\rr\e^{-\rmi\,\kk\cdot\rr}\,\,\,\opcal{G}^{01}_{<S>}
:\Pcal^{11}:\opcal{G}^{11}_{\ll SV\gg}\,\nonumber \\
\fl  & \qquad \qquad \qquad :\Pcal^{11}:<\vec{E}^{1\,t}\otimes\vec{E}^{1\,t}>_{S}({\RR+\frac{\rr}{2}},{\RR-\frac{\rr}{2}})\,.\label{I00incohLadder}
\end{eqnarray}
Here 
\begin{equation}
e_0=\frac{\ep_{vide}\,c_{vide}\,n_0}{2}\,,
\end{equation}
and $\E_{S}^{0s}(\rr)$, $\E_{S}^{1t}(\rr)$ are, respectively, the field scattered by the boundaries without any interaction with the particles and the field transmitted by the boundaries inside the slab before any scattering by the particles:
\begin{eqnarray}
\E_{S}^{0s}(\rr)&=\rmi\,\omega\,\mu_{vac}\int_{V_0} \rmd^3\rr_0
\,\op{G}_{S}^{0+0-}(\rr,\rr_0)\cdot\vec{j}_{source}(\rr_0)\,,\label{Chap3ES0rbis2}\,\\
\E_{S}^{1t}(\rr)&=\rmi\,\omega\,\mu_{vac}\int_{V_0} \rmd^3\rr_0
\,\op{G}_{S}^{10}(\rr,\rr_0)\cdot\vec{j}_{source}(\rr_0)\,.\label{Chap3ESetbis}
\end{eqnarray}
From the relationships in \eref{defIncoh} and \eref{I00incohR00incoh0} and the decomposition \eref{decIncoh}, we can also decompose the incoherent scattering cross-section in three parts:
\begin{equation}
\op{\sigma}^{incoh}=\op{\sigma}^{incoh}_{L=0}+\op{\sigma}^{incoh}_{L=1}+\op{\sigma}_{Ladder}^{incoh}\,.
\end{equation}
In section \eref{secCrossed}, we will add a fourth contribution to the cross-section  taking
into account the enhanced backscattering.
From  equation \eref{I00incoh0}, we notice that the term 
 $\op{\sigma}^{incoh}_{L=0}$ is determined by the scattering matrix $\op{S}^{0+0-}$
 since in using the results of I, we can write the field scattered by the rough surfaces $\E^{0s}_S$ as
 \begin{equation}
\fl\E^{0s}_S(\rr)=\intp{}\,\e^{\rmi\,\p{}\cdot\x+\rmi\,\alp{0}{}\,z}\,\op{S}^{0+,0-}(\p{}|\p{0})\cdot\E^{0\,i}(\p{0})\,,
\end{equation} 
instead of  equation \eref{Chap3ES0rbis2}.
The term $\op{\sigma}^{incoh}_{L=1}$ is determined by the tensor $\opcal{P}^{11}$ and by the scattering
 operator $\op{S}^{0+1\pm}$, $\op{S}^{1\pm 0-}$ since the Green function $\op{G}^{01}$ and the transmitted field $\vec{E}^{1t}$ appearing in the
 equation \eref{I00incoh1} are
 \begin{eqnarray}
\fl \op{G}_{S}^{01}(\rr,\rr_0)=\op{G}_{S}^{0+1-}(\rr,\rr_0)+\op{G}_{S}^{0+1+}(\rr,\rr_0)\,,\label{decompG0}
\\
\fl\op{G}_{S}^{0+\,1a_0}(\rr,\rr_0)=\frac{\rmi}{2}\iintpV{}{0}{}
\e^{\rmi\,\p{}\cdot\x-\rmi\,\p{0}\cdot\x_0+\rmi\,\alp{0}{}\,z-a_0\,\rmi\,\alp{e}{0}\,z_0}\nonumber
\\ \times \RpS{0+\,1a_0}{}{0}\cdot(\op{I}-\hvec{k}^{1\,a_0}_{\p{0}}\hvec{k}^{1\,a_0}_{\p{0}})\frac{1}{\alp{e}{0}}\,,\label{expG0+1a0}\\
\fl\E^{1\,t}_S(\x,z)=\intp{}\,\e^{\rmi\,\p{}\cdot\x}\,\left[\e^{\rmi\,\alp{1}{}\,z}\,\op{S}^{1+,0-}(\p{}|\p{0})+\e^{-\rmi\,\alp{1}{}\,z}\,\op{S}^{1-,0-}(\p{}|\p{0})\right]\cdot
\E^{0\,i}(\p{0})\,,\nonumber\\\label{expE1t}
\end{eqnarray}
where $a_0=\pm$, and $K'_e=\Re(K_{e})$.
 
 On the other hand, to get the term $\op{\sigma}_{Ladder}^{incoh}$, we need to calculate
  term $\opcal{G}^{11}_{\ll SV\gg}$ which appears in equation \eref{I00incohLadder}.
As this equation is a  Wigner transform, it is convenient to write the Bethe-Salpeter equation \eref{Betheeq2}, satisfied by $\opcal{G}^{11}_{\ll SV\gg}$, by introducing the Wigner transform of a tensor by
\begin{equation}
\fl \opcal{G}(\RR,\kk|\RR_0,\kk_0)=\int_{V_1}\rmd^3\rr\rmd^3\rr_0\,\,\e^{-\rmi\kk\cdot\rr+\rmi\kk_0\cdot\rr_0}
\Gcal(\RR+\frac{\rr}{2},\RR-\frac{\rr}{2}|\RR_0+\frac{\rr_0}{2},\RR_0-\frac{\rr_0}{2})\,,
\label{wignerop}
\end{equation}
and we obtain
\begin{eqnarray}
\fl \opcal{G}_{\ll SV\gg}^{11}(\RR,\kk|\RR_0,\kk_0)=\opcal{G}_{\infty\,<S>}^{11}(\RR,\kk|\RR_0,\kk_0) +\int_{V_1} \rmd^3\RR_1\rmd^3\RR_2\int\frac{\rmd^3\kk_1}{(2\pi)^3}\frac{\rmd^3\kk_2}{(2\pi)^3}
\nonumber \\ \fl \,\times \opcal{G}_{\infty\,<S>}^{11}(\RR,\kk|\RR_1,\kk_1)
:\opcal{P}^{11}(\RR_1,\kk_1|\RR_2,\kk_2):\opcal{G}_{\ll SV\gg}^{11}(\RR_2,\kk_2|\RR_0,\kk_0)\,.\label{Betheeq3}
\end{eqnarray}
In section \eref{LadderApprox}, we will write explicitly the different terms of this equation
under the ladder approximation, and in  section \eref{radiative}, we will derive the radiative
transfer equation from it.
\section{Intensity operator $\Pcal^{11}$ and modified ladder approximation}
\label{Pcaldef}
In the previous section, we have introduced in a formal way the intensity operator
$\opcal{P}^{11}$ to write the Bethe-Salpeter equation. 
This intensity operator can be determined in using the energy conservation principle.
In fact, by introducing  the Dyson equation, defined in \ref{AppDyson}, and the Bethe-Salpeter equation \eref{Betheeq}, in the 
energy conservation equation, we obtain a Ward identity which is a relationship between
the mass operator $\op{M}^{11}$ \eref{AppDyson} and the intensity operator $\opcal{P}^{11}$.
The mass operator is known since it is a function of the effective  permittivity $\ep_e$:
\begin{equation}
\op{M}^{11}(\rr,\rr_0)=(\ep_e-\ep_1)\,\delta(\rr-\rr_0)\op{I}\,.
\end{equation}
In paper I, we have shown that under the Quasi-Crystalline Coherent Potential
Approximation (QC-CPA), the scalar $\ep_e$ satisfies a non-linear system of equations.
In using these equations in the Ward identity, we obtain an expression, called the modified
ladder approximation,
for the intensity operator $\opcal{P}^{11}$ which satisfies the energy conservation~\cite{Kong2001-3,Tsang1}:
\begin{eqnarray}
\fl \Pcal^{11}(\rr,\rr'|\rr_0,\rr'_0)=n\,\int \rmd^3 \rr_j\,<\op{C}^{11}_{SV,\rr_j}(\rr|\rr_0)>_{V;\rr_j}\otimes<\op{C}^{11\,*}_{SV,\rr_j}(\rr'|\rr'_0)>_{V;\rr_j}\,\nonumber \\
\fl  + n^2\,\int\!\!\!\int \rmd^3 \rr_j\,\rmd^3 \rr_l\,h(\rr_j-\rr_l)\,<\op{C}^{11}_{SV,\rr_j}(\rr|\rr_0)>_{V;\rr_j}\otimes<\op{C}^{11\,*}_{SV,\rr_l}(\rr'|\rr'_0)>_{V;\rr_l}\,,\label{expP11_in}
\end{eqnarray}
where
\begin{equation}
h(\rr)=g(\rr)-1\,.
\end{equation}
The term $<\op{C}^{11}_{SV,\rr_j}(\rr|\rr_0)>_{V;\rr_j}$ has been defined in  paper I
and represents the transition operator for a scatterer located at $\rr_j$ which
takes into account the correlation with scatterers close to the point $\rr_j$ through the 
function $g(\rr)$.
Hence, the first term in  equation \eref{expP11_in} describes the scattering process by one scatterer located at $\rr_j$,
and the second term represents the interference process between a wave scattered at $\rr_j$ and 
another one scattered at $\rr_l$. 
We do not reproduce here the derivation of the Ward identity and  equation \eref{expP11_in}
since the demonstration is formally identical to the infinite random medium case and is well
documented~\cite{Kong,Kong2001-3,Tsang1,Roth1}. However, we must notice that  equation \eref{expP11_in} is valid only for the static case when the harmonic dependence $\omega$ is the same in the
left and right hand sides of the tensorial products of all equations previously written.
In the dynamic case, the Ward identity contains a new term, taking into account the time
delay that the wave undergoes during scattering, which modified equation 
\eref{expP11_in}. The derivation of this dynamic Ward identity and the modification 
on the radiative transfer equation are described in  references~\cite{Lag1,Tig1,Tig4,Kogan,Livdan,Bara3,Bara4,Bara5,Nieh1,Bara6,Nieh2}.

In I, we have introduced the following notation for the average transition operator of a scatterer located at the origin:
\begin{equation}
 \op{C}^{11}_{o}(\kk|\kk_0)=<\op{C}^{11}_{SV,\rr_i=\vec{0}}(\kk|\kk_0)>_{V;\rr_i=\vec{0}}\,,
\end{equation}
where  the Fourier transform of the transition operator is defined by
\begin{equation}
\fl <\op{C}^{11}_{SV,\rr_j}(\kk|\kk_0)>_{V;\rr_j}=\iintr{}{0}\exp(-\rmi\kk\cdot\rr+\rmi\kk_0\cdot\rr_0)\,<\op{C}^{11}_{SV,\rr_j}(\rr|\rr_0)>_{V;\rr_j}\,.\label{fourierC11}
\end{equation}
In I, we have also shown that the transition operator for  a scatterer located at  point $\rr_j$ can be expressed as a function of $\op{C}^{11}_{o}(\kk|\kk_0)$:
\begin{equation}
<\op{C}^{11}_{SV,\rr_j}(\kk|\kk_0)>_{V;\rr_j}=\e^{-\rmi(\kk-\kk_0)\cdot\rr_j}\,\op{C}^{11}_{o}(\kk|\kk_0)\,.\label{invtransC11}
\end{equation}
In introducing the Fourier transform \eref{fourierC11} and the properties \eref{invtransC11}
in  equation \eref{expP11_in} and taking the Wigner transform \eref{wignerop} of the intensity operator $\opcal{P}^{11}$, we obtain
\begin{eqnarray}
\fl \Pcal^{11}(\RR,\kk|\RR_0,\kk_0)=n\,w(\kk-\kk_0)\nonumber \\ \fl  
\times\intK{}\,\e^{\rmi\,\KK\cdot(\RR-\RR_0)}\,\op{C}^{11}_{o}(\frac{\KK}{2}+\kk|\frac{\KK}{2}+\kk_0)\otimes\op{C}^{11\,*}_{o}(-\frac{\KK}{2}+\kk|-\frac{\KK}{2}+\kk_0)\,,\nonumber\\ \label{exP11-wigner}
 \end{eqnarray}
 where $w(\kk-\kk_0)$ is the structure factor of the medium identical to the one  defined in scattering by X-rays~\cite{Kong2001-2,Jackson}:
 \begin{equation}
  w(\kk-\kk_0)=1+n\,\int \rmd^3 \rr \,\,\e^{-\rmi(\kk-\kk_0)\cdot\rr}\,[g(\rr)-1]\,,
\end{equation}
 with $g(\rr)$ the pair distribution function.
 As the system under study is invariant along the $x,y$ axis,
 we introduce the Fourier transform along these directions:
 \begin{equation}
\fl  \Pcal^{11}(Z,\kk|Z_0,\kk_0;\vec{P})=\int \rmd^2(\vec{X}-\vec{X}_0)\,\e^{-\rmi(\vec{X}-\vec{X}_0)\cdot\vec{P}}\,\Pcal^{11}(\RR,\kk|\RR_0,\kk_0)\,,
 \end{equation}
 with 
 $\RR=\vec{X}+Z\ez$, and $\RR_0=\vec{X}_0+Z_0\ez$.
 From equation \eref{exP11-wigner}, we deduce that
 \begin{eqnarray}
 \fl \Pcal^{11}(Z,\kk|Z_0,\kk_0;\vec{P})=n\,w(\kk-\kk_0)\nonumber \\ \fl  
\times\int \rmd K_z\,\e^{\rmi\,K_z(Z-Z_0)}\,\op{C}^{11}_{o}(\frac{\KK_{\vec{P}}}{2}+\kk|\frac{\KK_{\vec{P}}}{2}+\kk_0)\otimes\op{C}^{11\,*}_{o}(-\frac{\KK_{\vec{P}}}{2}+\kk|-\frac{\KK_{\vec{P}}}{2}+\kk_0),\label{exP11-wigner2}
 \end{eqnarray}
 where $\KK_{\vec{P}}=\vec{P}+K_z\ez$.
 Expression \eref{exP11-wigner2} is rather complicated since the operator $\opcal{P}^{11}$
 is non local ($\opcal{P}^{11}\neq 0\mbox{ for }Z\neq Z_0$), and the operators $\op{C}^{11\,*}_{o}$ are off-shell evaluated  since we do not have necessarily
 \begin{equation}
\fl||\vec{K}_{\vec{P}}/2+\kk||=||\vec{K}_{\vec{P}}/2+\kk_0||\,,\quad \mbox{nor}\quad ||-\vec{K}_{\vec{P}}/2+\kk||=||-\vec{K}_{\vec{P}}/2+\kk_0||\,,
 \end{equation}
  in  equation \eref{exP11-wigner2}.
Accordingly, we use an on-shell approximation~\cite{Bara1} on the transition operator 
$\op{C}_o^{11}$:
\begin{equation} \op{C}^{11}_{o}(\pm\frac{\KK_{\vec{P}}}{2}+\kk|\pm\frac{\KK_{\vec{P}}}{2}+\kk_0)\simeq\op{C}^{11}_{o}(\kk^{1}_{\vec{p}^{\pm}}|\kk^{1}_{\vec{p}_0^{\pm}})\,,
\end{equation}
with
\begin{eqnarray}
\fl \p{}^{\pm}=\pm\frac{\vec{P}}{2}+\vec{k}_\perp\,\quad \p{0}^{\pm}=\pm\frac{\vec{P}}{2}+\vec{k}_{0\perp}\,,\\
\fl\kk^{1}_{\p{}^{\pm}}=\p{}^{\pm}+\sgn(k_z)\alpha_e'(\p{}^{\pm})\ez\,,\quad
\kk^{1}_{\p{0}^{\pm}}=\p{0}^{\pm}+\sgn(k_{0z})\alpha_e'(\p{0}^{\pm})\ez\,,
\end{eqnarray}
where $\kk_{\perp}=(\kk\cdot\ex)\ex+(\kk\cdot\ey)\ey$, and $\sgn(k_z)$ is the sign of $k_z$. In this way, the incident and scattered wave vector on the particles have the same norm $||\kk^{1}_{\vec{p}^{\pm}}||=||\kk^{1}_{\vec{p}_0^{\pm}}||\simeq K_e'$.
With this on-shell approximation, the intensity operator $\opcal{P}^{11}$ becomes 
localized at the point $Z=Z_0$, and we have
\begin{eqnarray}
&\Pcal^{11}(Z,\kk|Z_0,\kk_0;\vec{P}) =  \delta(Z-Z_0)\,\opcal{P}^{11}(\vec{k}|\vec{k}_0;\vec{P})\,,
\label{P11local}\\
&\opcal{P}^{11}(\vec{k}|\vec{k}_0;\vec{P}) =  n\,w(\kk-\kk_0)\,\op{C}^{11}_{o}(\kk^{1}_{\vec{p}^{+}}|\kk^{1}_{\vec{p}_0^{+}})\otimes\op{C}^{11\,*}_{o}(\kk^{1}_{\vec{p}^{-}}|\kk^{1}_{\vec{p}_0^{-}})\,. \label{P11}
 \end{eqnarray}
 We must also emphasize  that the correlations between the scatterers appear not only in the structure factor 
$w(\kk-\kk_0)$ but also in the scattering operator $\op{C}^{11}_{o}$, which satisfies
\begin{equation}
\fl \op{C}^{11}_{o}(\kk|\kk_0)=\op{t}^{11}_{o}(\kk|\kk_0)+n\,\intk{1}\,h(\kk-\kk_1)
\,\op{t}_{o}^{11}(\kk|\kk_1)\cdot\op{G}_{1}^{\infty}(\kk_1)\cdot\op{C}^{11}_{o}(\kk_1|\kk_0)
\,,\label{expC11o}\end{equation}
with 
\begin{equation}
h(\kk-\kk_1)=\intr \,\,\e^{-\rmi(\kk-\kk_1)\cdot\rr}\,[g(\rr)-1]\,.
\end{equation}
Here,
$\op{t}^{11}_{o}$ is the scattering operator for a particle located at the origin which, contrary to $\op{C}^{11}_{o}$, does not take into account the correlation with the other
particles.
\section{Bethe-Salpeter equation and ladder approximation for the rough surfaces}
\label{LadderApprox}
\begin{figure}[htbp]
   \centering
   \psfrag{z}{$z$}
      \psfrag{k}{$\kk$}
     \psfrag{k0}{$\kk_0$}
\psfrag{G11}{$\opcal{G}^{11}_{\ll SV\gg}$}
 \psfrag{G11S}{$\opcal{G}^{11}_{\infty\, <S>}$}
 \psfrag{G10}{$\scriptstyle \opcal{G}^{10}_{<S>}=<\op{G}^{10}_{S}\otimes
\op{G}^{10\,*}_{S}>_{S}$}
      \psfrag{G01}{$\scriptstyle \opcal{G}^{01}_{<S>}=<\op{G}^{01}_{S}\otimes
\op{G}^{01\,*}_{S}>{S}$}
      \psfrag{P}{$\scriptstyle \Pcal^{11}$}
     \epsfig{file=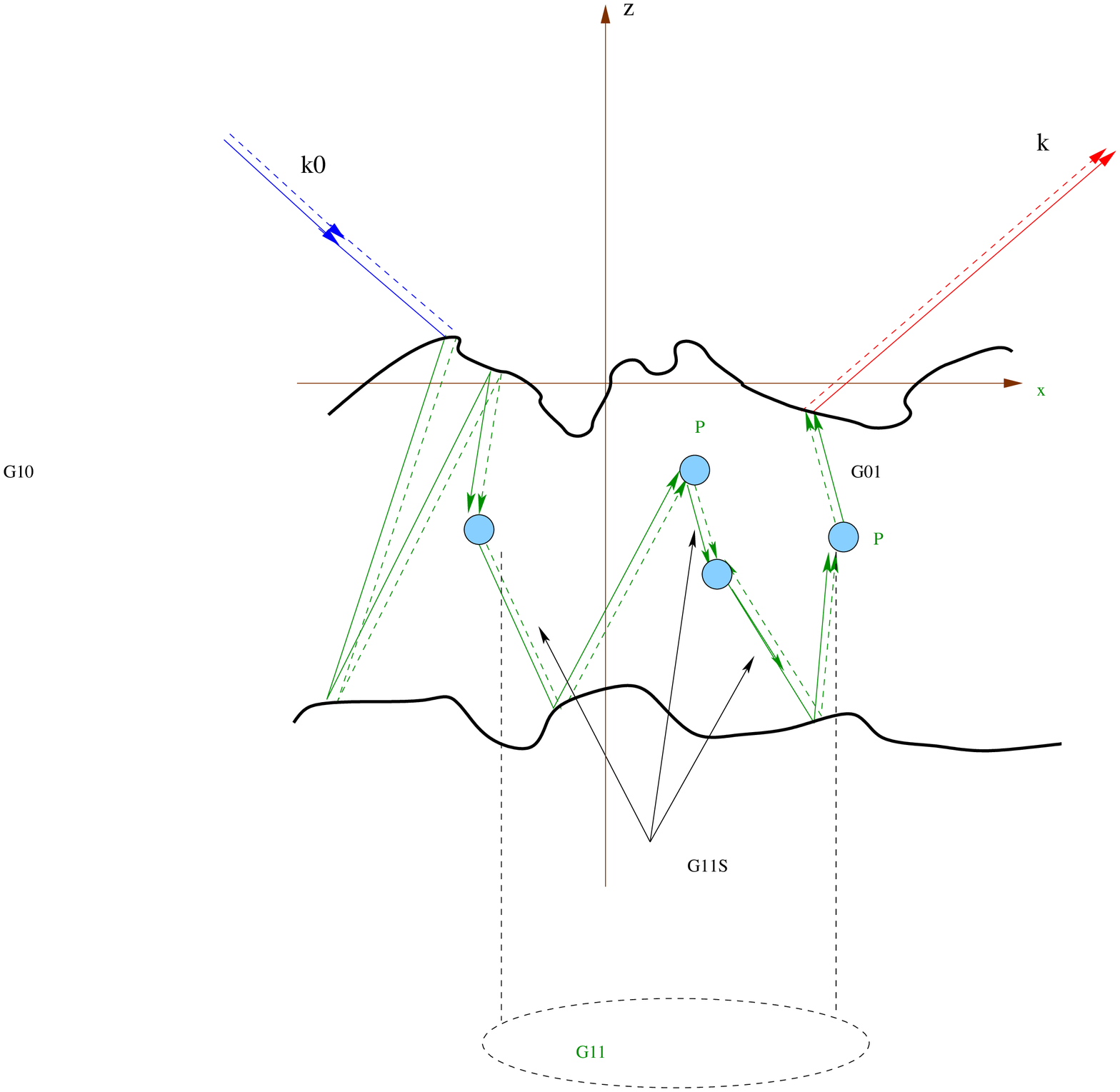,width=8cm}
      \caption{Scattering processes under the ladder approximation. The solid line and the dashed line represent, respectively,
      the wave on the left and right hand side of tensorial products such as $\opcal{G}^{11}_{\ll SV\gg}=\scriptstyle \ll\op{G}^{11}_{SV}\otimes
\op{G}^{11\,*}_{SV}\gg_{SV}$.}
      \label{FigLadder1}
 \end{figure}

\begin{figure}[htbp]
   \centering
   			\psfrag{G11S}{$\opcal{G}^{11}_{\infty\, <S>}$}
       \psfrag{Pe}{$\scriptstyle
       \opcal{P}^{11}$}
       \psfrag{Gam}{$\scriptstyle
       \opcal{G}^{11}_{\ll SV \gg }$}
       \psfrag{Cj}{$\scriptstyle \op{C}^{11}_{o}$}
       \psfrag{Cjetoile}{$\scriptstyle \op{C}^{11\,*}_{o}$}
       \psfrag{Cjd}{$\scriptstyle \op{C}^{11}_{o}$}
       \psfrag{Cjetoiled}{$\scriptstyle \op{C}^{11}_{o}$}
       \psfrag{Cj2}{$\scriptstyle \op{C}^{11}_{o}$}
       \psfrag{Cjetoile2}{$\scriptstyle \op{C}^{11\,*}_{o}$}
       \psfrag{Cj2d}{$\scriptstyle \op{C}^{11}_{o}$}
       \psfrag{Cjetoile2d}{$\scriptstyle \op{C}^{11\,*}_{o}$}
       \psfrag{Cj3}{$\scriptstyle \op{C}^{11}_{o}$}
       \psfrag{Cjetoile3}{$\scriptstyle \op{C}^{11\,*}_{o}$}
       \psfrag{Gk}{$\scriptstyle \op{G}^{11}_{S}$}
       \psfrag{Gketoile}{$\scriptstyle \op{G}^{11\,*}_{S}$}
			 \psfrag{Gk2}{$\scriptstyle \op{G}^{11\,*}_{S}$}
       \psfrag{Gketoile2}{$\scriptstyle \op{G}^{11\,*}_{S}$}
			\epsfig{file=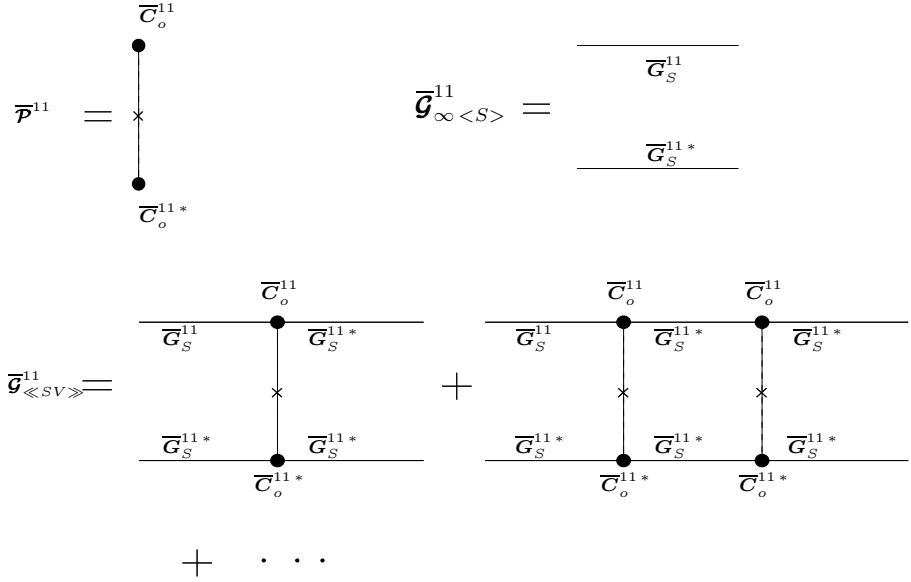,width=12cm}
      \caption{Diagrammatic representation of the ladder approximation.}
      \label{FigLadder2}
 \end{figure}

As discussed in  section \eref{Bethe1}, we need to know the tensor 
 \begin{equation}
 \Gcal^{11}_{\ll SV\gg}=\ll \op{G}^{11}_{SV}\otimes\op{G}^{11\,*}_{SV}\gg_{SV}\,
\end{equation}
 to determine the specific intensity $\veccal{I}^{incoh}_{Ladder}$, according to  equation \eref{I00incohLadder}.
We know that this tensor  verifies the integral Bethe-Salpeter equation \eref{Betheeq3}:
\begin{eqnarray}
\fl \opcal{G}_{\ll SV\gg}^{11}(\RR,\kk|\RR_0,\kk_0)=\opcal{G}_{\infty\,<S>}^{11}(\RR,\kk|\RR_0,\kk_0) +\int_{V_1} \rmd^3\RR_1\rmd^3\RR_2\int\frac{\rmd^3\kk_1}{(2\pi)^3}\frac{\rmd^3\kk_2}{(2\pi)^3}
 \nonumber \\ \fl \qquad   \times\opcal{G}_{\infty\,<S>}^{11}(\RR,\kk|\RR_1,\kk_1)
:\opcal{P}^{11}(\RR_1,\kk_1|\RR_2,\kk_2):\opcal{G}_{\ll SV\gg}^{11}(\RR_2,\kk_2|\RR_0,\kk_0)\,.\label{Betheeq4}
\end{eqnarray}
Between two scattering processes on the rough surfaces, the wave interacts with the particles,
and accordingly, it is reasonable to admit that the scattering events on the rough surfaces
are uncorrelated. Hence, the propagator $\opcal{G}_{<S>}^{11}=<\op{G}^{11}_{S}\otimes
\op{G}^{11\,*}_{S}>_S$ in equation \eref{Betheeq4} is approximated by
\begin{eqnarray}
\fl <\op{G}^{11}_{S}\otimes
\op{G}^{11\,*}_{S}>_{S}=\op{G}^{\infty}_1\otimes\op{G}_1^{\infty\,*}+ <\op{G}^{1+1+}_{S}\otimes
\op{G}^{1+1+\,*}_{S}>_{S}+<\op{G}^{1-1+}_{S}\otimes
\op{G}^{1-1+\,*}_{S}>_{S}\nonumber \\+<\op{G}^{1+1-}_{S}\otimes
\op{G}^{1+1-\,*}_{S}>_{S}+<\op{G}^{1-1-}_{S}\otimes
\op{G}^{1-1-\,*}_{S}>_{S}\,\label{LadderG11S}
\end{eqnarray}
since the Green function $\op{G}_{S}^{11}$ has a development in five components:
\begin{equation}
\op{G}^{11}_{S}=\op{G}^{\infty}_1+\op{G}^{1+1+}_{S}+\op{G}^{1-1+}_{S}+\op{G}^{1+1-}_{S}+\op{G}^{1-1-}_{S}\,.
\end{equation}
In using condensed notation, equation \eref{LadderG11S} is written
\begin{equation}
\Gcal^{11}_{\infty\,<S>}=\Gcal^{11}_{\infty}+\sum_{a,a_0=\pm}\,\Gcal^{1a1a_0}_{<S>}\,,\label{sumGcalGcal}
\end{equation}
with
\begin{eqnarray}
\Gcal^{11}_{\infty}=\op{G}^{\infty}_{1}\otimes\op{G}^{\infty\,*}_{1}\,,\\
\Gcal^{1a1a_0}_{<S>}=<\op{G}^{1a1a_0}_{S}\otimes
\op{G}^{1a1a_0\,*}_{S}>_S\,,\label{GcalGSGS} \,,\\
\opcal{G}_{\infty <S>}^{11}=<\op{G}^{11}_{S}\otimes
\op{G}^{11\,*}_{S}>_S\,,
\end{eqnarray}
and  $a$, $a_0$ are the sign $+$ or $-$.
Furthermore, the Green functions $\op{G}^{1a1a_0}_{S}$ are defined by using the scattering operator $\op{S}^{1a1a_0}$. Each operator $\op{S}^{1a1a_0}$ can be decomposed
by using the operators $\op{R}^{01}$ and $\op{R}^{H 21}$ describing, respectively,  the scattering by the upper and the lower rough surface. 
Then, the hypothesis of non-correlated diffusion on the rough surfaces must also 
be applied on each term of the operators $\op{S}^{1a1a_0}$.
For example, the Green function $\op{G}^{1-1-}_{S}$ depends on $\op{S}^{1-1-}$,
and this operator has the following development:
\begin{equation}
\op{S}^{1-1-}=\op{R}^{01}\cdot\op{R}^{H\,21}+\op{R}^{01}\cdot\op{R}^{H\,21}\cdot\op{R}^{01}\cdot\op{R}^{H\,21}+\dots\,.
\end{equation}
With our hypothesis, the tensorial product $\Gcal^{1-1-}_{<S>}=<\op{G}^{1-1-}_{S}\otimes
\op{G}^{1-1-\,*}_{S}>_S$ contains the following terms:
\begin{eqnarray}
\fl<\op{S}^{1-1-}\otimes
\op{S}^{1-1-\,*}>_S=<\op{R}^{01}\otimes\op{R}^{01\,*}>_S:<\op{R}^{H\,21}\otimes\op{R}^{H\,21\,*}>_S\nonumber\\
\fl+<\op{R}^{01}\otimes\op{R}^{01\,*}>_S:<\op{R}^{H\,21}\otimes\op{R}^{H\,21\,*}>_S:<\op{R}^{01}\otimes\op{R}^{01\,*}>_S:<\op{R}^{H\,21}\otimes\op{R}^{H\,21\,*}>_S
\nonumber \\ \fl+\dots\,.
\end{eqnarray}
Under these approximations, that corresponds to the ladder approximation for the rough surface
contributions; the Bethe-Salpeter equation \eref{Betheeq4} 
describes the scattering processes depicted
in \Fref{FigLadder1}, where the waves on the left and right sides of the tensorial
products follow the same path. 
As shown in  \Fref{FigLadder2}, we can also use Feynman diagrams~\cite{Kong,Sheng1,Frish} to represent the different terms of $\opcal{G}^{11}_{\ll SV \gg}$ in the ladder approximation obtained by  iteration of the Bethe-Salpeter
equation \eref{Betheeq4}:
\begin{eqnarray}
 \fl \opcal{G}^{11}_{\ll SV\gg} =\opcal{G}^{11}_{\infty\,<S>}+\opcal{G}^{11}_{\infty\,<S>}:\Pcal^{11}:\opcal{G}^{11}_{\infty\,<S>}\nonumber\\ \lo+ \opcal{G}^{11}_{\infty\,<S>}:\Pcal^{11}:\opcal{G}^{11}_{\infty\,<S>}:\Pcal^{11}:\opcal{G}^{11}_{\infty\,<S>}+\dots\,.
\label{PcalGcalbis}
\end{eqnarray}
We now examine in detail the different terms in the Bethe-Salpeter equation \eref{Betheeq4}.
Since, by hypothesis, the medium and the surfaces are statistically homogeneous, we  introduce a Fourier transform along the $X$, $Y$ axis: 
\begin{equation}
\fl \opcal{G}(Z,\kk|Z_0,\kk_0;\vec{P})=\int \rmd^2(\vec{X}-\vec{X}_0)\,\,\e^{-\rmi(\vec{X}-\vec{X}_0)\cdot\vec{P}}\,\,\opcal{G}(\RR,\kk|\RR_0,\kk_0)\,,\label{fourierXX0}
\end{equation}
 with 
 \begin{eqnarray}
 \RR=X\ex+Y\ey+Z\ez=\vec{X}+Z\ez\,,\\
 \RR_0=X_0\ex+Y_0\ey+Z_0\ez=\vec{X}_0+Z_0\ez\,,
 \end{eqnarray}
 and where $\opcal{G}$ is either $\opcal{G}^{11}_{\ll SV\gg}$, $\opcal{G}^{11}_{\infty\,<S>}$,
 $\opcal{G}^{1a1a_0}_{<S>}$, or $\opcal{G}^{11}_{\infty}$.
 With this Fourier transform,  equation \eref{sumGcalGcal} is now
 \begin{equation}
 \fl \Gcal^{11}_{\infty\,<S>}(Z,\kk|Z_0,\kk_0;\Ps{})=\Gcal_{\infty}^{11}(Z,\kk|Z_0,\kk_{0};\Ps{})
+\sum_{a,a_0=\pm}\,\Gcal^{1a1a_0}_{<S>}(Z,\kk|Z_0,\kk_{0};\Ps{})\,.\label{GcalRRkk}
 \end{equation}
Furthermore, in calculating the terms $\Gcal^{1a1a_0}_{<S>}$ and $\opcal{G}^{\infty}_{1}$  
in  equation \eref{GcalRRkk} and  using the Weyl representation of $\op{G}^{\infty}_1$ and the Green function decomposition in terms
of the scattering operator $\op{S}^{1a1a_0}$ given in I and mentioned at the end of  section \eref{Bethe1}, we obtain the following decomposition: 
\begin{eqnarray}
\fl\Gcal^{1a1a_0}_{<S>}(Z,\kk|Z_0,\kk_{0};\Ps{})=(2\pi)^2\,\delta\left(k_{z}-a\,\frac{\alpha_e(\p{}^+)+\alpha^*_e(\p{}^-)}{2}\right)\nonumber\\
\lo\times \delta\left(k_{0z}-a_0\,\frac{\alpha_e(\p{0}^+)+\alpha^*_e(\p{0}^-)}{2}\right)\,\Gcal^{1a1a_0}_{<S>}(Z,\p{}|Z_0,\p{0};\Ps{})\,,\label{cons1}\\ 
\fl\Gcal^{11}_{\infty}(Z,\vec{k}|Z_0,\vec{k}_{0};\Ps{})=\sum_{a,a_0=\pm}(2\pi)^2\,\delta\left(k_{z}-a\,\frac{\alpha_e(\p{}^+)+\alpha^*_e(\p{}^-)}{2}\right)\nonumber\\ \lo\times\delta\left(k_{0z}-a_0\,\frac{\alpha_e(\p{0}^+)+\alpha^*_e(\p{0}^-)}{2}\right)\,\Gcal^{1a1a_0}_{\infty}(Z,\p{}|Z_0,\p{0};\Ps{})\,,\nonumber\\
\label{cons2}
\end{eqnarray}
where $a$ and $a_0$ are the signs $+$ or $-$, $\kk=\p{}+k_z\,\ez$, $\kk_0=\p{0}+k_{0z}\,\ez$,
and 
\begin{equation}
\vec{p}^{\pm}=\vec{p}\pm\Psd{}\,,\qquad\vec{p}_0^{\pm}=\vec{p}_0\pm\Psd{}\,.
\end{equation}
Under the quasi-uniform field approximation~\cite{Apresyan}, we approximate $\alpha_e(\p{0}^+)+\alpha^*_e(\p{0}^-)$ by $\alpha_e(\p{})+\alpha^*_e(\p{})=2\Re\, \alpha_e(\p{})$, 
and equations \eref{cons1} and \eref{cons2} become
\begin{eqnarray}
\fl\Gcal^{1a1a_0}_{<S>}(Z,\kk|Z_0,\kk_{0};\Ps{})\nonumber\\
\lo =(2\pi)^2\delta(k_z-a\,\alpime{e}{})\,\delta(k_{0z}-a_0\,\alpime{e}{0})\,\Gcal^{1a1a_0}_{<S>}(Z,\p{}|Z_0,\p{0};\Ps{})\,,\label{cons1bis}\\ 
\fl\Gcal^{11}_{\infty}(Z,\vec{k}_{}|Z_0,\vec{k}_{0};\Ps{})\nonumber\\ \lo=\!\!\sum_{a,a_0=\pm}\!\!(2\pi)^2\delta(k_z-a\,\alpime{e}{})\,\delta(k_{0z}-a_0\,\alpime{e}{0})\,\Gcal^{1a1a_0}_{\infty}(Z,\p{}|Z_0,\p{0};\Ps{})\,,\nonumber\\
\label{cons2bis}\,,
\end{eqnarray}
where $\alpime{e}{}=\Re[\alp{e}{}]$.
The Dirac distributions insure that the wave vector directions $\vec{k}$  and $\vec{k}_0$ have  constant norms $||\vec{k}||\simeq K_e'$, $||\vec{k}_0||\simeq K_e'$. Furthermore,
we found the following expression for the tensors  $\Gcal^{1a1a_0}_{<S>}$ and $\Gcal^{1a1a_0}_{\infty}$:
\begin{eqnarray}
\fl\Gcal^{1a1a_0}_{<S>}(Z,\p{}|Z_0,\p{0};\Ps{})&=\opcal{S}^{1a1a_0}_{\otimes\,<S>}(\p{}|\p{0};\Ps{})\frac{\e^{\rmi\,a\,\Delta
  \alpha(\p{};\Ps{})\,Z-\rmi\,a_0\,\Delta
  \alpha(\p{0};\Ps{})\,Z_0}}{4\,\alpha_e(\p{0}^+)\alpha_e^*(\p{0}^-)}\,\,,\label{defGS}\\
\fl\Gcal^{1a1a_0}_{\infty}(Z,\p{}|Z_0,\p{0};\Ps{})&=(2\pi)^2\,\delta(\p{}-\p{0}) \op{I}_\perp^{1a}(\p{}^+)\otimes\op{I}_\perp^{1a\,*}(\p{}^-)\,\delta_{a,a_0}\,\delta_{a,sgn(Z-Z_0)}\nonumber\ \\
\fl &\times\frac{\,\e^{\rmi\,a\,\Delta
  \alpha_e(\p{};\Ps{})\,(Z-Z_0)}}{4\,\alpha_e(\p{}^+)\alpha_e^*(\p{}^-)}\,,\label{defGinft}
\end{eqnarray}
where 
\begin{equation}
\delta_{a,\sgn(Z-Z_0)}=\left\{\begin{array}{ll}
1 & \mbox{if } a=\sgn(Z-Z_0)\\
0 & \mbox{if } a\neq \sgn(Z-Z_0)\,, 
\end{array}\right.
\end{equation}
with $\sgn(Z-Z_0)$ the sign of $Z-Z_0$, and
\begin{eqnarray}
& \Delta\alpha_e(\p{};\Ps{})=
  \alpha_e(\p{}^+)-\alpha^*_e(\p{}^-)\,,\\
  &\op{I}_\perp^{1a}(\p{})=
(\op{I}-\hvec{k}^{1a}_{\p{}}\hvec{k}^{1a}_{\p{}})=\evp{1a}{}\evp{1a}{}+\ehp{}\ehp{}\,,\label{Cahp4defDeltaalp}\\
&\vec{k}^{1a}_{\p{}}= \p{}+a\,\alpha_{e}(\p{})\,\ez\,,\quad \hvec{k}^{1a}_{\p{}}=\vec{k}^{1a}_{\p{}}/||\vec{k}^{1a}_{\p{}}||\,.
\end{eqnarray}
The tensors $\opcal{S}_{\otimes\,<S>}^{1a\,1a_0}$ describe the intensity scattered by the boundaries and are defined by
\begin{eqnarray}
\fl (2\pi)^2\,\delta(\vec{0})\,\opcal{S}_{\otimes\,<S>}^{1a\,1a_0}(\p{}|\p{0};\Ps{0})=<\op{S}^{1a1a_0}\left(\p{}^+|\p{0}^+\right)\otimes\op{S}^{1a\,1a_0\,*}\left(\p{}^-|\p{0}^-\right)>_{S}\,.\label{Chap3LadderNotaN}
\end{eqnarray}
In introducing properties \eref{cons1bis} and \eref{cons2bis} in the Bethe-Salpeter equation \eref{Betheeq4}, we demonstrate that $\Gcal^{11}_{\ll SV\gg}$ verifies a development
similar to equations \eref{cons1bis} and \eref{cons2bis}:
\begin{eqnarray}
\fl \Gcal^{11}_{\ll SV\gg}(Z,\p{}|Z_0,\p{0};\Ps{})\nonumber \\\fl=\sum_{a,a_0=\pm}(2\pi)^2\delta(k_z-a\,\alpime{e}{})\,\delta(k_{0z}-a_0\,\alpime{e}{0})\,\Gcal^{1a1a_0}_{\ll SV\gg}(Z,\p{}|Z_0,\p{0};\Ps{})
\,.
\label{GcalRRkkbis}
\end{eqnarray}
Accordingly, we can write  equation \eref{Betheeq4} for each sign $a$ and $a_0$,
and we have
\begin{eqnarray}
\fl\Gcal^{1a1a_0}_{\ll SV\gg}(Z,\hvec{k}|Z_0,\hvec{k}_{0};\Ps{})=\Gcal^{1a1a_0}_{\infty<S>}(Z,\p{}|Z_0,\p{0};\Ps{})+\sum_{a_1,a_2}\,\int
\frac{\rmd^2 \vec{p}_{1}}{(2\pi)^2} \frac{\rmd^2
  \vec{p}_{2}}{(2\pi)^2}\int_{-H}^0\rmd Z_{21}\,\,\nonumber\\
  \fl \times
\Gcal^{1a1a_2}_{\infty<S>}(Z,\p{}|Z_{21},\p{2};\Ps{})
:\Pcal^{1a_2 1a_1}(\p{2}|\p{1};\Ps{}):\Gcal^{1a_1a_0}_{\ll SV\gg}(Z_{21},\p{1}|Z_0,\p{0};\Ps{})\,,\label{Betheeq5}
\end{eqnarray}
with
\begin{eqnarray}
\fl \Gcal^{1a1a_0}_{\infty<S>}(Z,\p{}|Z_0,\p{0};\Ps{})=\Gcal^{1a1a_0}_{\infty}(Z,\p{}|Z_0,\p{0};\Ps{})+\Gcal^{1a1a_0}_{<S>}(Z,\p{}|Z_0,\p{0};\Ps{})\,,\label{defG11infS}\\
\fl\Pcal^{1a_2 1 a_1}(\p{2}|\p{1};\Ps{})= 
n\,w(\hvec{k}^{1a_2}_{\p{2}}-\hvec{k}^{1a_1}_{\p{1}}) \,\op{C}^{11}_{o}(\hvec{k}^{1a_2}_{\p{2}^+}|\hvec{k}^{1a_1}_{\p{1}^+})\otimes\op{C}^{11\,*}_{o}(\hvec{k}^{1a_2}_{\p{2}^-}|\hvec{k}^{1a_1}_{\p{1}^-})\
\,\\
\vec{p}_{1}^{\pm}=\vec{p}_1\pm\Psd{}\,,\qquad\vec{p}_2^{\pm}=\vec{p}_2\pm\Psd{}\,,\\
\hvec{k}^{1a}_{\p{}}= \p{}+a\,\alpha'_{e}(\p{})\,\ez\,.\label{Chap3LadderNota1_bis}
\end{eqnarray}
The numerical solution of  integral equation \eref{Betheeq5} is a difficult task, and 
in the next section, we will give a differential form of this equation more appropriate  in this case.
However, equation \eref{Betheeq5} is well suited if we want to obtain an iterative
solution for $\Gcal^{1a1a_0}_{\ll SV\gg}$. For example, in introducing the  first term of this iteration,
$$\Gcal^{1a1a_0}_{\infty<S>}(Z,\p{}|Z_0,\p{0};\Ps{})=\Gcal^{1a1a_0}_{\infty}(Z,\p{}|Z_0,\p{0};\Ps{})+\Gcal^{1a1a_0}_{<S>}(Z,\p{}|Z_0,\p{0};\Ps{})$$
in  equation \eref{I00incohLadder}, we obtain the double scattering 
contribution by the particles. The term $\Gcal^{\infty\,aa_0}_{1}$ describes the direct  propagation of the wave between the two scatterers, and $\Gcal^{1a1a_0}_{<S>}$ describes
the interaction with the boundaries between the two scattering processes by the particles.
\section{Radiative transfer equation and boundaries conditions}
\label{radiative}
To derive the radiative transfer equation verified by $\Gcal^{1a1a_0}_{\ll SV\gg}(Z,\p{}|Z_0,\p{0};\Ps{})$, we must first  notice the following properties of $\Gcal_{\infty}^{1a1a_0}$ and $\Gcal^{1a1a_0}_{<S>}$:
\begin{eqnarray}
\fl \Der{}{\scriptstyle{Z}}\Gcal^{1a1a_0}_{\infty}(Z,\p{}|Z_0,\p{0};\Ps{})&=\frac{(2\pi)^2\,\delta(\p{}-\p{0})}{4\,\alpha_e(\p{}^+)\alpha_e^*(\p{}^-)}\,\op{I}_{\perp\,\otimes}^{1\,a}(\p{};\Ps{})\,\delta_{a,a_0}\,a\,\delta(Z-Z_0)\,,\nonumber\\
\fl &+\rmi\,a\,\Delta\alpha_e(\p{};\Ps{})\,\Gcal^{1a1a_0}_{\infty}(Z,\p{}|Z_0,\p{0};\Ps{})
\label{der1}\\
\fl \Der{}{\scriptstyle{Z}}\Gcal^{1a1a_0}_{<S>}(Z,\p{}|Z_0,\p{0};\Ps{})&=\rmi\,a\,\Delta\alpha_e(\p{};\Ps{})\,\Gcal^{1a1a_0}_{<S>}(Z,\p{}|Z_0,\p{0};\Ps{})\,,\label{der2}
\end{eqnarray}
with
\begin{eqnarray}
\op{I}^{1a}_{\perp\,\otimes}(\p{};\Ps{})=\op{I}_\perp^{1a}(\p{}^+)\otimes\op{I}_\perp^{1a\,*}(\p{}^-)\,.\\
\p{}^{\pm}=\p{}\pm\Psd{}\,.
\end{eqnarray}
To obtain equation \eref{der1}, we have to introduce in  equation \eref{defGinft} the identity
\begin{equation}
\lo \delta_{a,\sgn(Z-Z_0)}=\delta_{a,+}\Theta(Z-Z_0)+\delta_{a,-}\Theta(Z_0-Z)\,,
\end{equation}
with $\Theta$ as the Heaviside function, and then use the property:
\begin{equation}
\frac{\rmd \Theta}{\rmd z}(z)=\delta(z)\,.
\end{equation}
Hence, in taking the derivative of the Bethe-Salpeter equation \eref{Betheeq5}, we obtain the radiative transfer equation satisfied by $\Gcal^{1a1a_0}_{\ll SV\gg}(Z,\p{}|Z_0,\p{0};\Ps{})$~\footnote{To get this equation, we have supposed that the operator $\op{C}^{11}_{o}(\hvec{k}|\hvec{k}_0)$, which defined the operator $\opcal{P}^{11}$,
is transverse to the propagation direction defined by $\hvec{k}$: 
$\hvec{k}\cdot\op{C}^{11}_{o}(\hvec{k}|\hvec{k}_0)=0$. In this case, we have:
\begin{equation}
\op{I}_\perp^{1a}(\p{})\cdot\op{C}^{11}_{o}(\hvec{k}^{1a}_{\p{}}|\hvec{k}^{1a_0}_{\p{0}})=
\op{C}^{11}_{o}(\hvec{k}^{1a}_{\p{}}|\hvec{k}^{1a_0}_{\p{0}})\,.
\end{equation}
This hypothesis is satisfied in the far-field approximation.
}:
\begin{eqnarray}
\fl &\Der{}{\scriptstyle{Z}}\Gcal^{1a1a_0}_{\ll SV\gg}(Z,\p{}|Z_0,\p{0};\Ps{})=\frac{(2\pi)^2\,\delta(\p{}-\p{0})}{4\,\alpha_e(\p{}^+)\alpha_e^*(\p{}^-)}\,\op{I}_{\perp\,\otimes}^{1\,a}(\p{};\Ps{})\,\delta_{a,a_0}\,a\,\delta({Z-Z_0})\no\\
\fl &\quad+\frac{a}{4\,\alpha_e(\p{}^+)\alpha_e^*(\p{}^-)}\,\sum_{a_1=\pm}\intp{1}\,
\Pcal^{1a1a_1}(\p{}|\p{1};\Ps{}):\Gcal^{1a_1a_0}_{\ll SV\gg}(Z_{},\p{1}|Z_0,\p{0};\Ps{})\no\\
\fl &\quad+\rmi\,a\,\Delta\alpha_e(\p{};\Ps{})\,\Gcal^{1a_1a_0}_{\ll SV\gg}(Z_{},\p{}|Z_0,\p{0};\Ps{})\,.
\label{Rad}
\end{eqnarray}
To solve this differential equation, we need boundary conditions on the interface. 
In I, we have shown that the scattering operator $\op{S}^{1\pm1\pm}$ can be decomposed as a function of the scattering operators $\op{R}^{01}$ and $\op{R}^{H\,21}$ of the upper and lower rough surfaces. In using these equations, we easily show the following relationships between the scattering operators:
\begin{eqnarray}
\op{S}^{1-1-}=\op{R}^{01}\cdot\op{S}^{1+1-}\,,\label{RSdef1}\\
\op{S}^{1-1+}=\op{R}^{01}+\op{R}^{01}\cdot\op{S}^{1+1+}\,,\\
\op{S}^{1+1-}=\op{R}^{H\,21}+\op{R}^{H\,21}\cdot\op{S}^{1+1-}\,,\\
\op{S}^{1+1+}=\op{R}^{H\,21}\cdot\op{S}^{1-1+}\,.\label{RSdef4}
\end{eqnarray}
where we have used the notation,
\begin{equation}
[\op{f}\cdot\op{g}](\p{}|\p{0})\equiv\intpV{1}{}\,\op{f}(\p{}|\p{1})\cdot\op{g}(\p{1}|\p{0})\,.
\end{equation}
With equations (\ref{defGS},\ref{defGinft},\ref{RSdef1}-\ref{RSdef4}) and the definition of  $\op{R}^{H\,21}$ given in I:
\begin{equation}
\op{R}^{H\,21}(\p{}|\p{0})=\e^{\rmi\,(\alp{1}{}+\alp{1}{0})\,H}\,\op{R}^{21}(\p{}|\p{0})\,,
\end{equation}
we obtain
\begin{eqnarray}
\fl  \Gcal^{1-1a_0}_{\infty<S>}(Z=0,\p{}|Z_0,\p{0};\Ps{})\nonumber\\
=\intp{1}\,\opcal{R}^{01}_{\otimes\,<S>}(\p{}|\p{1};\Ps{}):\Gcal^{1+1a_0}_{\infty<S>}(Z=0,\p{1}|Z_0,\p{0};\Ps{})\,,\label{PropGinfty<S>1}\\
\fl \Gcal^{1+1a_0}_{\infty<S>}(Z=-H,\p{}|Z_0,\p{0};\Ps{})\nonumber\\ =\intp{1}\,\opcal{R}^{21}_{\otimes\,<S>}(\p{}|\p{1};\Ps{}):\Gcal^{1-1a_0}_{\infty<S>}(Z=-H,\p{1}|Z_0,\p{0};\Ps{})\,,\label{PropGinfty<S>2}
\end{eqnarray}
where
we have defined 
\begin{eqnarray}
\fl(2\pi)^2\,\delta(\vec{0})\,\opcal{R}^{01}_{\otimes\,<S>}(\p{}|\p{0};\vec{P})& =<\op{R}^{01}(\p{}^+|\p{0}^+)\otimes\op{R}^{01}(\p{}^-|\p{0}^-)>_S\,,\label{Radiacond1}\\
\fl(2\pi)^2\,\delta(\vec{0})\,\opcal{R}^{21}_{\otimes<S>}(\p{}|\p{0};\vec{P})&=<\op{R}^{21}(\p{}^+|\p{0}^+)\otimes\op{R}^{21}(\p{}^-|\p{0}^-)>_S\,.\label{Radiacond2}
\end{eqnarray}
Then, with the help of the Bethe-Salpeter equation \eref{Betheeq5}, we easily demonstrate that $\Gcal^{1a1a_0}_{\ll SV\gg}(Z,\p{}|Z_0,\p{0};\Ps{})$  verifies the following 
boundary conditions on the upper and lower rough surfaces:
\begin{eqnarray}
\fl \Gcal^{1-1a_0}_{\ll SV\gg}(Z=0,\p{}|Z_0,\p{0};\Ps{})\\
=\int\frac{\rmd^2\p{1}}{(2\pi)^2}\,\opcal{R}^{01}_{\otimes\,<S>}(\p{}|\p{1};\Ps{}):\Gcal^{1+1a_0}_{\ll SV\gg}(Z=0,\p{1}|Z_0,\p{0};\Ps{})\,,\label{RadBound1-0}\\
\fl \Gcal^{1+1a_0}_{\ll SV\gg}(Z=-H,\p{}|Z_0,\p{0};\Ps{})\\ =\int\frac{\rmd^2\p{1}}{(2\pi)^2}\,\opcal{R}^{21}_{\otimes\,<S>}(\p{}|\p{1};\Ps{}):\Gcal^{1-1a_0}_{\ll SV\gg}(Z=-H,\p{1}|Z_0,\p{0};\Ps{})\,,\label{RadBound2-0}
\end{eqnarray}
With  equation \eref{Rad} and the boundary conditions \eref{RadBound1-0}, \eref{RadBound2-0}
we  thus have obtained a closed system of equations sufficient to calculate the  tensors
$\Gcal^{1a1a_0}_{\ll SV\gg}$. In section \eref{Simplification}, we will show how to rewrite the radiative
transfer equation \eref{Rad} in a more classical form.
\section{Calculation of $\op{\sigma}^{incoh}_{L=0}$}
\label{secL=0}
In  sections \eref{MullerDef} and \eref{Bethe1}, we have shown that the first term of the incoherent
scattering cross-section is given by
\begin{equation}
\fl\frac{\alpha_0(\p{})^2}{\pi}\,\veccal{I}^{incoh}_{L=0}(\RR,\kk)=(2\pi)\,\delta(k_z-\alp{0}{})\,\op{\sigma}^{incoh}_{L=0}(\p{}|\p{0}):\veccal{J}^{0\,i}(\p{0})\,,\label{L=0}
\end{equation}
with $\vec{k}=\p{}+k_z\ez$ and
\begin{eqnarray}
\fl\veccal{I}^{incoh}_{L=0}(\RR,\kk)= & e_0\int_{V_0}\rmd^3\rr\e^{-\rmi\,\kk\cdot\rr}\,
\left[<\E^{0\,s}_{S}(\RR+\frac{\rr}{2})\otimes\E^{0\,s\,*}_{S}(\RR-\frac{\rr}{2})>_{S}\right.\nonumber\\
\fl &  \left.\qquad \qquad \qquad -<\E^{0\,s}_{S}(\RR+\frac{\rr}{2})>\otimes<\E^{0\,s\,*}_{S}(\RR-\frac{\rr}{2})>_{S}
\right]\,.\label{I00incoh0-2}
\end{eqnarray}
Furthermore, we know that the scattering intensity $\E^{0\,s}_{S}$ can be expressed as a function
of the scattering operator $\op{S}^{0+0-}$ by
\begin{equation}
\fl\E^{0s}_S(\x,z)=\intp{}\,\e^{\rmi\,\p{}\cdot\x+\rmi\,\alp{0}{}\,z}\,\op{S}^{0+,0-}(\p{}|\p{0})\cdot\E^{0\,i}(\p{0})\,.
\end{equation}
Then, we find
\begin{equation}
\fl\op{\sigma}^{incoh}_{L=0}(\p{}|\p{0})=\frac{\alpha_0(\p{})^2}{\pi}\,\opcal{S}^{0+0-incoh}_{\otimes <S>}(\p{}|\p{0})\,,\nonumber\\
\end{equation} 
where  $\opcal{S}^{0+0-incoh}_{\otimes <S>}$ is defined by
\begin{eqnarray}
\fl(2\pi)^2\delta(\vec{0})\,\opcal{S}^{0+0-incoh}_{\otimes<S>}(\p{}|\p{0})=& <\op{S}^{0+0-}(\p{}|\p{0})\otimes \op{S}^{0+0-}(\p{}|\p{0})>_S\nonumber\\
\fl & -<\op{S}^{0+0-}(\p{}|\p{0})>_{S}\otimes <\op{S}^{0+0-}(\p{}|\p{0})>_{S}\,.
\label{devsigma}
\end{eqnarray}
\section{Calculation of $\op{\sigma}^{incoh}_{L=1}$}
\label{secL=1}
For the first order scattering contribution by the particles, the scattered specific
intensity is
\begin{equation}
\fl \veccal{I}^{incoh}_{L=1}(\RR,\kk)= e_0\,\int_{V_0}\rmd^3\rr\e^{-\rmi\,\kk\cdot\rr}\,<\Gue{01}\otimes\Gue{01\,*}>_S:\Pcal^{11}:<\E^{1\,t}_{S}\otimes\E^{1\,t\,*}_{S}>_{S}\,,\label{I00incoh1bis2}
\end{equation}
and the cross-section verifies the following equation:
\begin{equation}
\fl\frac{\alpha(\p{})^2}{\pi}\,\veccal{I}^{incoh}_{L=1}(\RR,\kk)=(2\pi)\,\delta(k_z-\alp{0}{})\,\op{\sigma}^{incoh}_{L=1}(\p{}|\p{0}):\veccal{J}^{0\,i}(\p{0})\label{L=0bis}
\, .\label{Chap3gaincohterme1}
\end{equation}
In introducing equations (\ref{decompG0}, \ref{expG0+1a0}) and \ref{expE1t},which connect the Green function $\op{G}^{01}$ and the transmitted field $\vec{E}^{1t}$ to the scattering operators $\op{S}^{0+1\pm}$ and $\op{S}^{1\pm 0-}$, in \eref{I00incoh1bis2}
we obtain
\begin{eqnarray}
\fl\op{\sigma}^{incoh}_{L=1}(\p{}|\p{0})=\frac{\alpha_0(\p{})^2}{\pi}\,\!\!\!\!\!\sum_{a_1,a_2=\pm}\int\frac{\rmd^2\p{1}}{(2\pi)^2}\,\frac{\rmd^2\p{2}}{(2\pi)^2}\frac{1}{4\,|\alp{e}{2}|^2}\,\frac{\left( 
 \e^{\rmi[a_2\Delta\alpha_e(\p{2})-
 a_1\Delta\alpha_e(\p{1})]\,H}-1
\right)}{\rmi[a_2\Delta\alpha_e(\p{2})-a_1
\Delta\alpha_e(\p{1})]} \nonumber \\
\fl \times \opcal{S}^{0+;1a_2}_{\otimes\,<S>}(\p{}|\p{2};\Ps{}=0):\Pcal^{1a_21a_1}(\p{2}|\p{1};\vec{P}=\vec{0}):\opcal{S}^{1a_1;0-}_{\otimes\,<S>}(\p{1}|\p{0};\Ps{}=\vec{0})\,,\label{Chap4ordreL=1}
\end{eqnarray}
where 
\begin{eqnarray}
\fl (2\pi)^2\,\delta(\vec{0})\,\opcal{S}_{\otimes\,<S>}^{0+;1a_1}(\p{}|\p{0};\Ps{0})=<\op{S}^{0+1a_1}\left(\p{}^+|\p{0}^+\right)\otimes\op{S}^{0+1a_1}\left(\p{}^-|\p{0}^-\right)>_{S}\,,
\label{SwithP}\\
\fl(2\pi)^2\,\delta(\vec{0})\opcal{S}_{\otimes\,<S>}^{1a_2;0-}(\p{}|\p{0};\Ps{0})
=<\op{S}^{1a_2\,0-}\left(\p{}^+|\p{0}^+\right)\otimes\op{S}^{1a_2\,0-}\left(\p{}^-|\p{0}^-\right)>_{S}\,,\label{SwithP2}\\
\fl \Pcal^{1a_11a_2}(\p{2}|\p{1})= n\,w(\kk^{1a_2}_{\p{2}}-\kk^{1a_1}_{\p{1}})\,\op{C}^{11}_{o}(\kk^{1a_2}_{\p{2}^+}|\kk^{1a_1}_{\p{1}^+})\otimes\op{C}^{11}_{o}(\kk^{1a_2}_{\p{2}^-}|\kk^{1a_1}_{\p{1}^-})\,,\label{P11withP}
\end{eqnarray}
with
\begin{eqnarray}
\fl\vec{p}^{\pm}=\vec{p}\pm\Psd{}\,,\quad\vec{p}_2^{\pm}=\vec{p}_2\pm\Psd{}\,,\quad\vec{p}_{1}^{\pm}=\vec{p}_1\pm\Psd{}\,,\quad\vec{p}_0^{\pm}=\vec{p}_0\pm\Psd{}\,,\\
\vec{k}^{1a}_{\p{}}= p{}+a\,\alpha'_{e}(\p{})\,\ez\,,
\end{eqnarray}
and
\begin{eqnarray}
\Delta\alpha_e(\p{}) &=
\alp{e}{}-[\alp{e}{}]^*\,\\
&=2\rmi\,\Im[\alp{e}{}].
\end{eqnarray}
\section{Calculation of $\op{\sigma}^{incoh}_{Ladder}$}
\label{secLadder}
Similarly to the previous section, the multiple scattering contribution is given by
\begin{eqnarray}
\fl \veccal{I}^{incoh}_{Ladder}(\RR,\vec{k})=e_0\,\int_{V_0}\rmd^3\rr\,\e^{-\rmi\,\kk\cdot\rr}\,<\Gue{01}\otimes\Gue{01\,*}>_S:\Pcal^{11}:\opcal{G}^{11}_{\ll SV\gg}\no\\
\lo :\Pcal^{11}:<\E^{1\,t}_{S}\otimes\E^{1\,t\,*}_{S}>_{S}\,,\label{I00incohLadderbis}
\end{eqnarray}
and the cross-section is
\begin{equation}
\fl\frac{\alpha(\p{})^2}{\pi}\,\veccal{I}^{incoh}_{Ladder}(\RR,\kk)=(2\pi)\,\delta(k_z-\alp{0}{})\,\op{\sigma}^{incoh}_{Ladder}(\p{}|\p{0}):\veccal{J}^{0\,i}(\p{0})\,.\label{Ladder}
\end{equation}
From equations (\ref{decompG0}, \ref{expG0+1a0}, \ref{expE1t}) and  properties (\ref{fourierXX0}, \ref{GcalRRkkbis}) of the tensor $\Gcal^{1a 1a_0}_{\ll SV\gg}$, we find
\begin{eqnarray}
\fl\op{\sigma}^{incoh}_{Ladder}(\p{}|\p{0})=\frac{\alpha(\p{})^2}{\pi}\,\!\!\!\!\!\sum_{a_1,a_2,a_3,a_4=\pm}\int_{-H}^0\,\rmd 
Z_{43}\,\rmd
Z_{21}\int\frac{\rmd^2\p{1}}{(2\pi)^2}\,\frac{\rmd^2\p{2}}{(2\pi)^2}\,\frac{\rmd^2\p{3}}{(2\pi)^2}\frac{\rmd^2\p{4}}{(2\pi)^2}\nonumber\\
\fl 
\frac{\e^{-\rmi\,a_4\,\Delta\alpha_e(\p{4})\,Z_{43}+\rmi\,a_1\,\Delta\alpha_e(\p{1})\,Z_{21}}}{4\,|\alpha_e(\p{4})|^2}
\opcal{S}_{\otimes\,<S>}^{0+\,1a_4}(\p{}|\p{4};\Ps{}=\vec{0}):\Pcal^{1a_4 1
  a_3}(\p{4}|\p{3};\Ps{}=\vec{0})\nonumber\\\fl\times \Gcal^{1a_3 1a_2}_{\ll SV\gg}(Z_{43},\p{3}|Z_{21},\p{2};\Ps{}=0)
:\Pcal^{1a_2 1a_1}(\p{2}|\p{1};\Ps{}=\vec{0}):\opcal{S}_{\otimes\,<S>}^{1a_1\,0-}(\p{1}|\p{0};\Ps{}=0)\,.
\nonumber\\ \label{Chap3Ladder}
\end{eqnarray}
\section{Enhanced backscattering and reciprocity}
\label{secCrossed}
\begin{figure}[htbp]
   \centering
      \psfrag{k}{$\kk$}
     \psfrag{k0}{$\kk_0$}
     \psfrag{x}{$x$}
    
      \epsfig{file=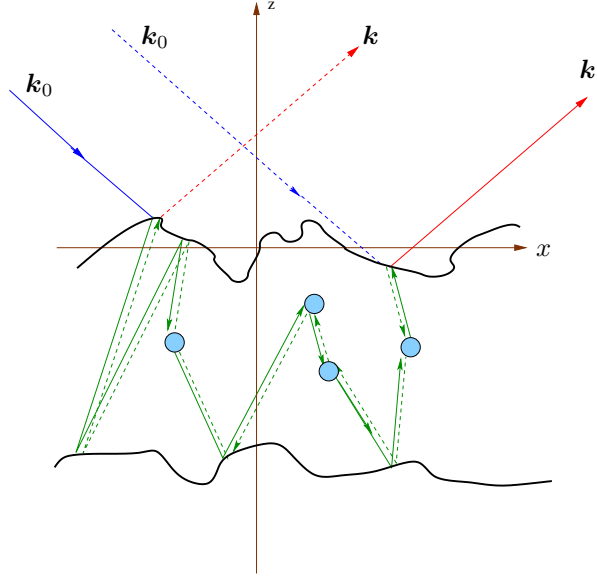,width=8cm}
      \caption{Scattering processes which are taken into account by the most-crossed contribution.}
      \label{Chap3Fig12}
 \end{figure}

During the last decade, several studies have been concerned with the enhanced backscattering
~\cite{Kong,Apresyan,Sheng1,Sheng2,POAN,Rossum,Bara1,Mark,Bara2,Mark2,Tsang2,Tsang3,Tsang4,Akkermans}. This phenomenon produces
a peak in the backscattering direction ($\hvec{k}=-\hvec{k}_0$) due to the interference of waves following the same
path but in opposite directions~(Figure \eref{Chap3Fig12}).
The most elegant way to introduce the enhanced backscattering in the theory is to use the reciprocity principle~\cite{Maynard1,Tig3}. Heuristically, it means that when we exchange the source and the detector position (respectively given by $\rr_0$ and $\rr$) and also their polarization, the field measured is the same. In using our Green functions, we translate this statement by
\begin{equation}
\op{G}^{00}_{SV}(\rr,\rr_0)=[\op{G}^{00}_{SV}(\rr_0,\rr)]^T\,,\label{ProprecipG00}
\end{equation}
where the transpose of dyad $\op{f}$, which exchanges the polarization, is defined by $\op{f}^T\cdot\vec{E}=\vec{E}\cdot\op{f}$.
If we decompose the dyad $\op{G}^{00}_{SV}(\rr,\rr_0)$ in the fixed basis $[\ex,\ey,\ez]$,
\begin{equation}
\op{G}^{00}_{SV}(\rr,\rr_0)=\sum_{i,j=x,y,z}\,{G}^{00}_{SV}(\rr,\rr_0)_{ij}\,\hvec{e}_i\hvec{e}_j\,,
\end{equation}
we re-write equation \eref{ProprecipG00} under the following form:
\begin{equation}
{G}^{00}_{SV}(\rr,\rr_0)_{ij}={G}^{00}_{SV}(\rr_0,\rr)_{ji}\,.
\end{equation}
The properties in \eref{ProprecipG00} can be demonstrated by using the Green theorem~\cite{Tai}.
By applying these properties on each element of the tensor, $$\opcal{G}^{00}_{\ll SV\gg}(\rr,\rr'|\rr_0,\rr_0')=\ll\op{G}^{00}_{SV}(\rr,\rr_0)\otimes\op{G}^{00\,*}_{SV}(\rr',\rr_0')\gg_{SV}\,,$$
we obtain three identities that must be satisfied:
\begin{eqnarray}
\fl\op{G}^{00}_{SV}\otimes\op{G}^{00\,*}_{SV}(\rr,\rr'|\rr_0,\rr_0')=[\op{G}^{00}_{SV}\otimes\op{G}^{00\,*}_{SV}(\rr_0,\rr_0'|\rr,\rr')]^{T_{LR}}\,,\label{Recipenergie}\\
\fl\op{G}^{00}_{SV}\otimes\op{G}^{00\,*}_{SV}(\rr,\rr'|\rr_0,\rr_0')=[\op{G}^{00}_{SV}\otimes\op{G}^{00\,*}_{SV}(\rr,\rr_0'|\rr_0,\rr')]^{T_{R}}\,,\label{Recipbackscattering}\\
\fl\op{G}^{00}_{SV}\otimes\op{G}^{00\,*}_{SV}(\rr,\rr'|\rr_0,\rr_0')=[\op{G}^{00}_{SV}\otimes\op{G}^{00\,*}_{SV}(\rr_0,\rr'|\rr,\rr'_0)]^{T_{L}}\,,\label{Recipbackscattering2}
\end{eqnarray}
where we have defined three transposes of a tensor $\opcal{M}$ with the following
decomposition (see \ref{AppTensor}):
\begin{displaymath}
\opcal{M}=\sum_{i,j,i',j'=x,y,z}\,\mathcal{M}_{ii';jj'}\left(\hvec{e}_i\otimes\hvec{e}_{i'}\right)\left(\hvec{e}_j\otimes\hvec{e}_{j'}\right)\,,
\end{displaymath}
by
\begin{eqnarray}
\lo[\mathcal{M}_{ii';jj'}]^{T_{LR}}=\mathcal{M}_{jj';ii'}\,,\\
\lo[\mathcal{M}_{ii';jj'}]^{T_{L}}=\mathcal{M}_{ji';ij'}\,,\\
\lo[\mathcal{M}_{ii';jj'}]^{T_{R}}=\mathcal{M}_{ij';ji'}\,.\label{Chap4DefTransR}
\end{eqnarray}
We can easily check that if two of the conditions (\ref{Recipenergie}-\ref{Recipbackscattering2}) are satisfied, the third condition comes automatically.
These properties can be reformulated by using the tensor  $\op{\Gamma}^{11}_{S}=<\op{T}^{11}_{SV}\otimes\op{T}^{11}_{SV}>_V$:
\begin{eqnarray}
\op{\Gamma}^{11}_{S}(\rr,\rr'|\rr_0,\rr_0')=[\op{\Gamma}^{11}_{S}(\rr_0,\rr_0'|\rr,\rr')]^{T_{LR}}\,,\label{RecipenergieG}\\
\op{\Gamma}^{11}_{S}(\rr,\rr'|\rr_0,\rr_0')=[\op{\Gamma}^{11}_{S}(\rr,\rr_0'|\rr_0,\rr')]^{T_{R}}\,,\label{RecipbackscatteringG_0}\\
\op{\Gamma}^{11}_{S}(\rr,\rr'|\rr_0,\rr_0')=[\op{\Gamma}^{11}_{S}(\rr_0,\rr'|\rr,\rr'_0)]^{T_{L}}\,,\label{RecipbackscatteringG}
\end{eqnarray}
In fact, if we suppose that the scattering operators $\op{S}^{0+0-}$, $\op{S}^{1\pm 0}$, and $\op{S}^{0 1\pm }$
verify the reciprocity principle, it can be easily checked that the Green functions $\op{G}^{00}$, $\op{G}^{10}$, and $\op{G}^{01}$ verify a reciprocity condition similar to  
property \eref{ProprecipG00}. Now, in section \eref{Bethe1} we have demonstrated  that
\begin{eqnarray}
\fl \GVu{00}\otimes\GVu{00\,*}=\Gue{00}\otimes\Gue{00\,*} 
+\Gue{01}\otimes\Gue{01\,*}:
\op{\Gamma}^{11}_S:\Gue{10}\otimes\Gue{10\,*}\,, \label{G00eq-2}
\end{eqnarray}
and from the reciprocity of the Green functions $\op{G}^{00}$, $\op{G}^{10}$, $\op{G}^{01}$,  we deduce properties (\ref{RecipenergieG}-\ref{RecipbackscatteringG}).
The first condition \eref{RecipenergieG} is a reciprocity condition on energy. If we write
it using a Wigner transform, we have
\begin{equation}
\op{\Gamma}^{11}_{S}(\RR,\kk|\RR_0,\kk_0)=[\op{\Gamma}^{11}_{S}(\RR_0,-\kk_0|\RR,-\kk)]^{T_{LR}}\,.\label{GamareciprEner}
\end{equation}
Equation \eref{GamareciprEner} signifies that if the source and
detector positions are exchanged ($\RR \leftrightarrow \RR_0$) with their polarization (transpose $T_{LR}$) 
 and if the incident and scattered wave  directions are exchanged and inversed ($\kk \leftrightarrow -\kk_0$), then intensity measured is the same.
In section \eref{Bethe1}, we have also demonstrated that the tensor $\Gamcal^{11}_{S}$
is given by  equation \eref{gamcalGueGue}:
\begin{eqnarray}
\fl\Gamcal^{11}_{S}(\RR,\kk|\RR_0,\kk_0)=\Pcal^{11}(\RR,\kk|\RR_0,\kk_0)+\int_{V_1}\rmd^3\RR_1\rmd^3\RR_2\int\frac{\rmd^3\kk_1}{(2\pi)^3}\frac{\rmd^3\kk_2}{(2\pi)^3}\Pcal^{11}(\RR,\kk|\RR_2,\kk_2)\nonumber\\
\lo:<\op{G}^{11}_{SV}\otimes\op{G}^{11\,*}_{SV}>_V(\RR_2,\kk_2|\RR_1,\kk_1):\Pcal^{11}(\RR_1,\kk_1|\RR_0,\kk_0)\,.\label{gamcalGueGue2}
\end{eqnarray}
Furthermore, with the help of the Wigner representation \eref{exP11-wigner} of 
 $\opcal{P}^{11}$ and the  Bethe-Salpeter equation \eref{Betheeq4}, we demonstrate that
 \begin{eqnarray}
\fl\opcal{P}^{11}(\RR,\kk|\RR_0,\kk_0)=[\opcal{P}^{11}(\RR_0,-\kk_0|\RR,-\kk)]^{T_{LR}}\,,\label{P11wignerrecipr}\\
\fl<\op{G}^{11}_{SV}\otimes\op{G}^{11\,*}_{SV}>_V(\RR,\kk|\RR_0,\kk_0)=[<\op{G}^{11}_{SV}\otimes\op{G}^{11\,*}_{SV}>_V(\RR_0,-\kk_0|\RR,-\kk)]^{T_{LR}}\,,\label{G11wignerrecip}
\end{eqnarray}
if we suppose that the Green function $\op{G}_S^{11}$ and the scattering operator
$\op{C}^{11}_{o}(\kk|\kk_0)$ verify the reciprocity principle.
For the operator $\op{C}^{11}_{o}$, this principle can be formulated as 
\begin{equation}
\op{C}^{11}_{o}(\kk|\kk_0)=[\op{C}^{11}_{o}(-\kk_0|-\kk)]^{T}\,.\label{reciprP}
\end{equation}
In I, we 
have  shown that this property is effectively  satisfied since the scattering operator $\op{t}^{11}_o$ for one particle verifies
a similar property~\cite{Hulst1,Bohren,Mish1}. 
Accordingly, from  properties \eref{P11wignerrecipr} and \eref{G11wignerrecip} and  equation 
\eref{gamcalGueGue2}, we demonstrate that 
 $\op{\Gamma}^{11}_{S}(\RR,\kk|\RR_0,\kk_0)$ satisfies  condition \eref{GamareciprEner}.
 
 However, tensor $\op{\Gamma}^{11}_{S}(\RR,\kk|\RR_0,\kk_0)$ does not satisfy conditions \eref{RecipbackscatteringG_0} and \eref{RecipbackscatteringG}. In fact, these conditions describe the scattering process where
 two waves follow the same path but in opposite directions as shown in figure \eref{Chap3Fig12}.
 If we decompose  tensor $\op{\Gamma}^{11}_{S}$ under the form:
\begin{equation}
\Gamcal^{11}_{S}=\Pcal^{11}+\Gamcal^{11}_{S\,Ladder}\,,\label{gamcalGueGue2bis}
\end{equation}
with
\begin{equation}
\Gamcal^{11}_{S\,Ladder}=\Pcal^{11}:<\GVu{11}\otimes\GVu{11\,*}>_V:\Pcal^{11}\,,
\end{equation}
it can be checked that only $\Gamcal^{11}_{S\,Ladder}$ does not verify equations 
\eref{RecipbackscatteringG_0} and \eref{RecipbackscatteringG}. 
In fact, we show, in using the inverse Wigner transform $\Pcal^{11}(\rr,\rr'|\rr_0,\rr_0')$ of the tensor 
$\Pcal^{11}(\RR,\kk|\RR_0,\kk_0)$, that all the conditions (\eref{RecipenergieG}-\eref{RecipbackscatteringG}) are satisfied by $\Pcal^{11}(\rr,\rr'|\rr_0,\rr_0')$.

\begin{figure}[htbp]
   \centering
     
      \psfrag{Gam}{$\scriptstyle
        \op{\Gamma}^{11}_{S\,Crossed}$}
      \psfrag{Cj}{$\scriptstyle \op{C}^{11}_{o}$}
      \psfrag{Cjetoile}{$\scriptstyle \op{C}^{11\,*}_{o}$}
      
       \psfrag{Cj2}{$\scriptstyle \op{C}^{11}_{o}$}
      \psfrag{Cjetoile2}{$\scriptstyle \op{C}^{11\,*}_{o}$}

       \psfrag{Gk}{$\scriptstyle \op{G}^{11}_{S}$}
       \psfrag{Gketoile}{$\scriptstyle \op{G}^{11\,*}_{S}$}
			\epsfig{file=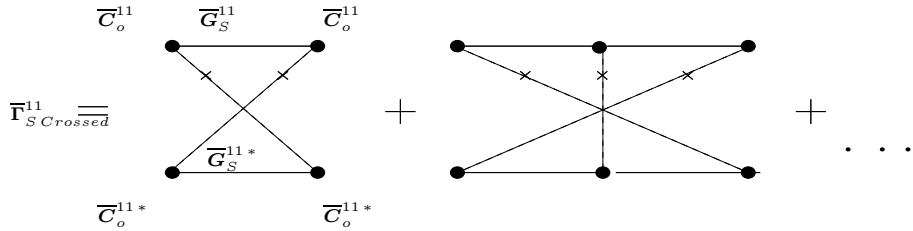,width=12cm}
      \caption{"Most-crossed"' diagrams.}
      \label{FigCrossed2}
 \end{figure}

When we applied the reciprocity condition
\eref{RecipbackscatteringG_0} to the tensor $\Gamcal^{11}_{S\,Ladder}$, we obtain a new tensor $\Gamcal^{11}_{S\,Crossed}$:
\begin{equation}
\Gamcal^{11}_{S\,Crossed}(\rr,\rr'|\rr_0,\rr_0')=[\Gamcal^{11}_{S\,Ladder}(\rr,\rr_0'|\rr_0,\rr')]^{T_R}\,.\label{DefGammaCroosed}
\end{equation}
To insure that $\Gamcal^{11}_{S}$ verifies all the reciprocity conditions,
we add the contribution $\Gamcal^{11}_{S\,Crossed}$ to $\Gamcal^{11}_{S}$:
\begin{equation}
\Gamcal^{11}_{S}=\Pcal^{11}+\Gamcal^{11}_{S\,Ladder}+\Gamcal^{11}_{S\,Crossed}\,.\label{gamcalGueGue2_2}
\end{equation}
This new contribution can be represented with the help of diagrams as the set of "most-crossed  diagrams"~\cite{Apresyan,Sheng1,Sheng2,Bara1}, as shown on \Fref{FigCrossed2}.
 In using the identity \eref{DefGammaCroosed} in the calculation of the ladder approximation, we obtain
the  contribution  of $\Gamcal^{11}_{S\,Crossed}$ to the cross-section:
\begin{eqnarray}
\fl\op{\sigma}^{incoh}_{Crossed}(\p{}|\p{0})=\frac{\alpha_0(\p{})\alpha_0(\p{0})}{\pi} \sum_{a_1,a_2,a_3,a_4=\pm} 
\int_{-H}^0\,\rmd 
Z_{43}\,\rmd
Z_{21}\int\frac{\rmd^2\p{1}}{(2\pi)^2}\,\frac{\rmd^2\p{2}}{(2\pi)^2}\,\frac{\rmd^2\p{3}}{(2\pi)^2}\frac{\rmd^2\p{4}}{(2\pi)^2}\,
\nonumber\\\fl \frac{\e^{-\rmi\,a_4\,\Delta\alpha_e(\p{4};\Ps{})\,Z_{43}+\rmi\,a_1\,\Delta\alpha_e(\p{1};\Ps{})\,Z_{21}}}{4\,\alpha_e(\Psd{}+\p{4})\,\alpha^*_e(-\Psd{}+\p{4})}
\left[\opcal{S}_{\otimes\,<S>}^{0+\,1a_4}(\Delta \p{}|\p{4};\Ps{}):\Pcal^{1a_4
  1 a_3}(\p{4}|\p{3};\Ps{})\right.\nonumber\\  \fl\left.:\Gcal^{1a_3
  1a_2}_{\ll SV\gg}(Z_{43},\p{3}|Z_{21},\p{2};\Ps{})
:\Pcal^{1a_21a_1}(\p{2}|\p{1};\Ps{}):\opcal{S}_{\otimes\,<S>}^{1a_1\,0-}(\p{1}|-\Delta \p{};\Ps{})\right]^{T_R}\,,
\label{cont crossed}
\end{eqnarray}
where we have used notations (\ref{SwithP}-\ref{P11withP}) and defined:
\begin{eqnarray}
&\Ps{}=\p{}+\p{0}\,,\label{defPC}\\
&\Delta \p{}=\frac{\p{}-\p{0}}{2}\\
&\Delta\alpha(\p{};\vec{P})=\alpha_e(\p{}+\vec{P}/2)-\alpha^*_e(\p{}-\vec{P}/2)\,.\label{defalpP}
\end{eqnarray}
The definition of the right transpose $^{T_R}$ when the tensors are expressed in the 
basis $[\evp{0\pm}{},\ehp{}]$ and not in the the fixed basis $[\ex,\ey,\ez]$ is defined in \ref{AppTran}.
It is worth mentioning that to calculate the contribution, we need to know the tensor
$\Gcal^{1a_31a_2}_{\ll SV\gg}$, as in the ladder approximation, 
and thus, we can use the same radiative transfer equation to calculate the ladder and the crossed
terms. The difference between the two calculations  is  the value of $\vec{P}$ which is the null vector for the ladder approximation and is given by equation \eref{defPC} for the crossed contribution.
\section{Another formulation}
\label{anotherFor}
In the previous development, the main quantity that has to be determined in using the radiative transfer
equation is the tensor $\Gcal^{1a_31a_2}_{\ll SV\gg}(Z_{43},\p{3}|Z_{21},\p{2};\Ps{})$. The main advantage of this formulation is that the ladder contribution and the crossed contribution
can be evaluated as soon as $\Gcal^{1a_31a_2}_{\ll SV\gg}(Z_{43},\p{3}|Z_{21},\p{2};\Ps{})$ is known. Nevertheless, this approach is rather unusual, and in this section, we introduce the  specific intensity $\opcal{I}(\RR,\p{})$, and we show that this specific intensity satisfies 
the classical radiative transfer equation.
First,  we introduce the reduced intensity by
\begin{equation}
\opcal{I}_{\otimes\,red}^{1a_1}(Z_1,\p{1};\vec{P})=\opcal{S}^{1a_10-}_{\otimes<S>}(\p{1}|-\Delta \p{};\vec{P})\,\e^{\rmi a_1\,\Delta\alpha_e(\p{1};\vec{P})Z_1}\,.\label{Ired}
\end{equation}
It can be easily shown that this intensity satisfies the radiative transfer equation \eref{Rad}
without the source and the scattering terms:
\begin{equation}
\fl \Der{}{\scriptstyle{Z_1}}\opcal{I}_{\otimes\,red}^{1a_1}(Z_1,\p{1};\vec{P})=
\rmi\,a_1\,\Delta\alpha_e(\p{1};\Ps{})\opcal{I}_{\otimes\,red}^{1a_1}(Z_1,\p{1};\vec{P})\,.
\label{RadRed}
\end{equation}
We now define  the tensor $\opcal{J}^{1a}_{\otimes}(Z,\p{})$ by
\begin{eqnarray}
\fl \opcal{J}^{1a}_{\otimes}(Z,\p{})=\sum_{a_1=\pm}\int\frac{\rmd^2\p{1}}{(2\pi)^2}\opcal{P}^{1a1a_1}(\p{}|\p{1};\vec{P}):\opcal{I}^{1a_1}_{\otimes\,red}(Z,\p{1};\vec{P})\nonumber\\
\fl + \!\!\!\!\sum_{a_1,a_2,a_3=\pm}\int_{-H}^0\rmd Z_{21}\int\frac{\rmd^2\p{1}}{(2\pi)^2}\frac{\rmd^2\p{2}}{(2\pi)^2}\frac{\rmd^2\p{3}}{(2\pi)^2}\,\opcal{P}^{1a1a_3}(\p{}|\p{3};\vec{P}):\opcal{G}^{1a_31a_2}_{\ll SV\gg}(Z_{},\p{3}|Z_{21},\p{2};\vec{P})\nonumber\\
\lo:\opcal{P}^{1a_21a_1}(\p{2}|\p{1};\vec{P}):\opcal{I}^{1a_1}_{\otimes\,red}(Z_{21},\p{1};\vec{P})\,,\label{defJ}
\end{eqnarray}
and the diffuse specific intensity $\opcal{I}_{\otimes\,d}^{1a}(Z,\p{})$ by
\begin{equation}
\fl \opcal{I}_{\otimes\,d}^{1a}(Z,\p{})=\!\!\!\!\sum_{a_1=\pm}\int_{-H}^0\rmd Z_1\int\frac{\rmd^2\p{1}}{(2\pi)^2}\,\opcal{G}^{1a1a_1}_{\infty <S>}(Z_{},\p{}|Z_1,\p{1};\vec{P}):\opcal{J}^{1a_1}_{\otimes}(Z_1,\p{1})\,,\label{defId}
\end{equation}
where $\opcal{I}_{\otimes\,d}^{1+}(Z,\p{})$ corresponds to upward wave propagating in the direction
$\vec{k}=\p{}+\alpha'_e(\p{})\ez$, and $\opcal{I}_{\otimes\,d}^{1-}(Z,\p{})$ to down-going wave propagating in the direction $\vec{k}=\p{}-\alpha'_e(\p{})\ez$. 
We notice that in the definition of $\opcal{J}_{\otimes}^{1a}(Z,\p{})$,
we have added the first order scattering term to the high-order scattering contribution.
With these definitions, we can express the scattering cross-section as a function of $\opcal{I}_{\otimes\,d}^{1a}(Z,\p{})$. First, 
in noticing the following properties of the scattering operators $\op{S}^{0+1\pm}$:
\begin{equation}
\op{S}^{0+1-}=\op{T}^{01}\cdot\op{S}^{1+1-}\,,\qquad \op{S}^{0+1+}=\op{T}^{01}+\op{T}^{01}\cdot\op{S}^{1+1+}\,,
\end{equation}
that we rewrite in one equation:
\begin{eqnarray}
\fl\op{S}^{0+1a_0}(\p{}|\p{0})=\intp{1}\op{T}^{01}(\p{}|\p{1})\nonumber\\
\lo\cdot \left[(2\pi)^2\delta(\p{1}-\p{0})\delta_{a_0,+}
\,\op{I}_{\perp}^{1+}(\p{0})+\op{S}^{1+1a_0}(\p{1}|\p{0})\right]\,.
\end{eqnarray}
In using our hypothesis of statistical independence of scattering processes on the boundaries,
we have
\begin{eqnarray}
\fl \opcal{S}^{0+1a_0}_{\otimes<S>}(\p{}|\p{0};\vec{P})=\intp{1}\opcal{T}^{01}_{\otimes<S>}(\p{}|\p{1};\vec{P})\nonumber\\
\fl \qquad \cdot[(2\pi)^2\delta(\p{1}-\p{0})\delta_{a_0,+}\,\op{I}_{\perp}^{1+}(\p{0})\otimes\op{I}_{\perp}^{1+\,*}(\p{0})+\opcal{S}^{1+1a_0}_{\otimes<S>}(\p{1}|\p{0};\vec{P})]\,,
\end{eqnarray}
where
\begin{equation}
\fl (2\pi)^2\delta(\vec{0})\,\opcal{T}^{01}_{\otimes<S>}(\p{}|\p{0};\vec{P})=<\op{T}^{01}(\p{}^+|\p{0}^+)\otimes\op{T}^{01\,*}(\p{}^-|\p{0}^-)>_S\,,
\end{equation}
with $\p{}^{\pm}=\p{}\pm\vec{P}/2$ and $\p{0}^{\pm}=\p{0}\pm\vec{P}/2$.
Then, from  definition \eref{defG11infS} of $\opcal{G}^{11}_{\infty<S>}$ and equations \eref{defGS} and \eref{defGinft},
we obtain
\begin{eqnarray}
\fl \opcal{S}_{\otimes<S>}^{0+1a_4}(\p{}|\p{4};\vec{P})\,\frac{\e^{-\rmi a_4\Delta\alpha_e(\p{4};\vec{P})Z_4}}{4\alpha_e(\vec{p}_4^+)\alpha^*_e(\vec{p}_4^-)}\nonumber\\
\lo=\intp{1}\opcal{T}^{01}_{\otimes<S>}(\p{}|\p{1};\vec{P}):\opcal{G}^{1+1a_4}_{\infty<S>}(Z=0,\vec{p}_1|Z_4,\p{4};\vec{P})\,,\label{relSTG}
\end{eqnarray}
with $\p{4}^{\pm}=\p{4}\pm\vec{P}/2$.
In comparing equations (\ref{Chap4ordreL=1}, \ref{Chap3Ladder}, \ref{cont crossed}) with
definitions (\ref{Ired}, \ref{defJ}, \ref{defId}) and  expression \eref{relSTG}, we obtain the following expression for
the scattering cross-sections:
\begin{eqnarray}
\fl \op{\sigma}_{L=1}^{incoh}(\p{}|\p{0})+\op{\sigma}_{Ladder}^{incoh}(\p{}|\p{0})=\frac{\alpha_0(\p{})^2}{\pi}\nonumber\\
\lo \times \intp{4}\opcal{T}^{01}_{\otimes<S>}(\p{}|\p{4};\vec{P}=\vec{0}):\opcal{I}^{1+}_{\otimes\,d}(Z=0,\p{4};\vec{P}=\vec{0}) \,,\nonumber\\\fl
\\
\fl \op{\sigma}_{Crossed}^{incoh}(\p{}|\p{0})=\frac{\alp{0}{}\alp{0}{0}}{\pi} \,\nonumber\\
\fl \times \Big[\intp{4}\opcal{T}^{01}_{\otimes<S>}(\Delta\p{}|\p{4};\vec{P})
:\opcal{I}^{1+}_{\otimes d}(Z=0,\p{4};\vec{P}) \Big]^{T_R}-\frac{\alp{0}{0}}{\alp{0}{}}\op{\sigma}_{L=1}^{incoh}(\p{}|\p{0};\vec{P})
\end{eqnarray}
with $\vec{P}=\p{}+\p{0}$, and $\Delta\p{}=(\p{}-\p{0})/2$. The term $\op{\sigma}_{L=1}^{incoh}(\p{}|\p{0};\vec{P})$ is obtained from $\op{\sigma}_{L=1}^{incoh}(\p{}|\p{0})$  in replacing in equation \eref{Chap4ordreL=1}, 
$\vec{P}=0$ by $\vec{P}=\p{}+\p{0}$ and $\Delta\alpha_e(\p{})$ by $\Delta\alpha_e(\p{};\vec{P})$
defined in \eref{defalpP}.
To get the radiative transfer equation satisfied by $\opcal{I}^{1a}(Z,\p{};\vec{P})$, we first derive  equation \eref{defId} by using (\ref{der1}, \ref{der2}):
\begin{equation}
\fl \frac{\partial \opcal{I}^{1a}_{\otimes d}}{\partial {\scriptstyle Z}}(Z,\p{};\vec{P})=\frac{a_4}{\alpha_e(\p{}^+)\alpha_e^*(\p{}^-)}\opcal{J}^{1a}_{\otimes}(Z,\p{};\vec{P})+\rmi a\Delta\alpha_e(\p{};\vec{P})\,\opcal{I}^{1a}_{\otimes d}(Z,\p{};\vec{P})\,.\label{derI}
\end{equation}
Next, we express the tensor $\opcal{J}^{1a}_{\otimes}$ as a function of $\opcal{I}^{1a}_{\otimes \,d}$. 
From definition \eref{defJ} and the integral equation \eref{Betheeq5} satisfied by $\opcal{G}^{1a_2a_1}_{\ll SV\gg}$, we obtain an integral equation on $\opcal{J}^{1a}_{\otimes}$:
\begin{eqnarray}
\fl \opcal{J}^{1a}_{\otimes}(Z,\p{})= \sum_{a_1=\pm}\int\frac{\rmd^2\p{1}}{(2\pi)^2}\opcal{P}^{1a1a_1}(\p{}|\p{1};\vec{P}):\opcal{I}^{1a_1}_{\otimes red}(Z,\p{1};\vec{P})\nonumber\\
\fl + \sum_{a_1,a_2=\pm}\int_{-H}^0\rmd Z_{1}\int\frac{\rmd^2\p{1}}{(2\pi)^2}\frac{\rmd^2\p{2}}{(2\pi)^2}\opcal{P}^{1a1a_2}(\p{}|\p{2};\vec{P}):\opcal{G}^{1a_21a_1}_{\infty<S>}(Z_{},\p{2}|Z_1,\p{1};\vec{P})\nonumber\\
\lo\qquad \qquad :\opcal{J}^{1a_1}_{\otimes}(Z_1,\p{1};\vec{P})\,,\label{eqsurJ}
\end{eqnarray}
and introducing definition \eref{defId} in the second term of right hand side of equation \eref{eqsurJ}, we express $\opcal{J}^{1a}_{\otimes}$ as a function of $\opcal{I}^{1a_1}_{\otimes d}$:
\begin{eqnarray}
\fl \opcal{J}^{1a}_{\otimes}(Z,\p{})= \sum_{a_1=\pm}\int\frac{\rmd^2\p{1}}{(2\pi)^2}\opcal{P}^{1a1a_1}(\p{}|\p{1};\vec{P}):\opcal{I}^{1a_1}_{\otimes red}(Z,\p{1};\vec{P})\nonumber\\
\lo+ \sum_{a_1=\pm}\int\frac{\rmd^2\p{1}}{(2\pi)^2}\opcal{P}^{1a1a_1}(\p{}|\p{1};\vec{P}):\opcal{I}^{1a_1}_{\otimes d}(Z,\p{1};\vec{P})\,.\label{expJfuncI}
\end{eqnarray}
By introducing equation \eref{expJfuncI} in \eref{derI}, we obtain a radiative
transfer equation on $\opcal{I}^{1a}_{\otimes\, d}$:
\begin{eqnarray}
\fl \frac{\partial}{\partial {\scriptstyle Z}} \opcal{I}^{1a}_{\otimes \,d}(Z,\p{};\vec{P})=
\frac{a_4}{\alpha_e(\p{}^+)\alpha_e^*(\p{}^-)}\sum_{a_1=\pm}\int\frac{\rmd^2\p{1}}{(2\pi)^2}\,\opcal{P}^{1a1a_1}(\p{}|\p{1};\vec{P}):\opcal{I}^{1a_1}_{\otimes\, d} (Z,\p{1};\vec{P})\,,\nonumber\\
\fl+\rmi a\Delta\alpha_e(\p{};\vec{P})\,\opcal{I}^{1a}_{\otimes\, d}(Z,\p{};\vec{P})+\opcal{S}^{1a}_{\otimes}(Z,\p{};\vec{P})\,,
\label{RadsurI}
\end{eqnarray}
where $\opcal{S}_{\otimes}^{1a}$ is the source term:
\begin{equation}
\fl \opcal{S}^{1a}_{\otimes}(z,\p{};\vec{P})=\frac{a_4}{\alpha_e(\p{}^+)\alpha_e^*(\p{}^-)}
\sum_{a_1=\pm}\int\frac{\rmd^2\p{1}}{(2\pi)^2}\opcal{P}^{1a1a_1}(\p{}|\p{1};\vec{P}):\opcal{I}^{1a_1}_{\otimes red}(Z,\p{1};\vec{P})\nonumber\\
\end{equation}
From definition \eref{defId} and the properties (\ref{PropGinfty<S>1}, \ref{PropGinfty<S>2}), we easily obtain
the boundary conditions satisfied by $\opcal{I}^{1\pm}_{\otimes d}$:
\begin{eqnarray}
\fl \opcal{I}^{1-}_{\otimes \,d}(Z=0,\p{};\vec{P})=\int\frac{\rmd^2\p{1}}{(2\pi)^2}\,\opcal{R}^{01}_{\otimes\,<S>}(\p{}|\p{1};\Ps{}):\opcal{I}^{1+}_{\otimes\, d}(Z=0,\p{1};\vec{P})\,,\label{RadBound1-2}\\
\fl\opcal{I}^{1+}_{\otimes\, d}(Z=-H,\p{};\vec{P})=\int\frac{\rmd^2\p{1}}{(2\pi)^2}\,\opcal{R}^{21}_{\otimes\,<S>}(\p{}|\p{1};\Ps{}):\opcal{I}^{1-}_{\otimes\, d}(Z=-H,\p{1};\vec{P})\,,\label{RadBound2-2}
\end{eqnarray}
We also notice that from  equation \eref{expJfuncI} and definition \eref{defId}, we obtain
the integral formulation of the radiative transfer equation \eref{RadsurI} and it's  boundary conditions (\eref{RadBound1-2}, \eref{RadBound2-2}):
\begin{eqnarray}
\fl \opcal{I}^{1a}_{\otimes\,d}(Z,\p{};\vec{P})=\!\!\! \sum_{a_1,a_2}\!\int_{-H}^0\!\!\!\rmd Z_{21}\!\int\frac{\rmd^2\p{1}}{(2\pi)^2}\,\frac{\rmd^2\p{2}}{(2\pi)^2}\,\opcal{G}^{1a1a_2}_{\infty <S>}(Z,\p{}|Z_{21},\p{2};\vec{P}):\opcal{P}^{1a_21a_1}(\p{2}|\p{1};\vec{P})\nonumber\\
\fl:\opcal{I}^{1a_1}_{\otimes red}(Z_{21},\p{1};\vec{P})
+ \sum_{a_1,a_2=\pm}\int_{-H}^0\rmd Z_{21}\int\frac{\rmd^2\p{1}}{(2\pi)^2}\,\frac{\rmd^2\p{2}}{(2\pi)^2}\,\opcal{G}^{1a1a_2}_{\infty <S>}(Z,\p{}|Z_{21},\p{2};\vec{P})\nonumber\\
\lo:
\opcal{P}^{1a_21a_1}(\p{2}|\p{1};\vec{P}):\opcal{I}^{1a_1}_{\otimes d}(Z_{21},\p{1};\vec{P})\,.\label{expJfuncIbis}
\end{eqnarray}

In the next section, we will show that this formulation is identical to the one usually used in the phenomenological radiative transfer theory.
\section{Simplifications}
\label{Simplification}
So far, we have used the two-dimensional vectors $\p{}$ to described the direction of propagation. To recover the usual formulation of the radiative transfer theory, we have to introduced the three-dimensional normalized vector $\hvec{k}$. For a vector in 
the medium~1 such that $||\p{}||<K_e'$ (hence, we neglect the evanescent waves), we have the following relationship between $\p{}$ and  $\hvec{k}$:
\begin{equation}
K_e'\,\hvec{k}=\p{}\pm\alpha'_e(\p{})\ez\quad \mbox{or}\quad  \p{}=K_e'\hvec{k}_{\perp}\quad \mbox{with} \quad \pm\alpha'_e(\p{})=K_e'\,\hat{k}_z\,\,,
\end{equation}
where the $+$ and $-$ signs correspond, respectively, to upward and downward waves, and $\hvec{k}_{\perp}=\hat{k}_x\ex+\hat{k}_y\ey$ is the projection of the vector $\hvec{k}$ on the plane $(Oxy)$. To achieve  the transformation of $\p{}$ into $\hvec{k}$, we have, in particular,  to express all the factors
involving  $\alpha_{e}(\p{}\pm\frac{\vec{P}}{2})$, 
occurring in the previous equations, as a function of $\alpha'_{e}(\p{})$.
We remark that $\vec{P}$ is different from zero only for the enhanced backscattering contribution. Furthermore, this contribution is important due to the exponential factor in  equation \eref{cont crossed}, only closed to the backscattering direction where $\vec{P}\simeq \vec{0}$. Thus, in a good approximation, we can use a Taylor development of $\alpha_{e}(\p{}\pm\frac{\vec{P}}{2})$, and we have
 \begin{eqnarray}
 \alpha_{e}(\p{}\pm\frac{\vec{P}}{2})&\simeq\alpha_{e}(\p{})\pm\frac{\vec{P}}{2}\cdot\nabla \alpha_{e}(\p{})\,,\\
 &=\alpha_{e}(\p{})\pm\frac{\vec{P}\cdot\vec{p}}{2\,\alpha_{e}(\p{})}\,,\label{devalp1}
 \end{eqnarray}
 since $\alp{e}{}=\sqrt{K_e^2-\p{}^2}$.
Moreover, in most cases, we have   $K_e''\ll K_e'$ for $K_e=K_e'+\rmi K_e''$, and a Taylor development of  $\alpha_{e}(\p{})$ which gives us
 \begin{equation}
\alp{e}{}\approx
\alpha_e'(\p{})+\rmi\,\frac{K_e'\,K_e''}{\alpha_e'(\p{})}\,,\label{devalp2}
\end{equation}
with $\sqrt{(\Re\,K_e)^2-\p{}^2}\simeq \Re \sqrt{K_e^2-\p{}^2}\equiv\alpha_e'(\p{})$.
In particular, from equations \eref{devalp1} and \eref{devalp2}  we have
\begin{equation}
\fl \Delta\alpha_e(\p{};\Ps{})\approx 2\,\rmi\,\frac{K_e'\,K_e''}{\alpha_e'(\p{})}-\frac{\vec{P}\cdot\p{}}{\alpha_e'(\p{})}\,.
\label{Dirackp}
\end{equation}
In using these approximations in the radiative transfer equation \eref{Rad}, we obtain
  \begin{eqnarray}
\fl &\frac{1}{K_e'}\left(\rmi\p{}\cdot\vec{P}+a\alpha_e'(\p{})\Der{}{\scriptstyle{Z}}\right)\,\opcal{G}^{1a1a_0}_{\ll SV\gg}(Z,\p{}|Z_0,\p{0};\vec{P})+\kappa_e\,\opcal{G}^{1aa_0}_{\ll SV\gg}(Z_{},\p{}|Z_0,\p{0})\nonumber\\
\fl & =\frac{(2\pi)^2\,\delta(\p{}-\p{0})}{4\,K_e'\alpha'_e(\p{})}\,
\op{I}_\perp^{1\,a}(\p{})\otimes\op{I}_\perp^{1\,a}(\p{})\,\delta_{a,a_0}\,a\,\delta(Z-Z_0)\no\\
\fl &+\frac{1}{4\,K_e'\,\alpha'_e(\p{})}\,\sum_{a_1=\pm}\intp{1}\,
\opcal{P}^{1a
  1a_1}(\p{}|\p{1}):\opcal{G}^{1a_1a_0}_{\ll SV\gg}(Z_{},\p{1}|Z_0,\p{0};\vec{P})\,,
\label{Radiatifgeneraliseebis_2}
\end{eqnarray}
 with $\kappa_e=2K_e''$, and
 \begin{equation}
 \opcal{P}^{1a1a_1}(\p{}|\p{1})=\opcal{P}^{1a1a_1}(\p{}|\p{1};\vec{P}=\vec{0})\,.
 \end{equation}
 In writing  $\vec{P}=\vec{0}$ in the tensors $\opcal{P}^{1a1a_1}(\p{}|\p{1};\vec{P})$ and $\op{I}_\perp^{1\,a}(\p{}+\vec{P}/2)\otimes\op{I}_\perp^{1\,a}(\p{}-\vec{P}/2)$, we impose that 
 in the enhanced backscattering contribution, the direction of propagation is the same for the two waves 
 which we consider to calculate the intensity. Hence, the contribution of $\vec{P}$ appears only in the first term of equation 
 \eref{Radiatifgeneraliseebis_2}.
 In the integral form of the radiative transfer equation given by equation \eref{Betheeq5}, these simplifications allow us
 to write the Green functions $\Gcal^{1a1a_0}_{<S>}$ and $\Gcal^{1a1a_0}_{\infty}$ given by (\ref{defGS}, \ref{defGinft}) as
\begin{eqnarray}
\fl\Gcal^{1a1a_0}_{\infty}(Z,\p{}|Z_0,\p{0};\Ps{})=\frac{(2\pi)^2\,\delta(\p{}-\p{0})}{4\,|\alpha_e(\p{0})|^2} \op{I}_\perp^{1a}(\p{0})\otimes\op{I}_\perp^{1a\,*}(\p{0})\,\delta_{a,a_0}\,\delta_{a,sgn(Z-Z_0)}\nonumber\ \\
\fl \qquad \times\,\exp\left(a\,\left(-\kappa_e\,\frac{K_e'}{\alpha'_e(\p{0})}+\rmi\frac{\vec{P}\cdot\vec{p}_0}{\alpha_e'(\p{0})}\right)\,(Z-Z_0)\right)\,,\label{defGinft2}\\
\fl\Gcal^{1a1a_0}_{<S>}(Z,\p{}|Z_0,\p{0};\Ps{})=\frac{\opcal{S}^{1a1a_0}_{\otimes\,<S>}(\p{}|\p{0};\Ps{}=\vec{0})}{4\,|\alpha_e(\p{0})|^2}\nonumber\\
\qquad \fl \times\exp\left(a\,\left(-\kappa_e\,\frac{K_e'}{\alpha'_e(\p{})}+\rmi\frac{\vec{P}\cdot\vec{p}}{\alpha_e'(\p{})}\right)\,Z-
a_0\,\left(-\kappa_e\,\frac{K_e'}{\alpha'_e(\p{0})}+\rmi\frac{\vec{P}\cdot\vec{p}_0}{\alpha_e'(\p{0})}\right)\,Z_0\right)\,,\label{defGS2}
\end{eqnarray}
where the factor $\vec{P}$ appears only in the exponential terms.

 Furthermore, if we remember that
 \begin{equation}
\fl \opcal{G}^{1a1a_0}_{\ll SV\gg}(\RR,\p{}|\RR_0,\p{0})=\int \frac{\rmd^2\vec{P}}{(2\pi)^2}\, \e^{\rmi\vec{P}\cdot(\vec{X}-\vec{X}_0)}\,\opcal{G}^{1a1a_0}_{\ll SV\gg}(Z_{},\p{}|Z_0,\p{0};\vec{P})\,,
 \end{equation}
 with $\RR=\vec{X}+Z\ez$, and $\RR_0=\vec{X}_0+Z_0\ez$, then
 equation  \eref{Radiatifgeneraliseebis_2} is now
  \begin{eqnarray}
\fl &\hvec{k}^{1a}_{p}\cdot\nabla\opcal{G}^{1a1a_0}_{\ll SV\gg}(\RR,\p{}|\RR_0,\p{0})+\kappa_e\,\opcal{G}^{1aa_0}_{\ll SV\gg}(\RR,\p{}|\RR_0,\p{0})\nonumber\\
\fl & =\frac{(2\pi)^2\,\delta(\p{}-\p{0})}{4\,K_e'\alpha'_e(\p{})}\,
\op{I}_\perp^{1\,a}(\p{})\otimes\op{I}_\perp^{1\,a}(\p{})\,\delta_{a,a_0}\,\delta(\RR-\RR_0)\no\\
\fl &+\frac{1}{4\,K_e'\,\alpha'_e(\p{})}\,\sum_{a_1=\pm}\intp{1}\,
\opcal{P}^{1a
  1a_1}(\p{}|\p{1}):\opcal{G}^{1a_11a_0}_{\ll SV\gg}(\RR,\p{1}|\RR_0,\p{0})\,,
\label{Radiatifgeneraliseebis_3}
\end{eqnarray}
 with $\vec{k}^{1a}_{\p{}}=\p{}+a\alpha'_e(\p{})\ez$ and $\hvec{k}^{1a}_{\p{}}=\vec{k}^{1a}_{\p{}}/K_e'$.
 
There is also a another way  to describe a vector $\kk$ of constant norm $K_e'$ other than using the decomposition $\kk_{\p{}}^{1a}=\p{}+a\,\alpha_e'(\p{})$ with $a=sgn(k_z)$.
\begin{figure}[!htbp]
   \begin{center}
      \psfrag{kx}{$k_x$}
      \psfrag{ky}{$k_y$}
      \psfrag{kz}{$k_z$}
      \psfrag{phi}{$\phi$}
      \psfrag{t}{$\theta$}
      \psfrag{t0}{$\theta_0$}
      \psfrag{p}{$\p{}$}
      \psfrag{alp}{$\alpha_e'(\p{})$}
      \psfrag{alp0}{$\alpha_e'(\p{0})$}
      \psfrag{kp}{$\vec{k}_{\p{}}^{1+}=\vec{k}=||\kk||\hvec{k}$}
      \epsfig{file=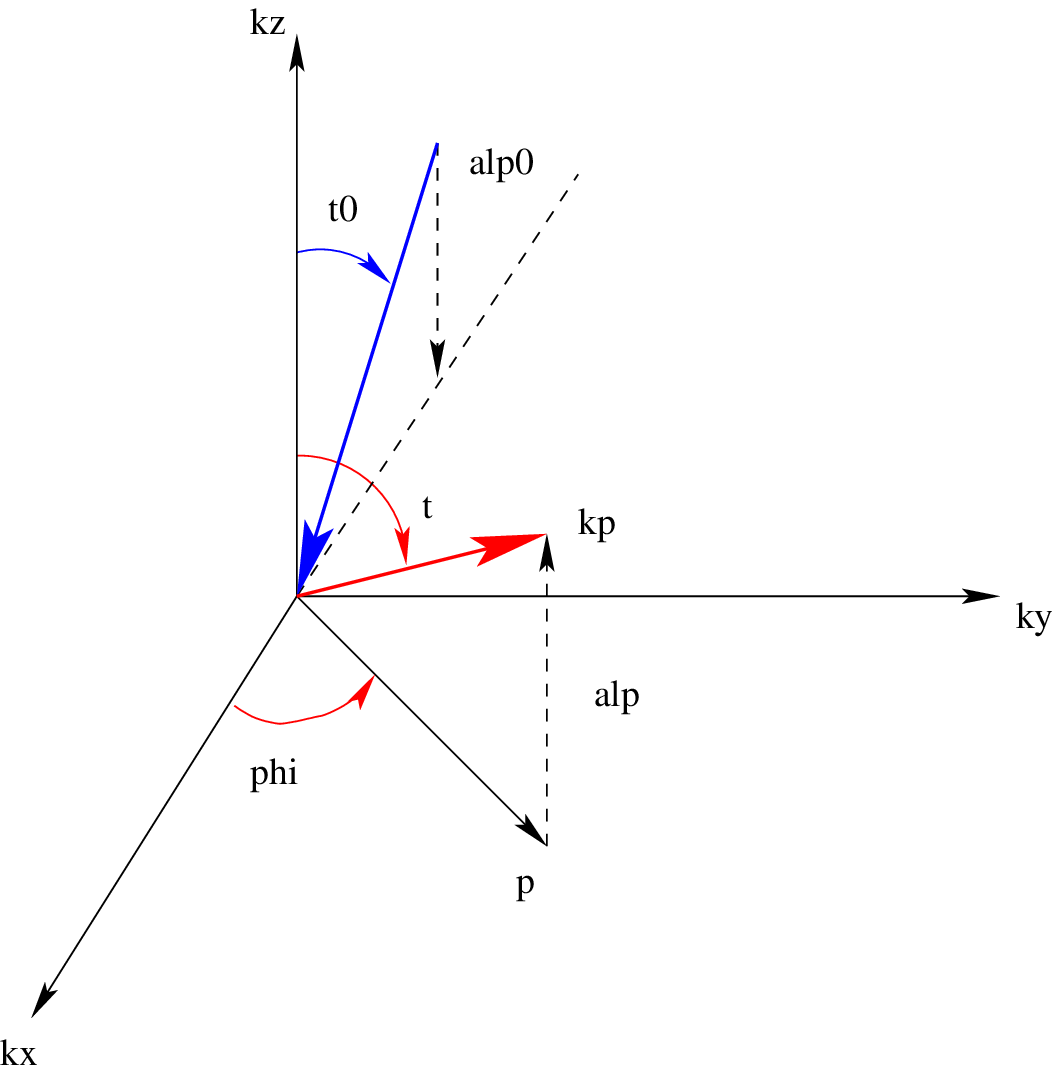,height=5cm}
      \caption{Wave vector decompositions. $\theta\in [0,\pi]$, and $\phi \in [0,2\pi]$.}
      \label{FigDecomp}
   \end{center}
\end{figure}
We can introduce the normalized vector $\hvec{k}$ such that $\vec{k}=||\vec{k}||\,\hvec{k}$,
and
 \begin{equation}
 \hvec{k}=\begin{vecteur}
 \sin \phi \cos \theta\\
 \sin \phi \sin \theta\\
 \cos \phi
 \end{vecteur}\,,
 \end{equation}
In this way, for every function $F(\vec{k})$ which contains a Dirac distribution fixing the norm of the vector $\kk$, we can define two new function $F(\hvec{k})$ and $F^a(\p{})$ (that we differentiate by their variable $\hvec{k}$ or $\p{}$) by
\begin{eqnarray}
F(\kk)=\frac{(2\pi)^3}{K_e^{'2}}\,\delta(||\kk||-K_e')\,F(\hvec{k})\,,\label{Fhvec}\\
F(\kk)=\sum_{a=\pm}(2\pi)\delta(k_z-a\alpha'_e(\p{}))\,F^a(\p{})\,,\label{Fp}
\end{eqnarray}
such that
\begin{equation}
\int \frac{\rmd^3 \kk}{(2\pi)^3}\,\,F(\kk)=\int \rmd^2\hvec{k}\,\,F(\hvec{k})=\sum_{a=\pm}\int_{||\p{}||<K_e'} \frac{\rmd^2 \p{}}{(2\pi)^2}\,\,F^a(\p{})\,,\label{Ftrois}
\end{equation}
where $\rmd^2\hvec{k}=\sin\theta \rmd\theta \rmd \phi$ is the integration on the solid angle.
We can connect the functions $F(\hvec{k})$ and $F(\p{})$ from  equation \eref{Ftrois} by
\begin{equation}
F^a(\hvec{k})\,\rmd^2\hvec{k}=F^a(\p{})\frac{\rmd^2 \p{}}{(2\pi)^2}\,,\label{Fentre}
\end{equation}
where we have defined for $a=\pm$
\begin{equation}
F^a(\hvec{k})=\Theta(ak_z)\,F(\hvec{k})\,.
\end{equation}
with $\Theta$ the Heaviside function.
From \Fref{FigDecomp}, we have
\begin{eqnarray}
||\p{}||=K_e'\sin \theta\,,\\
\alpha_e'(\p{})=K_e'|\cos \theta|\,,\label{Defalpcos}
\end{eqnarray}
since for a wave in the medium 1, we have $||\kk||\simeq K_e'$.
Then 
\begin{eqnarray}
\frac{\rmd^2 \p{}}{(2\pi)^2}&=\frac{1}{(2\pi)^2}||\p{}||\rmd||\p{}|| \rmd \phi\,,\\
&=\frac{K_e'\alpha'_e(\p{})}{(2\pi)^2}\,\rmd \hvec{k}\,,
\end{eqnarray}
so that equation \eref{Fentre} can be transformed into
\begin{equation}
F^a(\hvec{k})=\frac{K_e'\alpha_e'(\p{})}{(2\pi)^2}\,F^a(\p{})\,.\label{relaF}
\end{equation}
In particular, for the Dirac distribution $F(\kk)=(2\pi)^3\,\delta(\kk-\kk_0)$,
we have
\begin{eqnarray}
F(\hvec{k})=\delta(\hvec{k}-\hvec{k}_0)\,,\\
F(\p{})=(2\pi)^2\,\delta(\p{}-\p{0})\,,
\end{eqnarray}
and from  equation \eref{relaF}
\begin{equation}
\delta(\hvec{k}-\hvec{k}_0)=K_e'\alpha_e'(\p{})\,\delta(\p{}-\p{0})\,.\label{relaDirac}
\end{equation}
 Hence, rather than using the variable $\p{}$ and $\p{0}$ for $\opcal{G}^{1a_1a_0}_{\ll SV\gg}$, we can introduce the vectors $\hvec{k}$ and $\hvec{k}_0$ and, defined similarly to  equation \eref{Fentre}, new tensors 
 such that
 \begin{equation}
 \fl \opcal{G}^{1a1a_0}_{\ll SV\gg}(\RR,\hvec{k}|\RR_0,\hvec{k}_0)=\frac{K_e'\alpha'_e(\p{})}{(2\pi)^2}\,\opcal{G}^{1a1a_0}_{\ll SV\gg}(\RR,\vec{p}|\RR_0,\vec{p}_0)\,\frac{K_e'\alpha'_e(\p{0})}{(2\pi)^2}\,,
 \end{equation}
 with $\vec{k}=K_e' \hvec{k}=\p{}+a\alpha_e'(\p{})\ez$ and $\vec{k}_0=K_e' \hvec{k}_0=\p{0}+a_0\alpha_e'(\p{0})\ez$.
With this definition, the radiative transfer equation \eref{Radiatifgeneraliseebis_3} is now
\begin{eqnarray}
\fl &\hvec{k}\cdot\nabla\opcal{G}^{1a1a_0}_{\ll SV\gg}(\RR,\hvec{k}|\RR_0,\hvec{k}_{0})+\kappa_e\,\opcal{G}^{1a1a_0}_{\ll SV\gg}(\RR,\hvec{k}|\RR_0,\hvec{k}_{0})\nonumber\\
\fl & =\frac{\delta(\hvec{k}-\hvec{k}_{0})}{(4\pi)^2}\,\delta_{a,a_0}\,\delta(\RR-\RR_0)\,
\op{I}_\perp^{1\,a}(\hvec{k})\otimes\op{I}_\perp^{1\,a}(\hvec{k})\no\\\fl &+\sum_{a_1=\pm}\int_{a_1} \rmd^2\hvec{k}_1\,
\opcal{P}^{1a
  1a_1}(\hvec{k}|\hvec{k}_{1}):\opcal{G}^{1a_11a_0}_{\ll SV\gg}(\RR,\hvec{k}_1|\RR_0,\hvec{k}_{0})\,,
\label{Radiatifgeneraliseebis_4}
\end{eqnarray}
where $a=\sgn(\hvec{k}\cdot\ez)$, $a_0=\sgn(\hvec{k}_0\cdot\ez)$, and
\begin{eqnarray}
\int_{a_1} \rmd^2\hvec{k}_1=\int_{a_1\hvec{k}_{z}\cdot\ez>0} \rmd^2\hvec{k}_1\,,\\
\op{I}_\perp^{1\,a}(\hvec{k})=\op{I}_\perp^{1\,a}(\vec{p})\,,\\
\opcal{P}^{1a
  1a_1}(\hvec{k}|\hvec{k}_{1})=\frac{1}{(4\pi)^2}\,\opcal{P}^{1a
  1a_1}(\vec{p}|\vec{p}_{1})\,.
\end{eqnarray}
The integration on $\hvec{k}_1$ is performed on the hemisphere defined by $a_1\hvec{k}_1\cdot\ez>0$ with $a_1=\pm$:
\begin{equation}
\left\{\begin{array}{cc}
+\hvec{k}_1\cdot\ez>0 &\Longleftrightarrow \theta_1\in[0,\pi/2]\,,\\
-\hvec{k}_1\cdot\ez>0 &\Longleftrightarrow \theta_1\in[\pi/2,\pi]\,,\\
\end{array}
\right.
\end{equation}
with $\hvec{k}_1\cdot\ez=\cos \theta_1$.
If we sum over all the propagation directions and define the tensor,
\begin{equation}
\opcal{G}_{\ll SV\gg}(\RR,\hvec{k}|\RR_0,\hvec{k}_{0})=\sum_{a=\pm\,,a_0=\pm}\opcal{G}^{1a1a_0}_{\ll SV\gg}(\RR,\hvec{k}|\RR_0,\hvec{k}_{0})
\end{equation}
which does not distinguish if the wavevectors $\hvec{k}$ and $\hvec{k}_0$ are vertically directed upward or downward, the radiative transfer equation takes its classical form~\cite{Case,Bara1,Bara2}:
\begin{eqnarray}
\fl &\hvec{k}\cdot\nabla\opcal{G}_{\ll SV\gg}(\RR,\hvec{k}|\RR_0,\hvec{k}_{0})+\kappa_e\opcal{G}_{\ll SV\gg}(\RR,\hvec{k}|\RR_0,\hvec{k}_{0})\nonumber\\
\fl & =\frac{\delta(\hvec{k}-\hvec{k}_{0})}{(4\pi)^2}\,\delta(\RR-\RR_0)
\op{I}_\perp(\hvec{k})\otimes\op{I}_\perp(\hvec{k})+\int \rmd^2\hvec{k}_1\,
\opcal{P}(\hvec{k}|\hvec{k}_{1}):\opcal{G}_{\ll SV\gg}(\RR,\hvec{k}_1|\RR_0,\hvec{k}_{0})\no\,\\\fl
\label{Radiatifgeneraliseebis_5}
\end{eqnarray}
with 
\begin{eqnarray}
\fl \op{I}_\perp(\hvec{k})&=\op{I}-\hvec{k}\hvec{k}\,,\\
\fl \opcal{P}^{11}(\hvec{k}|\hvec{k}_1) &=\sum_{a=\pm\,,a_1=\pm} \opcal{P}^{1a
  1a_1}(\hvec{k}|\hvec{k}_{1})\,,\\ \fl &=\frac{1}{(4\pi)^2}\,n\,w(K_e'\hvec{k}-K_e'\hvec{k}_1)\,\op{C}^{11}_{o}(K_e'\hvec{k}|K_e'\hvec{k}_1)\otimes\op{C}^{11\,*}_{o}(K_e'\hvec{k}|K_e'\hvec{k}_1)\,.
\end{eqnarray}
Thus, the tensor $\opcal{G}_{\ll SV\gg}(\RR,\hvec{k}|\RR_0,\hvec{k}_{0})$ is the Green function of the radiative transfer equation where the source term is discrete,
localized on $\RR_0$, and emitting in the direction $\hvec{k}_0$~\cite{Case,Bara1,Bara2}.
However, it is practically better to use the Green function $\opcal{G}^{1a1a_0}_{\ll SV\gg}(\RR,\hvec{k}|\RR_0,\hvec{k}_{0})$ because we have to distinguish if the waves are upward or downward to write the boundary conditions. If we use the wave vectors $\hvec{k}$
and $\hvec{k}_0$ to write these boundary conditions, then
\begin{eqnarray}
\fl \Gcal^{1-1a_0}_{\ll SV\gg}(Z=0,\hvec{k}|Z_0,\hvec{k}_{0};\Ps{})\\
=\int_{\hvec{k}_1\cdot\ez>0} \rmd\hvec{k}_{1}\,\opcal{R}^{01}_{\otimes\,<S>}(\hvec{k}|\hvec{k}_{1}):\Gcal^{1+1a_0}_{\ll SV\gg}(Z=0,\hvec{k}_{1}|Z_0,\hvec{k}_{0};\Ps{})\,,\label{RadBound1bis}\\
\fl \Gcal^{1+1a_0}_{\ll SV\gg}(Z=-H,\hvec{k}|Z_0,\hvec{k}_{0};\vec{P})\\ =\int_{\hvec{k}_1\cdot\ez<0} \rmd\hvec{k}_{1}\,\opcal{R}^{21}_{\otimes\,<S>}(\hvec{k}|\hvec{k}_{1}):\Gcal^{1-1a_0}_{\ll SV\gg}(Z=-H,\hvec{k}_{1}|Z_0,\hvec{k}_{0};\vec{P})\,,\label{RadBound2bis}
\end{eqnarray}
with
\begin{eqnarray}
\opcal{R}^{01}_{\otimes\,<S>}(\hvec{k}|\hvec{k}_{1})=\frac{K_e'\alpha_e'(\p{})}{(2\pi)^2}\,\opcal{R}^{01}_{\otimes\,<S>}(\vec{p}|\vec{p}_{1};\vec{P}=\vec{0})\,,\\
\opcal{R}^{21}_{\otimes\,<S>}(\hvec{k}|\hvec{k}_{1})=\frac{K_e'\alpha_e'(\p{})}{(2\pi)^2}\,\opcal{R}^{21}_{\otimes\,<S>}(\vec{p}|\vec{p}_{1};\vec{P}=\vec{0})\,.
\end{eqnarray}
The relationship with the phenomenological radiative transfer theory is even clearer if we rewrite the radiative equation of section \eref{anotherFor} in terms
of $\hvec{k}$ by using the procedure that we have applied on $\Gcal^{1a1a_0}_{\ll SV\gg}$:
\begin{equation}
\fl \hvec{k}\cdot\nabla\opcal{I}^{1a}_{\otimes \,d}(\RR,\hvec{k})+\kappa_e\,\opcal{I}^{1a}_{\otimes \,d}(\RR,\hvec{k})=\sum_{a_1=\pm}
\int\rmd^2\hvec{k}_1\,\opcal{P}^{1a1a_1}(\hvec{k}|\hvec{k}_1):\opcal{I}^{1a_1}_{\otimes\, d} (\RR,\hvec{k}_1)+\veccal{S}^{1a}_{\otimes}(\RR,\hvec{k})\,,
\end{equation}
where $\veccal{S}_{\otimes}^{1a}(\RR,\hvec{k})$ is the source term defined by:
\begin{equation}
\veccal{S}^{1a}_{\otimes}(\RR,\hvec{k})=\sum_{a_1=\pm}\int\rmd^2\hvec{k}_1\opcal{P}^{1a1a_1}(\hvec{k}|\hvec{k}_1):\opcal{I}^{1a_1}_{\otimes red}(\RR,\hvec{k}_1)\,,
\end{equation}
and $\opcal{I}^{1a_1}_{\otimes\, red}(\RR,\hvec{k}_1)$ the reduced intensity.
\section{Muller matrix and tensorial product}
\label{secMuller}
Until now, to take into acount  the polarization of the incident and scattered waves, we have used the tensorial product $\otimes$. 
However, in the optical and radar communities, it is customary to use  Stokes vectors~\cite{Kong2001-1,Hecht}.
In introducing the product $\odot$ between two vectors $\E^1$ and $\E^2$ by
\begin{equation}
 \vec{E}^1\odot\vec{E}^2=
\begin{vecteur}
E^1_V\,{E}^{2*}_V\\
{E}^1_H\,{E}^{2*}_H\\
{E}^1_V\,{E}^{2*}_H+{E}^1_H\,{E}^{2*}_V\\
\left({E}^1_V\,{E}^{2*}_H-{E}^1_H\,{E}^{2*}_V\right)/\rmi
\end{vecteur}
\end{equation}
for $\vec{E}^{1,2}=E^{1,2}_V\hvec{e}_V+E^{1,2}_H\hvec{e}_H$,
the modified Stokes vectors of the incident
and scattered waves are given by~\cite{Ishi1}
\begin{eqnarray}
\fl [\vec{I}^{0i}(\p{0})]=e_0\,\vec{E}^{0i}(\p{0})\odot\vec{E}^{0i}(\p{0})=e_0\,
 \begin{vecteur}
|E^{0i}(\p{0})_V|^2\\
|E^{0i}(\p{0})_H|^2\\
2\Re(E^{0i}(\p{0})_V\,E^{0i}(\p{0})_H^*)\\
2\Im(E^{0i}(\p{0})_V\,E^{0i}(\p{0})_H^*)
\end{vecteur}\,,\label{Stokes1}\\
\fl \left[ \vec{I}^{0s}_{SV}(\rr)\right]=e_0\,\vec{E}^{0s}(\rr)\odot\vec{E}^{0s}(\rr)=e_0\,
\begin{vecteur}
|E^{0s}_{SV}(\rr)_V|^2\\
|E^{0s}_{SV}(\rr)_H|^2\\
2\Re(E^{0s}_{SV}(\rr)_V\,E^{0s}_{SV}(\rr)_H^*)\\
2\Im(E^{0s}_{SV}(\rr)_V\,E^{0s}_{SV}(\rr)_H^*)
\end{vecteur}\,.\label{Stokes2}
\end{eqnarray}
To transform our previous equation using the tensorial product, in equations using the Stokes vector, we introduce the transformation
$Tr^{\otimes\to \odot}$ by
\begin{equation}
\fl Tr^{\otimes\to \odot}: \veccal{I}=\sum_{\beta,\beta'=H,V}{\cal E}_{\beta\beta'}\hvec{e}_{\beta}\otimes\hvec{e}^*_{\beta'} \longrightarrow [I]=\begin{vecteur}
{\cal E}_{VV}\\
{\cal E}_{HH}\\
{\cal E}_{VH}+{\cal E}_{HV}\\
({\cal E}_{VH}-{\cal E}_{HV})/\rmi
\end{vecteur}\,.
\end{equation}
With this transformation, we easily check the following property:
\begin{equation}
Tr^{\otimes\to \odot}\left(\vec{E}^1\otimes\vec{E}^{2\,*}\right)=\vec{E}^1\odot\vec{E}^2\,,
\end{equation}
since for 
\begin{equation}
\vec{E}^{1,2}=\sum_{\beta=H,V}E^{1,2}_{\beta}\,\hvec{e}_{\beta}\,,\label{defE1E2}
\end{equation}
we have
\begin{equation}
\vec{E}^1\otimes\vec{E}^{2\,*}=\sum_{\beta,\beta'=H,V}E^1_{\beta}\,E^{2\,*}_{\beta'}\hvec{e}_{\beta}\otimes\hvec{e}^*_{\beta'}\,.
\end{equation}
We can generalize the product $\odot$ and the transformation $\Tr^{\otimes\to\odot}$ to dyadics operators $\op{f}$ and $\op{g}$ 
\begin{equation}
\op{f}=\sum_{\beta,\beta'=H,V}\,f_{\beta\beta'}\hvec{e}_{\beta}\hvec{e}_{\beta'}\,,\\
\op{g}=\sum_{\beta,\beta'=H,V}\,g_{\beta\beta'}\hvec{e}_{\beta}\hvec{e}_{\beta'}\,,
\end{equation}
acting on the vectors $\vec{E}^1$ and $\vec{E}^2$ defined by equation \eref{defE1E2},
in requiring the following properties:
\begin{eqnarray}
\left(\op{f}\odot\op{g}\right)\cdot\left(\E^1\odot\E^2\right)=\left(\op{f}\cdot\E^1\right)\odot\left(\op{g}\cdot\E^2\right)\,,\label{eqfE1fE2}\\
\op{f}\odot\op{g}=Tr^{\otimes\to\odot}\left(\op{f}\otimes\op{g}^*\right)\,,\label{propodot2}
\end{eqnarray}
where  the product $\cdot$ in the left hand side of  equation \eref{eqfE1fE2} is the usual product between a vector and a matrix.
Then, we obtain
\begin{eqnarray}
\fl  \op{f}\odot \op{g}  
= \left(\begin{array}{llll} \scriptstyle
 \scriptstyle{f_{VV}g^*_{VV}} & 
\scriptstyle{f_{VH} g_{VH}^*} & 
\scriptstyle{\frac{1}{2}[f_{VV}g_{VH}^*+f_{VH}g_{VV}^*]} &        \scriptstyle{\frac{\rmi}{2}[f_{VV}g_{VH}^*-f_{VH}g_{VV}^*]} \\
 \scriptstyle{f_{HV}g^*_{HV}} &
\scriptstyle{f_{HH} g_{HH}^*} &    
\scriptstyle{\frac{1}{2}[f_{HV}g_{HH}^*+f_{HH}g_{HV}^*]} & \scriptstyle{\frac{\rmi}{2}[f_{HV}g_{HH}^*-f_{HH}g_{HV}^*]}\\
 \scriptstyle{\left[ f_{VV}g^*_{HV}\right.} & 
\scriptstyle{\left[ f_{VH}g^*_{HH}\right. } &  
\scriptstyle{\frac{1}{2}[f_{VV} g_{HH}^*+f_{HH} g_{VV}^*} & 
\scriptstyle{\frac{\rmi}{2}[f_{VV} g_{HH}^*-f_{HH} g_{VV}^*}\\
\scriptstyle{\quad\left. +f_{HV}g^*_{VV}\right]} & 
\scriptstyle{\quad\left.+f_{HH}g^*_{VH}\right]} & 
\scriptstyle{\quad+f_{VH} g_{HV}^*+f_{HV} g_{VH}^*]} & 
\scriptstyle{\quad-f_{VH} g_{HV}^*+f_{HV} g_{VH}^*]}\\
 \scriptstyle{-\rmi[f_{VV}g^*_{HV}} & 
\scriptstyle{-\rmi[f_{VH}g^*_{HH}} &  
\scriptstyle{\frac{1}{2\rmi}[f_{VV} g_{HH}^*-f_{HH} g_{VV}^*} & 
\scriptstyle{\frac{1}{2}[f_{VV} g_{HH}^*+f_{HH} g_{VV}^*]}\\
\scriptstyle{ \quad-f_{HV}g^*_{VV}]}& 
\scriptstyle{ \quad-f_{HH}g^*_{VH}]}& 
\scriptstyle{\quad+f_{VH} g_{HV}^*-f_{HV} g_{VH}^*]} & 
\scriptstyle{\quad-f_{VH} g_{HV}^*-f_{HV} g_{VH}^*]}
\end{array} \right),\nonumber\\
\label{produito}
\end{eqnarray}
and 
\begin{equation} 
\fl Tr^{\otimes\to\odot} : \opcal{M}=\sum_{\beta,\beta',\beta_0,\beta_0'=H,V}\!\!\!\!\!\!
\mathcal{M}_{\beta\beta';\beta_0\beta_0'}(\hvec{e}_{\beta}\otimes\hvec{e}_{\beta'})(\hvec{e}_{\beta_0}\otimes\hvec{e}_{\beta_0'})
\longrightarrow [\op{M}]\,,\nonumber
\end{equation}
where
\begin{eqnarray}
\fl [\op{M}]=\nonumber\\
\fl \left(\begin{array}{cccc}
 \scriptstyle{\mathcal{M}_{VV;VV}} & 
\scriptstyle{\mathcal{M}_{VV;HH}} &
\scriptstyle{\frac{1}{2}\left[\mathcal{M}_{VV;VH}+\mathcal{M}_{VV;HV}\right]} &
\scriptstyle{\frac{\rmi}{2}\left[\mathcal{M}_{VV;VH}-\mathcal{M}_{VV;HV}\right]} \\
 \scriptstyle{\mathcal{M}_{HH;VV}} & 
\scriptstyle{\mathcal{M}_{HH;HH}} &
\scriptstyle{\frac{1}{2}\left[\mathcal{M}_{HH;VH}+\mathcal{M}_{HH;HV}\right]} &
\scriptstyle{\frac{\rmi}{2}\left[\mathcal{M}_{HH;VH}-\mathcal{M}_{HH;HV}\right]} \\
 \scriptstyle{\left[ \mathcal{M}_{VH;VV}\right.}&
\scriptstyle{\left[\mathcal{M}_{VH;HH}\right.}&
\scriptstyle{\frac{1}{2}\left[\mathcal{M}_{VH;VH}+\mathcal{M}_{HV;HV}\right.} &
\scriptstyle{\frac{\rmi}{2}\left[\mathcal{M}_{VH;VH}-\mathcal{M}_{HV;HV}\right.}\\
\scriptstyle{\quad\left.+\mathcal{M}_{HV;VV}  \right]} & 
\scriptstyle{\quad\left.+\mathcal{M}_{HV;HH} \right]}&
\scriptstyle{\quad\left. +\mathcal{M}_{VH;HV}+\mathcal{M}_{HV;VH}\right]} & 
\scriptstyle{\quad\left. -\mathcal{M}_{VH;HV}+\mathcal{M}_{HV;VH}\right]}\\
 \scriptstyle{-\rmi\left[\mathcal{M}_{VH;VV}\right.}&
\scriptstyle{-\rmi\left[\mathcal{M}_{VH;HH}\right.}&
\scriptstyle{\frac{1}{2\rmi}\left[\mathcal{M}_{VH;VH}-\mathcal{M}_{HV;HV}\right.} &
\scriptstyle{\frac{1}{2}\left[\mathcal{M}_{VH;VH}+\mathcal{M}_{HV;HV}\right.}\\
\scriptstyle{\quad\left.-\mathcal{M}_{HV;VV}\right]}& 
\scriptstyle{\quad\left.-\mathcal{M}_{HV;HH}\right]}&
\scriptstyle{\quad\left.+\mathcal{M}_{VH;HV}-\mathcal{M}_{HV;VH}\right]} &
\scriptstyle{\quad\left.-\mathcal{M}_{VH;HV}-\mathcal{M}_{HV;VH}\right]}
\end{array}\right)\label{tenseur-Muller}\,.\nonumber\\
\end{eqnarray}
As was demonstrated in  section \eref{MullerDef}, the scattered field $\vec{E}^{0s}$ in the medium 0 is expressed as a function
of the incident field $\vec{E}^{0i}$ with the help of the operators $\op{f}(\p{}|\p{0})$:
\begin{equation}
\vec{E}^{0s}(\rr)=\frac{\e^{\rmi K_0||\rr||}}{||\rr||}\op{f}(\p{}|\p{0})\cdot\vec{E}^{0i}(\p{0})\,.
\end{equation}
The Stokes vector of the scattered wave can be related to the Stokes vector of the incident wave by the Muller matrix $[\op{M}]$:
\begin{equation}
[\vec{I}^{0s}_{SV}(\rr)]=\frac{1}{||\rr||^2}[\op{M}(\p{}|\p{0})]\cdot[\vec{I}^{0i}(\p{0})]\,,
\end{equation}
and from the properties in  \eref{eqfE1fE2}, we show that the Muller matrix is given by the product $\odot$ of the operators $\op{f}$ with itself:
\begin{eqnarray}
 [\op{M}(\p{}|\p{0})]&=&\op{f}(\p{}|\p{0})\odot\op{f}(\p{}|\p{0})\,.\label{eqMff}
\end{eqnarray}
Furthermore, we have defined a tensorial cross-section which verifies
\begin{equation}
\op{\sigma}(\p{0}|\p{0})=\frac{4\pi}{A}\,\op{f}(\p{}|\p{0})\otimes\op{f}(\p{}|\p{0})\,.\label{eqsff}
\end{equation}
With the transformation $Tr^{\otimes\to\odot}$, we can define a matrix cross-section $[\op{\sigma}(\p{}|\p{0})]$ by
\begin{equation}
[\op{\sigma}(\p{}|\p{0})]=Tr^{\otimes\to\odot}\left\{\op{\sigma}(\p{}|\p{0})\right\}\,,
\end{equation}
and from the properties in \eref{propodot2} and  equations (\ref{eqMff},\ref{eqsff}) we have:
\begin{equation}
[\op{\sigma}(\p{}|\p{0})]=\frac{4\pi}{A}\,[\op{M}(\p{}|\p{0})]\,.
\end{equation}
Hence, the field scattered by the random medium is entirely characterized by its Muller matrix $[\op{M}(\p{}|\p{0})]$, 
or equivalently by $[\sigma(\p{}|\p{0})]$ which is the Muller matrix representation 
of the tensor $\sigma(\p{}|\p{0})$.
Moreover, we can rewrite all the equations previously obtained
with the help of Muller matrices in transforming
the product $\otimes$  by the product $\odot$ since we have
\begin{equation}
Tr^{\otimes\to\odot}\left(\opcal{M}^1:\opcal{M}^2 \right)=\op{M}^1\cdot\op{M}^2\,
\end{equation}
for
\begin{equation}
Tr^{\otimes\to\odot}\left(\opcal{M}^{1,2}\right)=\op{M}^{1,2}\,.
\end{equation}
The only non trivial transformation that we have to carry out  is the right transpose of a tensor  that we  must  replace by the right transpose  of a Muller matrix and which is  defined by
\begin{equation}
[Tr^{\otimes\to\odot}\left\{\opcal{M}(\p{}|\p{0})\right\}]^{T_R}=Tr^{\otimes\to\odot}\left\{[\opcal{M}(\p{}|\p{0})]^{T_R}\right\}\,.
\end{equation}
The explicit expression of the right transpose  of a Muller matrix is given in  \eref{AppRightT}.
\section{Recapitulation}
In this section, we rewrite all the equations previously obtained in using the product 
$\odot$ and the simplifications described in section \eref{Simplification}.
\subsection{Scattering operators for the particles and the rough surfaces}
For an incident wave inside  medium 1:
\begin{equation}
\E^{1i}(\rr)=\E^{1i}(\hvec{k}_0)\,\e^{\rmi\,K_e\,\hvec{k}_0\cdot\rr}\,,
\end{equation}
the scattered field produced by a particle of permittivity $\tilde \ep_s=\ep_s-\ep_1+\ep_e$
surrounded by a medium of permittivity $\ep_e$ is described by a T matrix $\op{t}^{11}_o$:
\begin{equation}
\vec{E}^{1s}(\rr)=\int \rmd^3\rr_1\,\rmd^3\rr_2\,\,\op{G}^{\infty}_1(\rr,\rr_1)\cdot
\op{t}^{11}_{o}(\rr_1,\rr_2)\cdot\vec{E}^{i}(\rr_2)\,,\label{App2-1}
\end{equation}
where the Green function $\op{G}_{1}^{\,\infty}$ is
\begin{eqnarray}
\op{G}_{1}^{\,\infty}(\rr-\rr_0)&=\left(\op{I}+\frac{1}{K_{e}^2}\nabla\nabla\right)G_{1}^{\infty}(\rr-\rr_0)\,,
\label{greenb}\\
G_{1}^{\infty}(\rr-\rr_0)&=P.V.\frac{\e^{\rmi\,K_e\,||\rr-\rr_0||}}{4\pi\,||\rr-\rr_0||}\,,
\end{eqnarray}
and $K^2_e=\ep_e\,K_{vac}^2$.
In the far-field, the scattered field is a spherical wave:
\begin{equation}
\E^{1s}(\rr)=\frac{\e^{\rmi K_e||\rr||}}{||\rr||}\,\op{f}(\hvec{k}|\hvec{k}_0)\,,
\end{equation}
and the operator $\op{f}(\hvec{k}|\hvec{k}_0)$ has the following relationship with the T-matrix:
\begin{equation}
4\pi\,\op{f}(\hvec{k}|\hvec{k}_0)=(\op{I}-\hvec{k}\hvec{k})\cdot\op{t}^{11}_o(K_e\hvec{k}|K_e\hvec{k}_0)\cdot(\op{I}-\hvec{k}_0\hvec{k}_0)\,.\label{frelt11}
\end{equation}
For a spherical particle, an exact expression of $\op{f}(\hvec{k}|\hvec{k}_0)$ is given by the Mie theory~\cite{Bohren,Hulst1}.
\subsection{Effective medium}
The effective permittivity $\ep_e$ is solution of the following system of equations:
\begin{eqnarray}
\fl\ep_e\,K_{vac}^2\,\op{I}=\ep_1\,K_{vac}^2\,\op{I}+n\,\op{C}^{11}_{o}(\kk_0|\kk_0)\,,\label{eff-2}
\\\fl \op{C}^{11}_{o}(\kk|\kk_0)=\op{t}^{11}_{o}(\kk|\kk_0)+n\,\intk{1}\,h(\kk-\kk_1)
\,\op{t}_{o}^{11}(\kk|\kk_1)\cdot\op{G}_{1}^{\infty}(\kk_1)\cdot\op{C}^{11}_{o}(\kk_1|\kk_0)\,,\label{eqC11-3}
\end{eqnarray}
where 
\begin{eqnarray}
h(\kk-\kk_1)=\intr\,\exp(-\rmi\,(\kk-\kk_1)\cdot\rr)\,[g(||\rr||)-1]\,,\\
\op{G}_{1}^{\infty}(\kk)=\intr{}\exp(-\rmi\kk\cdot\rr)\,\op{G}_{1}^{\infty}(\rr)\,.\label{fourierG1}
\end{eqnarray}
$\op{C}^{11}_{o}(\kk|\kk_0)$ is the scattering operators of a particle located at the origin which takes into account the correlations with the other particles with the help of the pair distribution function $g(\rr)$.
\subsection{Surface scattering operators}
The fields scattered by the upper rough surface are described by the operators
$\op{R}^{10}$, $\op{T}^{10}$, $\op{R}^{01}$, $\op{T}^{01}$, and the field scattered by the lower rough surface  is described by the operator $\op{R}^{H\,21}$ such that\\
\begin{tabular}{|c|c|c|}
\hline
& Incident field in the medium & Scattered field in the medium \\
\hline
$\op{R}^{10}$ & 0 & 0 \\
$\op{T}^{10}$ & 0 & 1 \\
$\op{R}^{01}$ & 1 & 1 \\
$\op{T}^{10}$ & 1 & 0 \\
$\op{R}^{H\,21}$ & 1 & 1 \\
\hline
\end{tabular}\\
For example, an upward incident plane wave in medium 1: $$\vec{E}^{1i}(\rr)=\vec{E}^{1i}(\p{0})\,\e^{\rmi\,\vec{p}_0\cdot\x+\rmi\,\alpha_e(\p{0})z}$$
is scattered into an upward  wave in medium 0 and a downward wave in  medium 1 and they are respectively given by
\begin{eqnarray}
\vec{E}^{0s}(\rr)=\intp{}\,\e^{\rmi\p{}\cdot\x+\rmi\,\alpha_0(\p{})z}\,\op{T}^{01}(\p{}|\p{0})\cdot\vec{E}^{1i}(\p{0})\,,\\
\vec{E}^{1s}(\rr)=\intp{}\,\e^{\rmi\p{}\cdot\x-\rmi\,\alpha_e(\p{})z}\,\op{R}^{01}(\p{}|\p{0})\cdot\vec{E}^{1i}(\p{0})\,,
\end{eqnarray}
where $\alpha_0(\p{})=\sqrt{\ep_0 K_{vac}^2-\p{}^2}$, and $\alpha_e(\p{})=\sqrt{\ep_e K_{vac}^2-\p{}^2}$.
Since the mean plane of the lower rough surface is located at $z=-H$, 
the scattering operator $\op{R}^{H\,21}$ describing this surface is the scattering
operator $\op{R}^{21}$ of a surface having its mean plane located at $z=0$ multiplied by a phase factor which depends on the thickness $H$:
\begin{equation}
\op{R}^{H\,21}(\p{}|\p{0})=\e^{\rmi(\alpha_e(\p{})+\alpha_e(\p{0}))\cdot H}\,\op{R}^{21}(\p{}|\p{0})\,.
\end{equation}
With these scattering operators, we can construct the scattering operators $\op{S}^{}$ of an homogeneous slab of permittivity $\ep_e$ having rough boundaries:
\begin{equation}
\begin{array}{ll}
\fl \RS{1+\,1-}=\op{R}^{H\,21}+\op{R}^{H\,21}\cdot\op{R}^{01}\cdot\op{R}^{H\,21}+\dots\label{S1+1-}\,, &
\RS{0+\,0-}=\op{R}^{10}+\op{T}^{01}\cdot\RS{1+\,1-}\cdot\op{T}^{10}\,,\label{S0+0-}\\
\fl \RS{1-\,1-}=\op{R}^{01}\cdot\RS{1+\,1-}\,,\label{S1-1-}&
\RS{1+\,0-}=\RS{1+\,1-}\cdot\op{T}^{10}\,, \\
\fl \RS{1-\,1+}=\op{R}^{01}+\cdot\op{R}^{01}\cdot\op{R}^{H\,21}\cdot\op{R}^{01}+\dots\,,&
\RS{1-\,0-}=\op{T}^{10}+\RS{1-\,1-}\cdot\op{T}^{10}\,,\\
\fl \RS{1+\,1+}=\op{R}^{H\,21}\cdot\RS{1-\,1+}\,,& \RS{0+\,1-}=\op{T}^{01}\cdot\RS{1+\,1-}\,,\\
\fl & \RS{0+\,1+}=\op{T}^{01}+\op{T}^{01}\cdot\RS{1+\,1+}\,.
\end{array}
\end{equation}
For example, $\RS{1+\,0-}$ described the upward wave scattered inside the slab (medium 1) for an incident downward wave in  medium 0.   
\subsection{Coherent field and coherent scattering cross-section}
For an incident field $\E^{0i}(\rr)$ in medium 0
\begin{equation}
\E^{0i}(\rr)=\E^{0i}(\hvec{k}_0)\,\e^{\rmi\,K_0\,\hvec{k}_0\cdot\rr}\,.
\end{equation}
The average scattered field over the random medium is specular and is given by
\begin{equation}
\ll \E^{0s}(\rr)\gg_{SV}=\op{S}^{coh}(\p{0})\cdot \E^{0i}(\hvec{k}_0)\,\e^{\rmi\,K_0\,\hvec{k}\cdot\rr}\,,
\end{equation}
with 
\begin{eqnarray}
\lo K_0\hvec{k}_0=\p{0}-\alpha_0(\p{0})\ez\,,\qquad K_0\hvec{k}=\p{0}+\alpha_0(\p{0})\ez\,,\\
\lo <\op{S}^{0+0-}(\p{}|\p{0})>_S=(2\pi)^2\delta(\p{}-\p{0})\,\op{S}^{coh}(\p{})\,.
\end{eqnarray}
Then, we found for the average cross-section
\begin{equation}
[\op{\sigma}^{coh}(\hvec{k}|\hvec{k}_0)]=4\pi\,|\cos \theta_0|\,\delta(\hvec{k}-\hvec{k}_0)\op{S}^{coh}(\p{0})\odot\op{S}^{coh}(\p{0})\,,
\end{equation}
where $\p{0}=K_0\,\hvec{k}_{0\,\perp}$ and $\cos \theta_0=\hvec{k}_0\cdot\ez$.
\subsection{Incoherent cross-section}
The incoherent Muller cross-section $[\op{\sigma}^{incoh}]$ is decomposed in four components:
\begin{equation}
[\op{\sigma}^{incoh}]=[\op{\sigma}^{incoh}_{L=0}]+[\op{\sigma}^{incoh}_{L=1}]+[\op{\sigma}_{Ladder}^{incoh}]+[\op{\sigma}_{Crossed}^{incoh}]\,.
\end{equation}
\subsection{Zero order scattering in volume}
The first term $[\op{\sigma}^{incoh}_{L=0}]$ described only the incoherent scattering contribution due to the boundaries; the random medium is replaced by an homogeneous medium with the permittivity $\ep_e$:
\begin{equation}
[\op{\sigma}^{incoh}_{L=0}(\hvec{k}|\hvec{k}_0)]=4\pi\cos\theta\,\opcal{S}^{0+0-incoh}_{\odot <S>}(\hvec{k}|\hvec{k}_0)\,,\nonumber\\
\end{equation} 
where  $\cos \theta=\hvec{k}\cdot\ez$ and $\opcal{S}^{0+0-incoh}_{\odot <S>}$ is defined by
\begin{eqnarray}
\fl(2\pi)^2\delta(\vec{0})\,\opcal{S}^{0+0-incoh}_{\odot<S>}(\hvec{k}|\hvec{k}_0)=\frac{K_0^2\,\cos\theta}{(2\pi)^2}\nonumber\\
\fl\times\Big\{ <\op{S}^{0+0-}(\p{}|\p{0})\odot \op{S}^{0+0-}(\p{}|\p{0})>_S -<\op{S}^{0+0-}(\p{}|\p{0})>_{S}\odot <\op{S}^{0+0-}(\p{}|\p{0})>_{S}\Big\}\,,\nonumber\\
\label{devsigmabis}
\end{eqnarray}
and $\p{}=K_0\,\hvec{k}_{\perp}$, $\p{0}=K_0\,\hvec{k}_{0\perp}$.

\subsection{First order scattering in volume}
\label{subFirst}
The second component $[\op{\sigma}^{incoh}_{L=1}]$ of  the incoherent scattering cross-section contains only the single scattering contribution by the particles with the multiple scattering
terms due to the boundaries:
\begin{eqnarray}
\fl [\op{\sigma}^{incoh}_{L=1}(\hvec{k}|\hvec{k}_0)]=
\sum_{a_1,a_2=\pm}\int_{a_1,a_2}\rmd^2 \hvec{k}_1\,\rmd^2 \hvec{k}_2\,
\frac{\left( 
 1-\e^{-(a_2|\sec \theta_2|-a_1|\sec \theta_1|)\kappa_e\,H}\right)}{(a_2|\sec\theta_2|-
 a_1|\sec \theta_1|)\kappa_e} 
 \no\\\lo \times\,\frac{4\pi\,\cos \theta}{|\cos \theta_2|}\, \opcal{S}^{0+;1a_2}_{\odot\,<S>}(\hvec{k}|\hvec{k}_{2})\cdot\opcal{P}_{\odot
}^{1a_21a_1}(\hvec{k}_{2}|\hvec{k}_{1})\cdot\opcal{S}^{1a_1;0-}_{\odot\,<S>}(\hvec{k}_{1}|\hvec{k}_{0})\,,\label{Chap4ordreL=1bis}
\end{eqnarray}
where $\sec \theta=1/\cos \theta$, and $\kappa_e=2\Im K_e$.
The Muller matrix $\opcal{P}_{\odot}^{1a_21a_1}$ describing the scattering by one particle is defined by
\begin{eqnarray}
\fl \opcal{P}_{\odot}^{1a_21a_1}(\hvec{k}_2|\hvec{k}_1)=\frac{1}{(4\pi)^2} n\,w(K_e'\hvec{k}_2-K_e'\hvec{k}_1)\Theta(a_2\hvec{k}_{2z})\Theta(a_1\hvec{k}_{1z})\no\\
\lo\qquad \qquad  \times\op{C}^{11}_{o}(K_e'\hvec{k}_2|K_e'\hvec{k}_1)\odot\op{C}^{11}_{o}(K_e'\hvec{k}_2|K_e'\hvec{k}_1)\,,  \label{opPodot}
\end{eqnarray}
where $\Theta$ is the Heaviside function, and 
 $w$ is the structure factor of the random medium:
\begin{equation} 
w(\vec{k}-\vec{k}_0)=1+n\,\intr{} \e^{-\rmi(\kk-\kk_0)\cdot\rr}\,[g(\rr)-1]\,.
\end{equation}
The Muller matrices describing the scattering at the boundaries are 
\begin{eqnarray}
\fl (2\pi)^2\,\delta(\vec{0})\,\opcal{S}_{\odot\,<S>}^{0+;1a_2}(\hvec{k}|\hvec{k}_{2})=\sqrt{\frac{\ep_0}{\ep'_e}}\,\frac{K^2_0 \cos \theta}{(2\pi)^2}\,<\op{S}^{0+1a_2}\left(\p{}|\p{2}\right)\odot\op{S}^{0+1a_2}\left(\p{}|\p{2}\right)>_{S}\,,\label{SL=1}\\
\fl(2\pi)^2\,\delta(\vec{0})\,\opcal{S}_{\odot\,<S>}^{1a_1;0-}(\hvec{k}_1|\hvec{k}_{0})
=\sqrt{\frac{\ep'_e}{\ep_0}}\,\frac{K_e^{'2}|\cos \theta_1|}{(2\pi)^2}\,<\op{S}^{1a_1\,0-}\left(\p{1}|\p{0}\right)\odot\op{S}^{1a_1\,0-}\left(\p{1}|\p{0}\right)>_{S}\,,\nonumber\\\label{SL=1bis}
\end{eqnarray}
with
\begin{equation}
\begin{array}{ll}
\p{}=K_0\hvec{k}_{\perp}\,,& \cos \theta=\hvec{k}\cdot\ez \\
\p{2}=K_e'\hvec{k}_{2\perp}\,, & \cos \theta_2=\hvec{k}_2\cdot\ez \\
\p{1}=K_e'\hvec{k}_{1\perp}\,, & \cos \theta_1=\hvec{k}_1\cdot\ez\\
\p{0}=K_0\hvec{k}_{0\perp}\,,& \cos \theta_0=\hvec{k}_0\cdot\ez
\end{array}\,.
\end{equation}
In \eref{AppdefRotimes}, we justify why in introducing  the factors $\sqrt{\ep_0/\ep'_e}$
and $\sqrt{\ep_e'/\ep_0}$ in definitions \eref{SL=1} and \eref{SL=1bis}, the operators  $\opcal{S}_{\odot\,<S>}^{0+;1a_2}$ and $\opcal{S}_{\odot\,<S>}^{1a_1;0-}$ match those   used in the phenomenological radiative transfer
theory.
The integral  in  equations \eref{Chap4ordreL=1} is defined by
\begin{equation}
\int_{a_1,a_2} \rmd^2 \hvec{k}_1\,\rmd^2 \hvec{k}_2=\int_{a_1\hvec{k}_{1}\cdot\ez>0} \rmd^2 \hvec{k}_1\,\int_{a_2\hvec{k}_{2}\cdot\ez>0}\rmd^2 \hvec{k}_2
\end{equation}
where for $a_1=+$ the integration is on upper hemisphere, and for $a_1=-$ on the lower 
hemisphere:
\begin{equation}
\int_{a\hvec{k}\cdot\ez>0}\rmd^2\hvec{k}=
\left\{\begin{array}{ll}
 \int_{0}^{\pi/2}  \sin \theta \rmd \theta\int_{0}^{2\pi} \rmd \phi & \mbox{if } a=+  \,,\\
 \int_{\pi/2}^{\pi} \sin \theta \rmd \theta \int_{0}^{2\pi} \rmd \phi &\mbox{if } a=-  \,,
\end{array}\right.
\end{equation}
\subsection{Definition of the Green function $\opcal{G}^{1a1a_0}_{\odot\ll SV\gg}$}
To calculate the two other contributions $[\op{\sigma}^{incoh}_{Ladder}]$ and $[\op{\sigma}^{incoh}_{Crossed}]$, we need the Green function $\opcal{G}^{11}_{\ll SV\gg}$ of the radiative transfer equation.
This Green function is the Wigner transform of the Green function $\op{G}^{11}$ describing the propagation of the electric field in the medium I (See paper I):
\begin{eqnarray}
\fl \opcal{G}^{11}_{\odot\ll SV\gg}(\RR,\kk|\RR_0,\kk_0)=\iintr{}{0}\,\e^{-\rmi\kk\cdot\rr+\rmi\kk_0\cdot\rr_0}\no\\
\fl\times\Big[\ll
\op{G}^{11}_{SV}(\RR+\frac{\rr}{2},\RR_0+\frac{\rr_0}{2})\odot\op{G}^{11\,*}_{SV}(\RR-\frac{\rr}{2},\RR_0-\frac{\rr_0}{2})\gg_{SV}\no\\
\fl -\ll
\op{G}^{11}_{SV}(\RR+\frac{\rr}{2},\RR_0+\frac{\rr_0}{2})\gg_{SV}\odot\ll\op{G}^{11\,*}_{SV}(\RR-\frac{\rr}{2},\RR_0-\frac{\rr_0}{2})\gg_{SV}\Big]\,.
\label{wignerop2}
\end{eqnarray}
Under the quasi-uniform field approximation, we have~\cite{Apresyan}
\begin{equation}
\fl\opcal{G}^{11}_{\odot\ll SV\gg}(\RR,\kk|\RR_0,\kk_0)=\frac{(2\pi)^6}{K_e^{'4}}\,\delta(||\kk||-K_e')\delta(||\kk_0||-K_e')\,\opcal{G}^{11}_{\odot\ll SV\gg}(\RR,\hvec{k}|\RR_0,\hvec{k}_0)\,.
\end{equation}
To write the boundary conditions, we have to distinguish between upward and downward waves, and we introduce
\begin{equation}
\fl \opcal{G}^{1a1a_0}_{\odot\ll SV\gg}(\RR,\hvec{k}|\RR_0,\hvec{k}_0)=\Theta(a\hvec{k}\cdot\ez)\Theta(a_0\hvec{k}_0\cdot\ez)\,\opcal{G}^{11}_{\odot\ll SV\gg}(\RR,\hvec{k}|\RR_0,\hvec{k}_0)\,,
\end{equation}
where $\Theta$ is the Heaviside function, and $a=\pm$, $a_0=\pm$.
Finally, as the random medium and the rough surfaces are statistically homogeneous, we introduce
the Fourier transform of the Green function along the horizontal coordinates $(X,Y)$:
\begin{equation}
\fl \opcal{G}^{1a1a_0}_{\odot\ll SV\gg}(Z,\hvec{k}|Z_0,\hvec{k};\vec{P})=\int \rmd^2(\vec{X}-\vec{X}_0)\, \e^{-\rmi\vec{P}\cdot(\vec{X}-\vec{X}_0)}\opcal{G}^{1a1a_0}_{\odot\ll SV\gg}(\RR,\hvec{k}|\RR_0,\hvec{k}_0)
 \end{equation}
 with $\RR=\vec{X}+Z\ez$ and $\RR_0=\vec{X}_0+Z_0\ez$.
\subsection{Radiative transfer equation}
The matrices $\opcal{G}^{1a1a_0}_{\odot\ll SV\gg}$ are the Green functions of the radiative transfer equation:
\begin{eqnarray}
\fl &\hvec{k}\cdot\nabla_{\vec{P}}\,\opcal{G}^{1a1a_0}_{\odot\ll SV\gg}(Z,\hvec{k}|Z_0,\hvec{k}_{0};\vec{P})+\kappa_e\,\opcal{G}^{1a_11a_0}_{\odot\,\ll SV\gg}(Z,\hvec{k}|Z_0,\hvec{k}_{0};\vec{P})\nonumber\\
\fl & =\frac{\delta(\hvec{k}-\hvec{k}_{0})}{(4\pi)^2}\,\delta_{a,a_0}\,\delta(Z-Z_0)\,
\op{I}_\perp^{1\,a}(\hvec{k})\odot\op{I}_\perp^{1\,a}(\hvec{k})\no\\
\fl &+\sum_{a_1=\pm}\int_{a_1} \rmd^2\hvec{k}_1\,
\opcal{P}_{\odot}^{1a
  1a_1}(\hvec{k}|\hvec{k}_{1})\cdot\opcal{G}^{1a_11a_0}_{\odot\ll SV\gg}(Z,\hvec{k}_1|Z_0,\hvec{k}_{0};\vec{P})\,,
\label{Radiatifgeneraliseebis_4bis}
\end{eqnarray}
where $a=\sgn(\hvec{k}\cdot\ez)$, $a_0=\sgn(\hvec{k}_0\cdot\ez)$, the differential operator $\nabla_{\vec{P}}$ is
\begin{equation}
\nabla_{\vec{P}}=\rmi\vec{P}+\ez\frac{\partial }{\partial Z}\,,
\end{equation}
and the integral is defined by:
\begin{eqnarray}
\int_{a_1} \rmd^2\hvec{k}_1=\int_{a_1\hvec{k}_{z}\cdot\ez>0} \rmd^2\hvec{k}_1\,.
\end{eqnarray}
The boundary conditions on the upper and lower surfaces are
\begin{eqnarray}
\fl \Gcal^{1-1a_0}_{\odot\,\ll SV\gg}(Z=0,\hvec{k}|Z_0,\hvec{k}_0;\Ps{})\nonumber\\
\fl \qquad\qquad=\int_{\hvec{k}_1\cdot\ez>0} \rmd\hvec{k}_{1}\,\opcal{R}^{01}_{\odot\,<S>}(\hvec{k}|\hvec{k}_{1})\cdot\Gcal^{1+1a_0}_{\odot\ll SV\gg}(Z=0,\hvec{k}_{1}|Z_0,\hvec{k}_{0};\Ps{})\,,\label{RadBound1}\\
\fl \Gcal^{1+1a_0}_{\odot\ll SV\gg}(Z=-H,\hvec{k}|Z_0,\hvec{k}_{0};\vec{P})\nonumber\\ \fl \qquad\qquad =\int_{\hvec{k}_1\cdot\ez<0} \rmd\hvec{k}_{1}\,\opcal{R}^{21}_{\odot\,<S>}(\hvec{k}|\hvec{k}_{1})\cdot\Gcal^{1-1a_0}_{\odot\ll SV\gg}(Z=-H,\hvec{k}_{1}|Z_0,\hvec{k}_{0};\vec{P})\,,\label{RadBound2}
\end{eqnarray}
with
\begin{eqnarray}
\fl(2\pi)^2\,\delta(\vec{0})\,\opcal{R}^{01}_{\odot\,<S>}(\hvec{k}|\hvec{k}_{1})&  =\frac{K_e^{'2}|\cos \theta|}{(2\pi)^2}\,<\op{R}^{01}(\p{}|\p{0})\odot\op{R}^{01\,*}(\p{}|\p{0})>_S\,,\label{Radiacond1bis}\\
\fl(2\pi)^2\,\delta(\vec{0})\,\opcal{R}^{21}_{\odot\,<S>}(\hvec{k}|\hvec{k}_{1})&=\frac{K_e^{'2}|\cos \theta|}{(2\pi)^2}\,<\op{R}^{21}(\p{}|\p{0})\odot\op{R}^{21\,*}(\p{}|\p{0})>_S\,,\label{Radiacond2bis}
\end{eqnarray}
and 
\begin{eqnarray}
\p{}=K_e'\,\hvec{k}_\perp\,, & \cos\theta=\hvec{k}\cdot\ez\,,\\
\p{0}=K_e'\,\hvec{k}_{0\perp}\,, & \cos\theta_0=\hvec{k}_0\cdot\ez\,.
\end{eqnarray}
\subsection{Integral equation}
The integral form of the radiative transfer equation is
\begin{eqnarray}
\fl\Gcal^{1a1a_0}_{\odot\,\ll SV\gg}(Z,\hvec{k}|Z_0,\hvec{k}_{0};\Ps{})=\Gcal^{1a1a_0}_{\odot\,\infty<S>}(Z,\hvec{k}|Z_0,\hvec{k}_{0};\Ps{})+\sum_{a_1a_2}\,(4\pi)^2\int
\rmd^2 \hvec{k}_{1} \rmd^2
  \hvec{k}_{2}\,\rmd Z_{21}\,\,\nonumber\\
  \fl \times
\Gcal^{1a1a_2}_{\odot\,\infty<S>}(Z,\hvec{k}|Z_{21},\hvec{k}_{2};\Ps{})
\cdot\Pcal^{1a_2 1a_1}_{\odot}(\hvec{k}_{2}|\hvec{k}_{1})\cdot\Gcal^{1a_1a_0}_{\odot\,\ll SV\gg}(Z_{21},\hvec{k}_{1}|Z_0,\hvec{k}_{0};\Ps{})\,,\label{Betheeq5-2}
\end{eqnarray}
where
\begin{equation}
\fl \Gcal^{1a1a_0}_{\odot\,\infty<S>}(Z,\hvec{k}|Z_0,\hvec{k}_{0};\Ps{})=\Gcal^{1a1a_0}_{\odot\,\infty}(Z,\hvec{k}|Z_0,\hvec{k}_{0};\Ps{})+\Gcal^{1a1a_0}_{\odot\,<S>}(Z,\hvec{k}|Z_0,\hvec{k}_{0};\Ps{})\,,\label{defsommeG}
\end{equation}
and
\begin{eqnarray}
\fl\Gcal^{1a1a_0}_{\odot\, \infty}(Z,\hvec{k}|Z_0,\hvec{k}_{0};\Ps{})=\frac{\delta(\hvec{k}-\hvec{k}_{0})}{(4\pi)^2\,|\cos \theta|} \op{I}^{1a}_\perp(\hvec{k})\odot\op{I}^{1a}_\perp(\hvec{k})\,\delta_{a,a_0}\,\delta_{a,sgn(Z-Z_0)}\nonumber\ \\
\lo \times \exp\left(-a( \kappa_e+\rmi\hvec{k}\cdot\vec{P})(Z-Z_0)/|\cos \theta|\right)\,,\label{defGinft3}\\
\fl\Gcal^{1a1a_0}_{\odot\,<S>}(Z,\hvec{k}|Z_0,\hvec{k}_{0};\Ps{})=\frac{1}{(4\pi)^2\,|\cos \theta_0|}\,\opcal{S}^{1a1a_0}_{\odot\,<S>}(\hvec{k}|\hvec{k}_{0})\no\\
\lo \times\exp\left(-a( \kappa_e+\rmi\hvec{k}\cdot\vec{P})Z/|\cos \theta|+a_0( \kappa_e+\rmi\hvec{k}_0\cdot\vec{P})Z_0/|\cos \theta_0|\right)\,,\label{defGS3}
\end{eqnarray}
and 
\begin{equation}
\fl(2\pi)^2\,\delta(\vec{0})\,\opcal{S}^{1a1a_0}_{\odot\,<S>}(\hvec{k}|\hvec{k}_0) =\frac{K_e^{'2}|\cos \theta|}{(2\pi)^2}\,<\op{S}^{1a1a_0}(\p{}|\p{0})\odot\op{S}^{1a1a_0\,*}(\p{}|\p{0})>_S\,,
\end{equation}
with 
\begin{eqnarray}
\p{}=K_e'\,\hvec{k}_\perp\,, & \cos\theta=\hvec{k}\cdot\ez\,,\\
\p{0}=K_e'\,\hvec{k}_{0\perp}\,, & \cos\theta_0=\hvec{k}_0\cdot\ez\,.
\end{eqnarray}
\subsection{Ladder terms}
The third contribution to the incoherent scattering cross-section, which contains second order and higher scattering contributions by the particles, is
\begin{eqnarray}
\fl [\op{\sigma}^{incoh}_{Ladder}(\hvec{k}|\hvec{k}_{0})]=
\sum_{a_1,a_2,a_3,a_4=\pm}\int_{-H}^0\rmd Z_{43}\rmd Z_{21}\int_{a_1,a_2,a_3,a_4}\!\!\! \!\!\!\!\!\!\!\!\rmd^2 \hvec{k}_1\,\rmd^2 \hvec{k}_2\,\rmd^2\hvec{k}_3\,\rmd^2 \hvec{k}_4\,\no\\
\fl \times \frac{(4\pi)^3\,\cos \theta}{|\cos \theta_4|}\,\exp\left(a_4\,\kappa_e Z_{43}|\sec \theta_4|-a_1\,\kappa_e Z_{21}|\sec \theta_1|\right)
\,\opcal{S}^{0+;1a_4}_{\odot\,<S>}(\hvec{k}|\hvec{k}_{4})\no\\\fl  
\cdot\Pcal_{\odot}^{1a_41a_3}(\hvec{k}_{4}|\hvec{k}_{3})\cdot\opcal{G}^{1a_31a_2}_{\odot\,\ll SV\gg}(Z_{43},\hvec{k}_3|Z_{21},\hvec{k}_2;\vec{P}=\vec{0}) \cdot\Pcal_{\odot}^{1a_21a_1}(\hvec{k}_{2}|\hvec{k}_{1})\cdot\opcal{S}^{1a_1;0-}_{\odot\,<S>}(\hvec{k}_{1}|\hvec{k}_{0})\,,\no\\\label{Chap4ordreLadderbis}
\end{eqnarray}
with $\cos \theta=\hvec{k}\cdot\ez$, $\cos \theta_1=\hvec{k}_1\cdot\ez$, $\cos \theta_4=\hvec{k}_4\cdot\ez$ and
\begin{equation}
\fl \int_{a_1,a_2,a_3,a_4}\!\!\!\!\!\!\!\!\!\!\!\!\rmd^2 \hvec{k}_1\,\rmd^2 \hvec{k}_2\,\rmd^2\hvec{k}_3\,\rmd^2\hvec{k}_4=
\int_{a_1\hvec{k}_z\cdot\ez>0}\!\!\!\!\!\!\!\!\!\!\rmd^2\hvec{k}_1
\int_{a_2\hvec{k}_z\cdot\ez>0}\!\!\!\!\!\!\!\!\!\!\rmd^2\hvec{k}_2
\int_{a_3\hvec{k}_z\cdot\ez>0}\!\!\!\!\!\!\!\!\!\!\rmd^2\hvec{k}_3
\int_{a_4\hvec{k}_z\cdot\ez>0}\!\!\!\!\!\!\!\!\!\!\rmd^2\hvec{k}_4\,.
\end{equation}
\subsection{Most-crossed contributions}
The fourth contribution of the incoherent scattering cross-section, which describes the 
enhanced backscattering phenomenon, is
\begin{eqnarray}
\fl [\op{\sigma}^{incoh}_{Crossed}(\hvec{k}|\hvec{k}_{0})]=
\sum_{a_1,a_2,a_3,a_4=\pm}\int_{-H}^0\rmd Z_{43}\rmd Z_{21}\int_{a_1,a_2,a_3,a_4}\!\!\! \!\!\!\!\!\!\!\!\rmd^2 \hvec{k}_1\,\rmd^2 \hvec{k}_2\,\rmd^2\hvec{k}_3\,\rmd^2 \hvec{k}_4\,\no\\
\fl \times \frac{(4\pi)^3\,|\cos \theta_0|}{|\cos \theta^-_4|}\,
\exp\left(\rmi\,a_4 Z_{43}K_e'(|\cos\theta_4^+|-|\cos\theta_4^-|) -\rmi\,a_1 Z_{21}K_e'(|\cos\theta_1^+|-|\cos\theta_1^-|)\right)\,\no\\
\fl \times\exp\Big(a_4\,\kappa_e Z_{43}(|\sec \theta_4^{+}|+|\sec \theta_4^{-}|)/2-a_1\,\kappa_e Z_{21}(|\sec \theta_1^{+}|+|\sec \theta_1^{-}|)/2\Big)
\no\\\fl  
\times\Big[\opcal{S}^{0+;1a_4}_{\odot\,<S>}(-\hvec{k}_0|\hvec{k}_{4})\cdot\Pcal_{\odot}^{1a_41a_3}(\hvec{k}_{4}|\hvec{k}_{3})\cdot\opcal{G}^{1a_31a_2}_{\odot\ll SV\gg}(Z_{43},\hvec{k}_3|Z_{21},\hvec{k}_2;\vec{P})\no\\
\fl \cdot\Pcal_{\odot}^{1a_21a_1}(\hvec{k}_{2}|\hvec{k}_{1})\cdot\opcal{S}^{1a_1;0-}_{\odot\,<S>}(\hvec{k}_{1}|\hvec{k}_0)\Big]^{T_R}\,,\label{Chap4ordreCrossedbis}
\end{eqnarray}
where the right transposed $^{T_R}$ of a Muller matrix is defined in  \ref{AppRightT} and
\begin{eqnarray}
	\vec{P}=K_0(\hvec{k}_{\perp}+\hvec{k}_{0\perp})\,,\qquad \cos\theta_0=\hvec{k}_0\cdot\ez\,.
\end{eqnarray}
The definition of the angles $\theta_4^{\pm}$ and $\theta_1^{\pm}$ depends  
on the matrix $\opcal{S}^{0+;1a_4}_{\odot\,<S>}$ and $\opcal{S}^{1a_1;0-}_{\odot\,<S>}$.
Each of these matrices has a coherent and an  incoherent part:
\begin{eqnarray}
\fl\opcal{S}^{0+;1a_0}_{\odot\,<S>}(\hvec{k}|\hvec{k}_{0})=\opcal{S}^{0+;1a_0\,incoh}_{\odot\,<S>}(\hvec{k}|\hvec{k}_{0})+\delta(\hvec{k}^t-\hvec{k}_0)\opcal{S}^{0+;1a_0\,coh}_{\odot\,<S>}(\hvec{k}_{0})\,,\\
\fl\opcal{S}^{1a;0-}_{\odot\,<S>}(\hvec{k}|\hvec{k}_0)=\opcal{S}^{1a;0-\,incoh}_{\odot\,<S>}(\hvec{k}|\hvec{k}_0)+\delta(\hvec{k}^t-\hvec{k}_0)\,\opcal{S}^{1a;0-\,coh}_{\odot\,<S>}(\hvec{k}_0)\,,
\end{eqnarray}
where $\hvec{k}^t$ and $\hvec{k}_0$ are related by the Fresnel law:
\begin{eqnarray}
\fl K_0\,\hvec{k}^t_{\perp}=K_e'\hvec{k}_{0\,\perp}\,\mbox{for}\quad \opcal{S}^{0+;1a_4}_{\odot\,<S>}(\hvec{k}|\hvec{k}_{0})\,,\\
\fl K_e'\,\hvec{k}^t_{\perp}=K_0\,\hvec{k}_{0\perp}\, \mbox{for}\quad \opcal{S}^{1a;0-}_{\odot\,<S>}(\hvec{k}|\hvec{k}_{0})\,.
\end{eqnarray}
For the incoherent part, the angles are defined by
\begin{eqnarray}
|\cos \theta_4^+|=\sqrt{1-(\hvec{k}_{4\perp}+\vec{P}/2K_e')^2}\,,\\
|\cos \theta_4^-|=\sqrt{1-(\hvec{k}_{4\perp}-\vec{P}/2K_e')^2}\,,\\
|\cos \theta_1^+|=\sqrt{1-(\hvec{k}_{1\perp}+\vec{P}/2K_e')^2}\,,\\
|\cos \theta_1^-|=\sqrt{1-(\hvec{k}_{1\perp}-\vec{P}/2K_e')^2}\,,
\end{eqnarray}
and for the coherent part we have:
\begin{eqnarray}
|\cos \theta_4^+|=\sqrt{1-(K_0\hvec{k}_{\perp}/K_e')^2}\,,\\
|\cos \theta_4^-|=\sqrt{1-(K_0\hvec{k}_{0\perp}/K_e')^2}\,,\\
|\cos \theta_1^+|=|\cos \theta_4^-|\,,\\
|\cos \theta_1^-|=|\cos \theta_4^+|\,.
\end{eqnarray}
We notice that in the backscattering direction, we have $\hvec{k}=-\hvec{k}_0$ and then $\vec{P}=\vec{0}$, $\cos \theta_4^+=\cos \theta_4^-$, $\cos \theta_1^+=\cos \theta_1^-$.
In this direction, the crossed contribution differs from the ladder approximation only by the 
presence of the right transpose $^{T_R}$ around the Muller matrix in the equation \eref{Chap4ordreCrossedbis}.
\subsection{Other Formulation}
We can also calculate the incoherent cross-section in introducing the specific diffusive specific intensity 
$\opcal{I}^{1a}_{\odot\,d}(Z,\hvec{k};\vec{P})$ which verifies the usual phenomenological 
radiative transfer equation:
\begin{eqnarray}
\fl \hvec{k}\cdot\nabla_{\vec{P}}\,\opcal{I}^{1a}_{\odot \,d}(Z,\hvec{k};\vec{P})+\kappa_e\,\opcal{I}^{1a}_{\odot \,d}(Z,\hvec{k};\vec{P})=\sum_{a_1=\pm}
\int\rmd^2\hvec{k}_1\,\opcal{P}_{\odot}^{1a1a_1}(\hvec{k}|\hvec{k}_1)\cdot\opcal{I}^{1a_1}_{\odot\, d} (Z,\hvec{k}_1;\vec{P})\nonumber\\
\lo +\veccal{S}^{1a}_{\odot}(Z,\hvec{k};\vec{P})\,,\label{ApproRad}
\end{eqnarray}
where $\vec{S}^{1a}_{\odot}(Z,\hvec{k};\vec{P})$ is the source term defined by:
\begin{equation}
\veccal{S}^{1a}_{\odot}(Z,\hvec{k};\vec{P})=\sum_{a_1=\pm}\int\rmd^2\hvec{k}_1\opcal{P}_{\odot}^{1a1a_1}(\hvec{k}|\hvec{k}_1)\cdot\opcal{I}^{1a_1}_{\odot red}(Z,\hvec{k}_1;\vec{P})\,,
\end{equation}
and $\opcal{I}^{1a_1}_{\odot red}(Z,\hvec{k}_1;\vec{P})$ the reduced specific intensity~\cite{Ishi1}:
\begin{equation}
\fl \opcal{I}_{\odot\,red}^{1a_1}(Z_1,\hvec{k}_{1};\vec{P})=\opcal{S}^{1a_10-}_{\odot\,<S>}(\hvec{k}_{1}|\hvec{k}_0)\,\exp\left(-a_1\,(\kappa_e+\rmi\,\hvec{k}\cdot\vec{P} )Z_1/|\cos \theta_1|\right)\,,
\end{equation}
with $\cos\theta_1=\hvec{k}_z\cdot\ez$.
The boundary conditions necessary to solve the radiative transfer equation are
\begin{eqnarray}
\fl \opcal{I}^{1-}_{\odot \,d}(Z=0,\hvec{k};\vec{P})
=\int_{\hvec{k}_1\cdot\ez>0} \rmd\hvec{k}_{1}\,\opcal{R}^{01}_{\odot\,<S>}(\hvec{k}|\hvec{k}_{1})\cdot\opcal{I}^{1+}_{\odot \,d}(Z=0,\hvec{k}_1;\vec{P})\,,\label{ApproBound1}\\
\fl \opcal{I}^{1+}_{\odot \,d}(Z=-H,\hvec{k};\vec{P})=\int_{\hvec{k}_1\cdot\ez<0} \rmd\hvec{k}_{1}\,\opcal{R}^{21}_{\odot\,<S>}(\hvec{k}|\hvec{k}_{1})\cdot\opcal{I}^{1-}_{\odot \,d}(Z=-H,\hvec{k}_1;\vec{P})\,.\label{ApproxBound2}
\end{eqnarray}
Furthermore, we can also determine the specific intensity  $\opcal{I}^{1a}_{\odot\,d}(Z,\hvec{k};\vec{P})$ by using the integral formulation of the radiative transfer equations and its boundary conditions:
\begin{eqnarray}
\fl \opcal{I}^{1a}_{\odot\,d}(Z,\hvec{k};\vec{P})=\!\!\! \sum_{a_1,a_2}\!\int_{-H}^0\!\!\!\rmd Z_{21}\!\int\rmd^2\hvec{k}_1\rmd^2\hvec{k}_2\,\opcal{G}^{1a1a_2}_{\odot\,\infty <S>}(Z,\hvec{k}|Z_{21},\hvec{k}_{2};\vec{P})\cdot\opcal{P}_{\odot}^{1a_21a_1}(\hvec{k}_{2}|\hvec{k}_{1};\vec{P})\nonumber\\
\fl\cdot\opcal{I}^{1a_1}_{\odot red}(Z_{21},\hvec{k}_{1};\vec{P})
+ \sum_{a_1,a_2=\pm}\int_{-H}^0\rmd Z_{21}\int\rmd^2\hvec{k}_1\rmd^2\hvec{k}_2\,\opcal{G}^{1a1a_2}_{\odot\,\infty <S>}(Z,\hvec{k}|Z_{21},\hvec{k}_{2};\vec{P})\nonumber\\
\lo\cdot
\opcal{P}_{\odot}^{1a_21a_1}(\hvec{k}_{2}|\hvec{k}_{1};\vec{P})\cdot\opcal{I}^{1a_1}_{\odot d}(Z_{21},\hvec{k}_{1};\vec{P})\,,\label{ApproInt}
\end{eqnarray}
where $\opcal{G}^{1a1a_2}_{\odot\,\infty <S>}$ is defined by equations \eref{defsommeG} and (\ref{defGinft3}, \ref{defGS3}).
Once we have determined the specific intensity $\opcal{I}^{1a}_{\odot\,d}(Z,\hvec{k}_{4};\vec{P})$, the scattering cross-sections are given by
\begin{eqnarray}
\fl [\op{\sigma}_{L=1}^{incoh}(\hvec{k}|\hvec{k}_{0})]+[\op{\sigma}_{Ladder}^{incoh}(\hvec{k}|\hvec{k}_{0})]=4\pi\cos \theta\nonumber\\
\lo \times \int\rmd^2 \hvec{k}_4\,\opcal{T}^{01}_{\odot<S>}(\hvec{k}|\hvec{k}_{4};\vec{P}=\vec{0})\cdot\opcal{I}^{1+}_{\odot\,d}(Z=0,\hvec{k}_{4};\vec{P}=\vec{0}) \,,\nonumber\\\fl\label{ApproLadder}
\\
\fl [\op{\sigma}_{Crossed}^{incoh}(\hvec{k}|\hvec{k}_{0})]=4\pi|\cos\theta_0|\,\nonumber\\
\fl \times \Big[\int \rmd^2\hvec{k}_4\,\opcal{T}^{01}_{\odot<S>}(-\hvec{k}_0|\hvec{k}_{4};\vec{P})
\cdot\opcal{I}^{1+}_{\odot d}(Z=0,\hvec{k}_{4};\vec{P}) \Big]^{T_R}-\frac{\cos\theta_0}{\cos\theta}[\op{\sigma}_{L=1}^{incoh}(-\hvec{k}_0|\hvec{k}_{0};\vec{P})]\,,\label{ApproCrossed}
\end{eqnarray}
with $\vec{P}=\hvec{k}_{\perp}+\hvec{k}_{0\perp}$, and $[\op{\sigma}_{L=1}^{incoh}(\hvec{k}|\hvec{k}_{0};\vec{P})]$ is obtained from  equation 
\eref{Chap4ordreL=1} in replacing the term 
\begin{equation}
 \frac{\left(1-\e^{-(a_2|\sec \theta_2|-a_1|\sec \theta_1|)\kappa_e\,H}\right)}{(a_2|\sec\theta_2|-
 a_1|\sec \theta_1|)\kappa_e} \,
\end{equation}
by
\begin{equation}
 \frac{\left(1-\e^{-a_2(\kappa_e+\rmi\hvec{k}_2\cdot\vec{P})|\sec \theta_2|\,H+a_1(\kappa_e+\rmi\hvec{k}_1\cdot\vec{P})|\sec \theta_1|\,H}\right)}{a_2(\kappa_e+\rmi\hvec{k}_2\cdot\vec{P})|\sec \theta_2|-a_1(\kappa_e+\rmi\hvec{k}_1\cdot\vec{P})|\sec \theta_1|} \,.
\end{equation}
It must be emphasized that the specific intensity $\opcal{I}^{1a}_{\odot\,d}(Z,\hvec{k};\vec{P})$ is not a Stokes vector but a Muller
matrix. This choice was mandatory since the right transpose $^{T_R}$ is defined only for
a Muller matrix. However, we can easily transformed the Muller matrix $\opcal{I}^{1a}_{\odot\,d}(Z,\hvec{k};\vec{P})$ into a Stokes vector $\opcal{ I}^{1a}_{Stokes\,d}(Z,\hvec{k};\vec{P})$ with the following definition :
\begin{equation}
\opcal{ I}^{1a}_{Stokes\,d}(Z,\hvec{k};\vec{P})=\opcal{ I}^{1a}_{\odot\,d}(Z,\hvec{k};\vec{P})\cdot\opcal{I}^{0i}_{\odot}(\hvec{k}_0)\,,
\end{equation}
where $\opcal{I}^{0i}_{\odot}(\hvec{k}_0)$ is the Stokes vector of the incident wave:
\begin{equation}
\opcal{ I}^{0i}_{\odot}(\hvec{k}_0)=\frac{\ep_{vac}c_{vac}n_0}{2}\,\vec{E}^{0i}(\hvec{k}_0)\odot\vec{E}^{0i\,*}(\hvec{k}_0)\,.
\end{equation}
It is easy to see that the radiative transfer equation \eref{ApproRad}, the boundaries condition (\ref{ApproBound1}, \ref{ApproxBound2}), the integral equation \eref{ApproInt} and the scattering cross-section \eref{ApproLadder} can be written only in function of the Stokes vector $\opcal{I}^{0i}_{\odot}(\hvec{k}_0)$. However, this cannot be done for the most-crossed contribution \eref{ApproCrossed} due to  the right transpose.
\section{Link with the scalar approach}
In most of the papers devoted to the enhanced backscattering phenomenon, the polarization and
the index mismatch between the scattering medium and the surrounding medium is often neglected~\cite{Bara1,Mark,Mark2,Tsang2,Tsang3,Tsang4,Akkermans,Goro1,Goro2}. We can recover 
the expression obtained in this case by replacing the vectorial components with scalars and the product $\odot$ with the ordinary product between two complex numbers ($\op{f}\odot\op{g}\to f\,g^*$). Without a permittiviy gradient between the scattering  and the surrounding media $(\ep_0=\ep_1=\ep_2)$ and if we suppose
that the random medium is sparse ($\ep_e\simeq\ep_0$), we can neglect the reflexion at the boundaries, and we have
\begin{eqnarray}
\opcal{S}_{\odot\,<S>}^{0+;1a_2}(\hvec{k}|\hvec{k}_2) &=\delta_{a_2,+}\,\delta(\hvec{k}-\hvec{k}_2)\quad \Longrightarrow \quad &\cos \theta =\cos \theta_2 \,,\\
\opcal{S}_{\odot\,<S>}^{1a_1;0-}(\hvec{k}_1|\hvec{k}_0) &= \delta_{a_1,-}\,\delta(\hvec{k}_1-\hvec{k}_0)\quad \Longrightarrow \quad &\cos \theta_1 =\cos \theta_0 \,.
\end{eqnarray}
For isotropic scatterers,  and in  neglecting  the correlations between the particles, the scattering operators for one particle are constant and are given, following equations (\ref{frelt11}, \ref{eqC11-3}, \ref{opPodot}) by
\begin{eqnarray}
&& \op{t}^{11}_{o}(\vec{k}|\vec{k}_0)=4\pi\,\op{f}(\hvec{k}|\hvec{k}_0)=4\pi\,f\,,\\
&& \op{C}_{o}^{11}(\vec{k}|\vec{k}_0) \simeq  \op{t}^{11}_{o}(\vec{k}|\vec{k}_0)\,,\\
&& \opcal{P}_{\odot}^{1a_21a_1}(\hvec{k}_2|\hvec{k}_1)\simeq \frac{n\,|\op{C}_{o}^{11}(\vec{k}|\vec{k}_0)|^2}{(4\pi)^2}=n\,|f|^2\,.
\end{eqnarray}
The effective permitivity $\ep_e$ and the effective constant $K_e^2=\ep_e\,K^2_{vac}$ are given by  equation \eref{eff-2}:
\begin{equation}
K_e^2=K_0^2+4\pi\,n\,f\,,\quad \mbox{and then}\quad
K_e\simeq K_1+\frac{2\pi\,n\,f}{K_0}\,.
\end{equation}
With this effective wavevector $K_e$, we can obtain the extinction coefficient $\kappa_e=2\,\Im K_e$.
With these approximations and by introducing $\mu_s=\cos \theta$, $\mu_i=|\cos \theta_0|$, and the albedo of one particle $a$ by
\begin{equation}
a=4\pi\,n|f|^2/\kappa_e\,,
\end{equation}
the single scattering contribution~\eref{Chap4ordreL=1bis} by the random medium is
\begin{equation}
\fl \gamma^{incoh}_{L=1}(\mu_s,\mu_i)=\frac{\sigma_{L=1}^{incoh}(\mu_s,\mu_i)}{\mu_i}=\frac{a\,\mu_s}{\mu_s+\mu_i}\,\left\{1-\exp\left(-\kappa_e\,H\left[\frac{1}{\mu_s}+\frac{1}{\mu_i}  \right] \right) \right\}\,,\label{ScalL=1}
\end{equation}
where we have introduced the bistatic coefficient $\gamma^{incoh}_{L=1}$ which is the scattering cross-section $\sigma_{L=1}^{incoh}(\mu_s,\mu_i)$ divided by the cosinus of the incident angle. 
Similarly, we derive the contribution for the ladder term  from  equation \eref{Chap4ordreLadderbis} (where $\cos \theta_4=\cos \theta$, $\cos \theta_1=\cos \theta_0$):
\begin{eqnarray}
\fl \gamma^{incoh}_{Ladder}(\mu_s,\mu_i) &=\frac{\sigma_{Ladder}^{incoh}(\mu_s,\mu_i)}{\mu_i}\\
&=\frac{1}{\mu_i}\,\int_{-\kappa_eH}^0\,\rmd \tau_{1}\,\rmd \tau_{2}\,\Gamma(\tau_1,\tau_2,\alpha=0)\,\exp\left(-\left[\frac{|\tau_1|}{\mu_s}+\frac{|\tau_2|}{\mu_i} \right] \right)\,,\label{ScalLadder}
\end{eqnarray}
where we have defined $\tau_{1}=\kappa_e\,Z_{43}$, $\tau_{1}=\kappa_e\,Z_{21}$, $\alpha=||\vec{P}||/\kappa_e$, and
\begin{equation}
\fl \Gamma(\tau_1,\tau_2,\alpha)=(4\pi)\,a^2\,\sum_{a_3,a_2=\pm}\int_{a_3,a_2}\!\rmd^2\hvec{k}_3\,\rmd^2\hvec{k}_2\,\opcal{G}^{1a_31a_2}_{\odot\ll SV\gg}(Z_{43},\hvec{k}_3|Z_{21},\hvec{k}_2;\vec{P})\,,
\end{equation}
From  the integral equation \eref{Betheeq5-2} on $\opcal{G}^{1a_31a_2}_{\odot\ll SV\gg}(Z_{43},\hvec{k}_3|Z_{21},\hvec{k}_2;\vec{P})$, we obtain an integral equation on $\Gamma(\tau_1,\tau_2,\alpha)$:
\begin{equation}
\fl\Gamma(\tau_1,\tau_2,\alpha)=\frac{a^2}{2}\,W(|\tau_1-\tau_2|,\alpha)+\frac{a}{2}\,\int_{-\kappa_eH}^0
\rmd \tau'\,W(|\tau_1-\tau'|,\alpha)\,\Gamma(\tau',\tau_2,\alpha)\,,
\end{equation}
where the function $W$ is defined by
\begin{equation}
\fl W(|\tau_1-\tau_2|,\alpha)=(8\pi)\sum_{a_3,a_2=\pm}\int_{a_3,a_2}\!\rmd^2\hvec{k}_3\,\rmd^2\hvec{k}_2\,\opcal{G}^{1a_31a_2}_{\odot\infty<S>}(Z_1,\hvec{k}_3|Z_2,\hvec{k}_2;\vec{P})\,.
\end{equation}
As we suppose that we do not have any reflections at the interfaces between the scattering medium and the surrounding medium, we have 
$\opcal{S}^{1a1a_0}_{\odot<S>}(\hvec{k}|\hvec{k}_0)\simeq 0$, and 
we deduce from equation (\ref{defsommeG}-\ref{defGS3}) that
\begin{equation}
\opcal{G}^{1a_31a_2}_{\odot\infty<S>}(Z_1,\hvec{k}_3|Z_2,\hvec{k}_2;\vec{P})=\opcal{G}^{1a_31a_2}_{\odot\infty}(Z_1,\hvec{k}_3|Z_2,\hvec{k}_2;\vec{P})\,.
\end{equation}
In neglecting the polarization in $\opcal{G}^{1a_31a_2}_{\odot\infty}$, the function $W(|\tau_1-\tau_2|,\alpha)$ becomes 
\begin{equation}
 \fl W(|\tau_1-\tau_2|,\alpha)=\sum_{a_2=\pm}\int\rmd^2\hvec{k}_2 \frac{\delta_{a_2,\sgn(\tau_1-\tau_2)}}{2\pi|\cos\theta_2|}\,\exp\left(-a_2\left(\kappa_e+\rmi\hvec{k}_2\cdot\vec{P}\right)(\tau_1-\tau_2)/\kappa_e|\cos \theta_2| \right)\,,\label{ScalInt}
\end{equation}
with 
\begin{equation}
\hvec{k}_2=\begin{vecteur}
\sin\theta_2\,\cos \phi_2\\
\sin \theta_2\,\sin \phi_2\\
\cos \theta_2
\end{vecteur}\,.
\end{equation}
Since the vector $\vec{P}$ is either the null vector for ladder contribution or $\vec{P}=(\hvec{k}_{0\perp}+\hvec{k}_{\perp})/2$ for the crossed term, then 
$\vec{P}$ do not have a vertical component, and $\vec{P}\cdot\hvec{k}_2=\vec{P}\cdot\hvec{k}_{2\perp}$. We define angle $\phi_2$ such that $\vec{P}\cdot\hvec{k}_{2\perp}=||\vec{P}|| ||\hvec{k}_{2\perp}||\cos \phi_2$.
Furthermore, we have $||\hvec{k}_{2\perp}||=\sin \theta_2=\sqrt{1-\cos^2\theta_2}$,
and finally we obtain:
\begin{equation}
\fl W(|\tau_1-\tau_2|,\alpha)=\int_{0}^1
\frac{\rmd \mu_2}{\mu_2}\,\exp\left(-\kappa_e|\tau_1-\tau_2|/\mu_2 \right)\,J_0\left(\alpha\,|\tau_1-\tau_2|\sqrt{1-\mu_2^2}/\mu_2\right)\,,
\end{equation}
where $\mu_2=|\cos \theta_2|$, and $J_0(x)$ is the Bessel function of order zero.

We can also express the crossed term contribution as a function of $W(|\tau_1-\tau_2|,\alpha)$ with the help of  equation \eref{Chap4ordreCrossedbis}. (In this case, we have $\cos \theta=|\cos \theta_4^+|=|\cos\theta_1^-|$, $\cos \theta_0=|\cos \theta_4^-|=|\cos\theta_1^+|$):
\begin{eqnarray} \fl\gamma^{incoh}_{Crossed}(\mu_s,\mu_i)&=\frac{\sigma_{Crossed}^{incoh}(\mu_s,\mu_i)}{\mu_i}\nonumber\\
&=\frac{1}{\mu_i}\,\int_{-\kappa_eH}^0\,\!\!\rmd \tau_{1}\,\rmd \tau_{2}\,\Gamma(\tau_1,\tau_2,\alpha)\,\exp\left(\rmi \frac{K_0}{\kappa_e}(\mu_i-\mu_s)(|\tau_1|-|\tau_2|) \right)\nonumber\\
& \qquad\times\exp\left(-\frac{1}{2}\left[ \frac{1}{\mu_s}+\frac{1}{\mu_i}\right](|\tau_1|+|\tau_2|) \right)\,,\\
&=\frac{1}{\mu_i}\,\int_{-\kappa_eH}^0\,\!\!\rmd \tau_{1}\,\rmd \tau_{2}\,\Gamma(\tau_1,\tau_2,\alpha)\,\cos\left(\rmi \frac{K_0}{\kappa_e}(\mu_i-\mu_s)(|\tau_1|-|\tau_2|) \right)\nonumber\\
& \qquad\times\exp\left(-\frac{1}{2}\left[ \frac{1}{\mu_s}+\frac{1}{\mu_i}\right](|\tau_1|+|\tau_2|) \right)\,,\label{ScalCrossed}
\end{eqnarray}
where to obtain the last equality, we have used the fact that $\gamma^{incoh}_{Crossed}$ must be real, and  $\Gamma(\tau_1,\tau_2,\alpha)=\Gamma^*(\tau_1,\tau_2,\alpha)$.
The factor $\alpha$ for the crossed contribution is given by
\begin{eqnarray}
\fl \alpha &=\frac{||\vec{k}_{\perp}+\vec{k}_{0\perp}||}{\kappa_e}\,,\\
\fl &= \left[(\sin \phi \cos \theta+\sin \phi_0 \cos \theta_0)^2 +(\sin \phi \sin \theta+\sin \phi_0 \sin \theta_0)^2\right]^{1/2}/\kappa_e\,.
\end{eqnarray}
It can be easily checked that  equations (\ref{ScalL=1}, \ref{ScalLadder}, \ref{ScalInt}, \ref{ScalCrossed}) are identical to those obtained by Van der Mark et al.~\cite{Mark,Mark2} and Tsang et al.~\cite{Kong2001-3,Tsang2,Kong}. Thus, our approach is a generalization of the theory developed in the 
scalar case where the polarization, the boundaries, the finite size, and the correlations between the particles are taken into account. We  could also have shown that the equations obtained by other groups~\cite{Rossum,Bara2,Tsang3,Tsang4,Akkermans,Goro1,Goro2,Luck,Amic2}
can be recovered by our approach.
\section{Conclusion}
We have shown how to derive, starting from the Maxwell equations, the radiative transfer equation and its boundary conditions for a rough inhomogeneous slab. By using Green functions, we have obtained an integral equation, called the Bethe-Salpeter equation, on the intensity inside the slab. Then, by applying the Wigner transform to this equation and by considering the ladder approximation, we have obtained the integral 
formulation of the radiative transfer theory. The usual radiative transfer equation is straightforwardly obtained by differentiation. 

The main goal of this paper was to introduce the boundaries in the derivation of the 
radiative transfer equation. We have shown that this could be done by replacing the Green function for an infinite medium, which describes the propagation between two scattering events by particles, with the Green functions which take into account the reflexion of the waves by
the boundaries. We have, in particular, demonstrated that the ladder contribution is identical to the phenomenological radiative transfer theory and   satisfies boundary conditions that are identical to those derived from using a geometrical argument~\cite{Kong,Kong2001-3}. By starting from the Maxwell
equations, we were also able to give an unambiguous definition of the specific intensity as a function of the electric field, and thus clarify the meaning of this quantity. (See also \cite{Mish}.)
Furthermore, we  were able  to incorporate the correlations between the scatterers by multiplying the scattering operator of one particle, which appears in the radiative transfer  equation, by a structure factor which is identical to the one used to describe the scattering of X-rays by crystals. Finally, we have also incorporated the most crossed contributions in our theory to incorporate  the enhanced backscattering phenomenon. This contribution also satisfies  a radiative transfer equation and boundary conditions which are  slightly modified compared to the usual phenomenological radiative transfer theory.

\appendix
\section{Dyadic and tensorial notations}
\label{AppTensor}
To describe the intensity of the electromagnetic field, we introduce the
tensorial product $\veccal{I}=\E^1 \otimes \E^2$ of two vectors by
\begin{equation}
  \E^1\otimes \E^2= \sum_{i,i'=1}^3\,E^1_i\, E^2_{i'}\,\,\hvec{e}_i\otimes\hvec{e}_{i'}\,.\label{E1otimesE2}
\end{equation}
where each vector has been decomposed on a three-dimensional orthonormal basis $[\hvec{e}_1,\hvec{e}_2,\hvec{e}_3]$:
\begin{equation}
\E^{1,2}=\sum_{i=1}^3\,E^{1,2}_i\,\hvec{e}_i\,,\label{NotDevE12}
\end{equation}
In this paper, the basis $[\hvec{e}_1,\hvec{e}_2,\hvec{e}_3]$ is either the fixed basis
$[\ex,\ey,\ez]$, or if  we know the wave propagation direction, the basis
$[\evp{0\pm}{},\ehp{},\hvec{k}_{\p{}}^{0\pm}]$ or $[\evp{1\pm}{},\ehp{},\hvec{k}_{\p{}}^{1\pm}]$, as described in  paper I.
Then, we define the tensorial product : between two tensors $\veccal{I}^1$ and $\veccal{I}^2$
having the following decompositions:
\begin{equation}
\veccal{I}^{1,2} =\sum_{i,i'}\,\mathcal{I}^{1,2}_{ii'}\,\hvec{e}_i\otimes\hvec{e}_{i'}\,,
\end{equation}
by
\begin{equation}
\veccal{I}^1:\veccal{I}^2=\sum_{i,i'=1}^3\mathcal{I}^{1}_{ii'}\mathcal{I}^{2}_{ii'}
\end{equation}
or in a equivalent way by
\begin{equation}
(\hvec{e}_i\otimes\hvec{e}_{i'}):(\hvec{e}_k\otimes\hvec{e}_{k'})=(\hvec{e}_i\cdot\hvec{e}_k)\,(\hvec{e}_{i'}\cdot\hvec{e}_{k'})=\delta_{i,k}\delta_{i',k'}\,,
\end{equation}
where $\cdot$ is the usual scalar product between two vectors, and $\delta_{i,k}$ is
the Kronecker symbol.
We also introduce the tensorial product between two dyads $\op{f}^1$ and $\op{f}^2$ 
having the following decomposition:
\begin{equation}
\op{f}^{1,2}=\sum_{i,j}\,{f}^{1,2}_{ij}\,\hvec{e}_i\hvec{e}_j\,,\label{NotDevA12}
\end{equation}
by
\begin{equation}
\opcal{M}=\op{f}^1\otimes\op{f}^2=\sum_{i,j,i',j'}\,\mathcal{M}_{ii';jj'}\,\left(\hvec{e}_i\hvec{e}_j\otimes\hvec{e}_{i'}\hvec{e}_{j'}\right)\,,\label{A1otimesA2}
\end{equation}
with 
\begin{equation}
\mathcal{M}_{ii';jj'}={f}^{1}_{ij}\,{f}^{2}_{i'j'}\,.
\end{equation}
If we introduce the generalized dyadic notation $\left(\hvec{e}_i\otimes\hvec{e}_{i'}\right)\left(\hvec{e}_j\otimes\hvec{e}_{j'}\right)$:
\begin{equation}
\left(\hvec{e}_i\otimes\hvec{e}_{i'}\right)\left(\hvec{e}_j\otimes\hvec{e}_{j'}\right)\equiv\left(\hvec{e}_i\hvec{e}_j\otimes\hvec{e}_{i'}\hvec{e}_{j'}\right)\,,
\end{equation}
we naturally introduce the following definition for the product between the tensor  $\opcal{M}$ and the tensor $\veccal{I}$:
\begin{equation}
\opcal{M}:\veccal{I}=\sum_{j,j'=1}^3\mathcal{M}_{ii';jj'}\mathcal{I}_{jj'}\left(\hvec{e}_i\otimes\hvec{e}_{i'}\right)\,,\label{decompAcal}
\end{equation}
for
\begin{eqnarray}
\veccal{I} =\sum_{j,j'}\,\mathcal{I}_{jj'}\,\hvec{e}_j\otimes\hvec{e}_{j'}\,,\label{NotEtenseur}\\
\opcal{M}=\sum_{i,i',j,j'}\,\mathcal{M}_{ii';jj'}\left(\hvec{e}_i\otimes\hvec{e}_{i'}\right)\left(\hvec{e}_j\otimes\hvec{e}_{j'}\right)\,.\label{NotAtenseur}
\end{eqnarray}
This definition is equivalent to set
\begin{eqnarray}
\fl \left[\left(\hvec{e}_i\otimes\hvec{e}_{i'}\right)\left(\hvec{e}_j\otimes\hvec{e}_{j'}\right)\right]:(\hvec{e}_k\otimes\hvec{e}_{k'})&=
\left(\hvec{e}_i\otimes\hvec{e}_{i'}\right)\,\left[\left(\hvec{e}_j\otimes\hvec{e}_{j'}\right):(\hvec{e}_k\otimes\hvec{e}_{k'})\right]\,\nonumber\\
\fl &=\left(\hvec{e}_i\otimes\hvec{e}_{i'}\right)\delta_{j,k}\delta_{j',k'}\,.
\end{eqnarray}
Similarly, the product between two tensorial operators $\opcal{M}^1$ and $\opcal{M}^2$ having a decomposition similar to the equation \eref{NotAtenseur} is
\begin{equation}
\opcal{M}^1:\opcal{M}^2=\sum_{j,j'=1}^3\,\mathcal{M}^1_{ii';jj'}\mathcal{M}^2_{jj';i_0i_0'}\left(\hvec{e}_i\otimes\hvec{e}_{i'}\right)\left(\hvec{e}_{i_0}\otimes\hvec{e}_{i_0'}\right)\,.
\end{equation}
\section{Statistical properties of the scattering operators}
\label{AppStat}
It can be easily demonstrated, by using the uniqueness of the solution of the Maxwell equations, that the scattering operator describing an inhomogeneous slab with rough boundaries has the following invariance properties under horizontal translation~\cite{Voro}:
\begin{equation}
\op{R}_{SV}(\p{}|\p{0})\big|_{h_{1,2}(\x-\vec{d}), \rr_{i}+\vec{d}}=\e^{-\rmi(\p{}-\p{0})\cdot\vec{d}}\,\op{R}_{SV}(\p{}|\p{0})\big|_{h_{1,2}(\x),\rr_i}\,,\label{App1eq1}
\end{equation}
where the indexes $h_{1,2}(\x-\vec{d}), \rr_{i}+\vec{d}$ mean that the rough surfaces $h_1(\x)$, $h_2(\x)$ and the particle positions $\rr_i(i=1,\dots,N)$ have been translated by an horizontal vector $\vec{d}=d_x\ex+d_y\ey$.
For statistical homogeneous random medium and rough boundaries,  we have
\begin{eqnarray}
\fl \ll\op{R}_{SV}(\p{}|\p{0})\big|_{h_{1,2}(\x-\vec{d}),\rr_i+\vec{d}}\gg_{SV}=\ll\op{R}_{SV}(\p{}|\p{0})\big|_{h_{1,2}(\x),\rr_i}\gg_{SV}\,,\\
\fl \ll\op{R}_{SV}(\frac{\vec{P}}{2}+\p{}|\frac{\vec{P}_0}{2}+\p{0})\big|_{h_{1,2}(\x-\vec{d}),\rr_i+\vec{d}}\otimes \op{R}^{\,*}_{SV}(-\frac{\vec{P}}{2}+\p{}|-\frac{\vec{P}_0}{2}+\p{0})\big|_{h_{1,2}(\x-\vec{d}),\rr_i+\vec{d}}\gg_{SV}\nonumber\\
=\ll\op{R}_{SV}(\frac{\vec{P}}{2}+\p{}|\frac{\vec{P}_0}{2}+\p{0})\big|_{h_{1,2}(\x),\rr_i}\otimes \op{R}^{\,*}_{SV}(-\frac{\vec{P}}{2}+\p{}|-\frac{\vec{P}_0}{2}+\p{0})\big|_{h_{1,2}(\x),\rr_i}\gg_{SV}\,,\nonumber \\
\end{eqnarray}
whatever the vector $\vec{d}$.
In using  equation \eref{App1eq1}, we find
\begin{eqnarray}
\fl(1-\e^{-\rmi(\p{}-\p{0})\cdot\vec{d}})\, \ll\op{R}_{SV}(\p{}|\p{0})\big|_{h_{1,2}(\x),\rr_i}\gg_{SV}=0\,,\\
\fl (1-\e^{-\rmi(\vec{P}-\vec{P}_0)\cdot\vec{d}})\,\no\\
\fl \times \ll\op{R}_{SV}(\frac{\vec{P}}{2}+\p{}|\frac{\vec{P}_0}{2}+\p{0})\big|_{h_{1,2}(\x),\rr_i}\otimes \op{R}^{\,*}_{SV}(-\frac{\vec{P}}{2}+\p{}|-\frac{\vec{P}_0}{2}+\p{0})\big|_{h_{1,2}(\x),\rr_i}\gg_{SV}
=0\,.\nonumber \\
\end{eqnarray}
These equalities can be satisfied, whatever the value of $\vec{d}$, only if the average
scattering operators are proportional to  Dirac distributions:
\begin{eqnarray}
\fl \ll \op{R}_{SV}(\p{}|\p{0})_{h_{1,2}(\x),\rr_i}\gg_{SV}=(2\pi)^2\delta(\p{}-\p{0})\,\op{R}_{\ll SV\gg}(\p{0})\,\\
\fl \ll\op{R}_{SV}(\frac{\vec{P}}{2}+\p{}|\frac{\vec{P}_0}{2}+\p{0}\Big|_{h_{1,2}(\x),\rr_i}\otimes \op{R}^{\,*}_{SV}(-\frac{\vec{P}}{2}+\p{}|-\frac{\vec{P}_0}{2}+\p{0})\Big|_{h_{1,2}(\x),\rr_i}\gg_{SV}\no\\
\lo =(2\pi)^2\delta(\vec{P}-\vec{P}_0)\,\opcal{R}_{\otimes\ll SV\gg}(\p{}|\p{0};\vec{P})\,.
\end{eqnarray}
In particular, we have for $\vec{P}=\vec{P}_0$:
\begin{eqnarray}
\fl \ll\op{R}_{SV}(\frac{\vec{P}}{2}+\p{}|\frac{\vec{P}}{2}+\p{0}\otimes \op{R}^{\,*}_{SV}(-\frac{\vec{P}}{2}+\p{}|-\frac{\vec{P}}{2}+\p{0})\gg_{SV}\no\\
\lo 
=(2\pi)^2\delta(0)\,\opcal{R}_{\otimes\,\ll SV\gg}(\p{}|\p{0};\vec{0})
\end{eqnarray}
Notice that if we suppose that the illuminated  surface  has a finite area A, then $\delta(0)$ must be defined as $A/(2\pi)^2$.
\section{Mass operator and Dyson equation}
\label{AppDyson}
In paper I, we have bypassed the introduction of the mass operator and the Dyson equation since we have introduced from the beginning the effective permittivity in the definition of the Green function $\op{G}_{S}$. This mass operator is defined as
\begin{eqnarray}
\fl\nabla\times\nabla\times<\op{G}^{11}_{SV}(\rr,\rr_0)>_{V}-\ep_1\,K_{vac}^2<\op{G}^{11}_{SV}(\rr,\rr_0)>_{V}=\delta(\rr-\rr_0)\op{I}\no\\
\lo+\intr{1}\op{M}^{11}(\rr,\rr_1)\cdot<\op{G}^{11}_{SV}(\rr_1,\rr_0)>_{V}\,.\label{eqDyson}
\end{eqnarray}
Equation \eref{eqDyson} is referred as the Dyson equation~\cite{Kong,Apresyan,Sheng1,Sheng2,Kong2001-3,Frish}.
The exact expression of this mass operator is the sum of all the irreducible diagrams
that we can write and which represent the single scattering and all higher scattering 
process where recurrent scattering or correlation between scatterers are involved~\cite{Frish,Tig3}.
In I, we have introduced the Coherent Potential Approximation (CPA) which states that 
\begin{equation}
<\op{G}^{11}_{SV}(\rr,\rr_0)>_{V}=\op{G}^{11}_{S}(\rr,\rr_0)\,,
\end{equation}
where $\op{G}^{11}_{S}(\rr,\rr_0)$ is defined by
\begin{equation}
\fl \nabla\times\nabla\times<\op{G}^{11}_{S}(\rr,\rr_0)>_{V}-\ep_e\,K_{vac}^2<\op{G}^{11}_{S}(\rr,\rr_0)>_V=\delta(\rr-\rr_0)\op{I}\,.
\end{equation}
In the (CPA) approach, the mass operator is, thus, given by
\begin{equation}
\op{M}^{11}(\rr,\rr_0)=(\ep_e-\ep_1)\,\delta(\rr-\rr_0)\op{I}\,,
\end{equation}
where the effective permittivity $\ep_e$ satisfied  equations
(177-178) in  paper I.
\section{Transpose of a tensor}
\label{AppTran}
In  equation \eref{cont crossed} of  section \eref{secCrossed}, we carry out 
the right transpose of a dyadic operator $\opcal{M}(\p{}|\p{0})$ defined by
\begin{eqnarray}
\fl \opcal{M}(\p{}|\p{0})=\opcal{S}_{\otimes\,<S>}^{0+\,1a_4}(\Delta
\p{}|\p{4};\Ps{}):\Pcal^{1a_41 a_3}(\p{4}|\p{3};\Ps{}):\Gcal^{1a_3
  1a_2}_{\ll SV\gg}(Z_{43},\p{3}|Z_{21},\p{2};\Ps{})\nonumber
\\
:\Pcal^{1a_2
  1a_1}(\p{2}|\p{1};\Ps{}):\opcal{S}_{\otimes\,<S>}^{1a_1\,0-}(\p{1}|-\Delta \p{};\Ps{})\,,
\end{eqnarray}
where $\vec{P}=\vec{p}+\vec{p}_0$, and $\Delta \p{}=(\vec{p}-\vec{p}_0)/2$.
Since we have
\begin{eqnarray}
\vec{P}/2+\Delta \p{}=\vec{p}\,,\quad -\vec{P}/2+\Delta \p{}=-\vec{p}_0\,,\\
\vec{P}/2-\Delta \p{}=\vec{p}_0\,,\quad -\vec{P}/2-\Delta \p{}=-\vec{p}\,,
\end{eqnarray}
we deduce from the definition \eref{SwithP} and \eref{SwithP2} of  $\opcal{S}_{\otimes<S>}^{0+\,1a_4}$, $\opcal{S}^{1a_10-}_{\otimes<S>}$ that the tensor $\opcal{M}(\p{}|\p{0})$ has the following decomposition:
\begin{equation}
\fl\opcal{M}(\p{}|\p{0})=\sum_{\beta,\beta',\beta_0,\beta'_0=H,V}\mathcal{M}_{\beta\beta';\beta_0\beta'_0}(\hvec{e}_{\beta}^{0+}(\p{})\otimes\hvec{e}^{0+}_{\beta'}(-\p{0}))(\hvec{e}^{0-}_{\beta_0}(\p{0})\otimes\hvec{e}^{0-}_{\beta'_0}(-\p{}))\,.
\end{equation}
From the definition of the right transpose \eref{Chap4DefTransR} in the fixed basis $[\ex,\ey,\ez]$
 and from the following properties,
\begin{eqnarray}
\hvec{e}^{0+}_{V}(\p{})&=\hvec{e}^{0-}_{V}(-\p{})\,,\\
\hvec{e}_{H}(\p{})&=-\hvec{e}_{H}(-\p{})\,,
\end{eqnarray}
we demonstrate that
\begin{equation}
\fl [\opcal{M}(\p{}|\p{0})]^{T_R}=\sum_{\beta,\beta',\beta_0,\beta'_0=H,V}\mathcal{M}^{T_R}_{\beta\beta';\beta_0\beta'_0}(\hvec{e}_{\beta}^{0+}(\p{})\otimes\hvec{e}^{0+}_{\beta'}(\p{}))(\hvec{e}^{0-}_{\beta_0}(\p{0})\otimes\hvec{e}^{0-}_{\beta'_0}(\p{0}))\,,\label{Chap4transposeTrtens}
\end{equation}
with
\begin{eqnarray}
\mathcal{M}^{T_R}_{\beta\,\beta';\beta_0\,\beta'_0}=\mathcal{M}_{\beta\,\beta'_0;\beta\,\beta'}\quad\mbox{if} \quad\beta'=\beta_0'\,,\\
\mathcal{M}^{T_R}_{\beta\,\beta';\beta_0\,\beta'_0}=-\mathcal{M}_{\beta\,\beta'_0;\beta\,\beta'}\quad\mbox{if} \quad\beta'\neq\beta_0'\,,
\end{eqnarray}
for $\beta,\beta',\beta_0,\beta_0'=H,V$.
\section{Transpose of  Muller matrices}
\label{AppRightT}
To write the enhanced backscattering contribution with the help of Muller matrices, we need
to define the right transpose of a Muller matrix. In using the transformation $Tr^{\otimes\to\odot}$, we introduce  the following definition of the right transpose of a Muller matrix $[\op{M}]$:
\begin{equation}
[\op{M}]^{T_R}=Tr^{\otimes\to\odot}\,\{\opcal{M}^{T_R}\}\,,
\end{equation}
with
\begin{equation}
[\op{M}]=Tr^{\otimes\to\odot}\,\{\opcal{M}\}\,,
\end{equation}
and where the right transpose of the tensor $\opcal{M}$ is defined in the previous appendix.
Since there is a univocal relationship between a tensor $\opcal{M}$ and its Muller matrix
$\op{M}$, we can find the inverse $Tr^{\odot\to\otimes}$ of the transformation $Tr^{\otimes\to\odot}$:
\begin{equation}
\opcal{M}=Tr^{\odot\to\otimes}\,[\op{M}]\,,
\end{equation}
and the right transpose of a Muller matrix is
\begin{equation}
[\op{M}]^{T_R}=Tr^{\otimes\to\odot}\,\Big\{\left[Tr^{\odot\to\otimes}\,[\op{M}] \right]^{T_R}\Big\}\,.
\end{equation}
If $M_{ii'}$ with $i,i'=1,\dots,4$ are the elements of the matrix $[\op{M}]$, then we obtain
the following expression for $[\op{M}]^{T_R}$:
\begin{eqnarray}
\fl [\op{M}]^{T_R}=\nonumber\\
\fl\left(\begin{array}{llll}
M_{11} & \frac{1}{2}[M_{44}-M_{33} & \frac{1}{4}[M_{13}-M_{31} & \frac{-\rmi}{4}[M_{31}+M_{13}\\
& \quad -\rmi\,M_{34}-\rmi\,M_{43}] & \quad -\rmi\,M_{41}+\rmi\,M_{14}] & \quad +\rmi\,M_{41}+\rmi\,M_{14}]\\
\frac{1}{2}[M_{44}-M_{33} & M_{22} & \frac{1}{4}[M_{23}-M_{32} & \frac{\rmi}{4}[M_{23}+M_{32}\\
\quad+\rmi\,M_{34}+\rmi\,M_{43}] & &\quad-\rmi\,M_{24}+\rmi\,M_{42}]& \quad-\rmi\,M_{24}-\rmi\,M_{42}]\\
\frac{1}{2}[M_{31}-M_{13} & \frac{1}{2}[M_{32}-M_{23} & \frac{1}{2}[M_{33}+M_{44} & \frac{1}{2}[M_{34}-M_{43} \\
\quad+\rmi\,M_{42}-\rmi\,M_{24}] & \quad-\rmi\,M_{41}+\rmi\,M_{14}] & \quad-\rmi\,M_{12}-\rmi\,M_{21}] & \quad+\rmi\,M_{12}-\rmi\,M_{21}]\\
\frac{\rmi}{2}[M_{31}+M_{13} & \frac{-\rmi}{2}[M_{32}+M_{23} & \frac{1}{2}[M_{43}-M_{34} & \frac{1}{2}[M_{33}+M_{43} \\
\quad-\rmi\,M_{41}-\rmi\,M_{14}] & \quad+\rmi\,M_{42}+\rmi\,M_{24}] & \quad+\rmi\,M_{12}-\rmi\,M_{21}] & \quad+\rmi\,M_{12}-\rmi\,M_{21}]
\end{array}\right)\,.\nonumber
\end{eqnarray}
\section{Definition of the scattering matrix $\op{S}^{0+1\pm}_{\odot<S>}$ and  $\op{S}^{1\pm0-}_{\odot<S>}$}
\label{AppdefRotimes}
In  equations \eref{SL=1} and \eref{SL=1bis}, of section (\eref{subFirst}), we have, respectively, introduced
the factors $(\ep_0/\ep'_e)^{1/2}$ and  $(\ep_e'/\ep_0)^{1/2}$ in the definition of $\op{S}^{0+1\pm}_{\odot<S>}$ and $\op{S}^{1\pm0-}_{\odot<S>}$. These factors insure that we recover the usual expression of the phenomenological radiative transfer theory.
In fact, for  a semi-infinite medium with plane boundaries, we have
\begin{equation}
\op{S}^{0+1-}(\p{}|\p{0})=(2\pi)^2\delta(\p{}-\p{0})\,\op{t}^{01}(\p{0})\,,
\end{equation}
with 
\begin{equation}
\op{t}^{01}(\p{0})=t^{01}_V(\p{0})\,\evp{0+}{0}\evp{1-}{0}+t^{01}_H(\p{0})\,\ehp{0}\ehp{0}\,,
\end{equation}
where $t^{01}_V(\p{0})$ and $t^{01}_H(\p{0})$ are, respectively, the Fresnel coefficients in transmission for
the polarization $TM$ and $TE$.
From  equations \eref{SL=1}, \eref{relaDirac} and \eref{Defalpcos}, we deduce that
\begin{equation}
\opcal{S}^{0+;1-}_{\odot\,<S>}(\hvec{k}|\hvec{k}_{0})=\delta(\hvec{k}-\hvec{k}_0)\frac{\ep_0}{\ep_e'}\,\op{T}_{\odot}^{0+1-}(\p{})
\end{equation}
where
\begin{equation}
\op{T}^{0+1-}_{\odot}(\p{})=\frac{\sqrt{\ep_0}\cos \theta}{\sqrt{\ep'_e}\cos \theta_0}\,\op{t}^{01}(\p{})\odot\op{t}^{01}(\p{})\,,
\end{equation}
and
\begin{eqnarray}
\p{}=K_0\,\hvec{k}_{\perp}\,,\quad \cos\theta=\hvec{k}\cdot\ez\,,\\
\p{}=K_e'\,\hvec{k}_{0\,\perp}\,,\quad \cos\theta_0=\hvec{k}_0\cdot\ez\,.
\end{eqnarray}
In particular, for the polarization $V\to V$ and $H\to H$, we have
\begin{eqnarray}
\op{T}^{0+1-}(\p{})_{VV;VV}=\frac{\sqrt{\ep_0}|\cos \theta|}{\sqrt{\ep'_e}\cos \theta_0}\,|\op{t}^{01}_V(\p{})|^2\,,\\
\op{T}^{0+1-}(\p{})_{HH;HH}=\frac{\sqrt{\ep_0}\cos \theta_0}{\sqrt{\ep'_e}\cos \theta_0}\,|\op{t}^{01}_H(\p{})|^2\,.
\end{eqnarray}
The terms $\op{T}^{0+1-}(\p{})_{VV;VV}$ and $\op{T}^{0+1-}(\p{})_{HH;HH}$ are the usual
transmission coefficients of intensity for the polarization (TM) and (TE)~\cite{Ishi1,Kong2001-1,Kong}.  
Thus, for an incident wave in  medium 0 characterized by the Stokes vector $[\op{I}^{0i}]$, the Stokes vector $[\op{I}^{0s}]$ of the transmitted wave by the plane surface is
\begin{eqnarray}
{I}^{0s}_V(Z=0,\hvec{k})=\frac{\ep_0}{\ep'_e}\op{T}_{\odot}^{0+1-}(\p{})_{VV;VV}{I}^{0i}_V(Z=0,\hvec{k}_0)\,,\\
{I}^{0s}_H(Z=0,\hvec{k})=\frac{\ep_0}{\ep'_e}\op{T}_{\odot}^{0+1-}(\p{})_{HH;HH}{I}^{0i}_H(Z=0,\hvec{k}_0)\,.
\end{eqnarray}
The factor $\ep_0/\ep_e'$ accounts for the spreading of the solid angle of the specific intensity transmitted from the medium 1 (with the permittivity $\ep'_e$) to  medium 0 (with permittivity $\ep_0$)~\cite{Ishi1,Kong2001-1}. Hence, we see that the introduction of the 
term $(\ep_0/\ep'_e)^{1/2}$ in definition \eref{SL=1} was mandatory to recover the usual 
expression of the transmission coefficients.
The same reasoning can be applied to justify the introduction of the factor $(\ep_e'/\ep_0)^{1/2}$ in the definition of $\op{S}^{1\pm0-}_{\odot<S>}$.
\section*{References}
\bibliographystyle{Wavesunsrt}
\bibliography{Livres_Bib,Articles_Bib}

\end{document}